\documentclass[12pt, preprint]{aastex}

\shorttitle{Physical Properties O-type Stars}
\shortauthors{Massey et al}
\slugcomment{ApJ, in press}

\begin{document}

\title{The Physical Properties and Effective Temperature Scale of O-type Stars as a Function of Metallicity. III. More Results from the Magellanic Clouds\altaffilmark{1}}

\author{
Philip Massey and Amanda M. Zangari\altaffilmark{2}}
\affil{Lowell Observatory, 1400 W Mars Hill Road, Flagstaff, AZ 86001;
phil.massey@lowell.edu, azangari@mit.edu}

\author{Nidia I. Morrell}
\affil{Las Campanas Observatory, The Carnegie Observatories, Colina El Pino s/n, Casilla 601, La Serena, Chile; nmorrell@lco.cl}

\author{Joachim Puls}

\affil{Universit\"{a}ts-Sternwarte M\"{u}nchen, Scheinerstr.\ 1, 81679 
M\"{u}nchen, Germany; uh101aw@usm.uni-muenchen.de}

\author{Kathleen DeGioia-Eastwood} 
\affil{Department of Physics and Astronomy, Northern Arizona University,
P.O. Box 6010, Flagstaff, AZ 86011-6010; kathy.eastwood@nau.edu}

\author{Fabio Bresolin and Rolf-Peter Kudritzki}
\affil{Institute for Astronomy, 2680 Woodlawn Drive, Honolulu, HI 26822-1839; bresolin@ifa.hawaii.edu, kud@ifa.hawaii.edu}

\altaffiltext{1}{This paper is based on data gathered with the 6.5 meter Magellan telescopes located at Las Campanas Observatory, Chile, and also on observations made with the NASA/ESA {\it Hubble Space Telescope,} obtained from
the Data Archive at the
Space Telescope Science Institute (STScI), which is operated by the Association of Universities for Research in Astronomy, Inc.,
under NASA contract NAS 5-26555.  These observations are associated with programs 9412, 9795, and 11270.}
\altaffiltext{2}{Astronomy Department, Wellesley College, Wellesley, MA 024871. Research Experiences for Undergraduates (REU) participant, Summer 2007.  Current address: Department of Earth, Atmospheric, and Planetary
Sciences, Massachusetts Institute of Technology, Cambrdige, MA 02139.}

\begin{abstract}
In order to better determine the physical properties of hot, massive
stars as a function of metallicity, we obtained very high SNR optical
spectra of 26 O and early B stars in the Magellanic Clouds.  These
allow accurate modeling even in cases where the He I $\lambda$4471
line has an equivalent width of only a few tens of m\AA. The spectra
were modeled with FASTWIND, with good fits obtained for 18 stars; 
the remainder show signatures of being binaries. 
We include stars in common to 
recent studies to investigate possible systematic differences.
The ``automatic" FASTWIND modeling
method of Mokiem and collaborators produced temperatures 1100~K
hotter on the average, presumably due to the different emphasis given
to various temperature-sensitive lines. More significant, however,
is that the automatic method always produced some ``best" answer,
even for stars we identify as composite (binaries).  The temperatures
found by the TLUSTY/CMFGEN modeling of Bouret, Heap, and collaborators
yielded temperatures 1000 K cooler than ours, on average.  Significant
outliers were due either to real differences in the data (some of
the Bouret/Heap data were contaminated by moonlight continua) or
the fact we could detect the He~I line needed to better constrain
the temperature. Our new data agrees well with the effective temperature
scale we presented previously. We confirm that the ``Of" emission-line
do not track luminosity classes in the exact same manner as 
in Milky Way stars. We revisit
the the issue of the ``mass discrepancy", finding that some of the
stars in our sample do have spectroscopic masses that are significantly smaller
than those derived from stellar evolutionary models. We do not find
that the size of the mass discrepancy is simply related to either
effective temperature or surface gravity.
\end{abstract}


\section{Introduction}

The highest mass stars spend their main-sequence lifetimes as O-type
dwarfs, giants, and supergiants, and as early B-type giants and 
supergiants\footnote{Unlike the case for stars of lower mass, ``dwarf" and 
``main-sequence" are not synonymous terms, as high mass stars will become O and B supergiants
while still on the hydrogen-burning main-sequence.}.
In order to determine their physical properties 
(effective temperatures, bolometric luminosities, stellar radii, surface gravities, etc.) we must rely upon modeling
their spectra, as these stars are so hot that their optical and UV colors have only slight dependence on their physical properties, even effective temperature (see Massey 1998a, 1998b).

The stellar atmospheres of such stars are physically complex, and producing realistic synthetic spectra requires  the
inclusion of NLTE, as well as careful treatment of the hydrodynamics of the stellar wind.  The recent inclusion of
line blanketing in these models has led to a significant lowering of the effective temperature scale for Galactic
O-type stars (Martins et al.\ 2002; Bianchi \& Garcia 2002; Repolust et al.\ 2004; Martins et al.\ 2005).  Such a change has a strong
effect on the expected amount of ionizing radiation for stars of a given spectral type; 
it also has (as yet largely unconsidered) implications for the
previous determinations of the ages of young Galactic clusters and OB associations,
and stellar evolutionary theory, as the theoretical zero-age main-sequence (ZAMS) may now be too hot
for the locations of stars in the H-R diagram. 

In two previous papers (Massey et al.\ 2004, 2005, hereafter, Papers I and II, respectively) we used {\it HST} and ground-based
optical and UV data to determine the effective temperatures of O stars in the LMC and SMC.  Included in this sample were
many O stars of the earliest type (O2-O4).  Our expectations were that since the metallicity is lower in the SMC and LMC
($Z/Z_\odot=0.2$ and $Z/Z_\odot=0.5$; Russell \& Dopita 1990 and discussion
in Paper I)
than in the Milky Way, the effects of stellar wind-blanketing and line-blanketing would be less, with the result that the
effective temperatures of O stars would be hotter in the Clouds than in their Galactic counterparts. Indeed, this was
what we found: SMC stars of spectral type O3-7~V were found to be 4000 K hotter than Galactic stars of the same
spectral type, with stars of slightly later type showing increasingly smaller effects until B0~V, at which type there was
no discernible effect.   The supergiants show a similar effect. The effective temperatures of LMC O stars, for which the metallicity
is intermediate between that of the SMC and the Milky Way, are
between that of the SMC
and Milky Way.  In contrast, Heap et al.\ (2006) find relatively low effective temperatures for a sample of SMC O stars. (Their study included
their earlier work, presented in Bouret et al.\ 2003.) In Paper~II we argue that this may have been due to a lack of nebular subtraction in their short-slit Echelle
data, particularly for stars
in NGC~346, one of the strongest H~II regions in the SMC, and/or a consequence
of the ``sky-subtraction" method used for their data as described by Walborn
et al.\   (2000), which would result in incorrect continuum levels being assigned.

A necessary input in such modeling efforts is the terminal velocity of the stellar wind, which we measure primarily from the
CIV $\lambda 1550$ doublet. By the time we finished Papers I and II, the astronomical community had lost its only
resource for such work, the Space Telescope Imaging Spectrograph (STIS) on {\it HST}, and it appeared that it would be
many years before such data could be obtained again\footnote{The resurrection of STIS is included on the agenda
for Servicing Mission 4, currently due to be launched in October 2008.}. However, we found that there were a significant
number of Magellanic Cloud O-type stars observed with STIS UV data available in the archives, but which had not
been included in Papers I or II.  

We therefore decided to obtain very high signal-to-noise ratio (SNR) optical data on this remaining subset of stars, and to determine
the physical properties of these stars via modeling with FASTWIND (Santolaya-Rey et al.\ 1997; Puls et al.\ 2005), performing a similar analysis to that of Papers I and II.  
 The sample not only included some of the NGC~346 stars analyzed by Bouret et al.\ (2003) and Heap et al.\ (2006),
allowing us to resolve that controversy, but also stars of spectral subtypes poorly represented in Papers~I and II.
Our analysis here will follow closely that of Papers I and II.  We use archival STIS  UV spectra to determine the
star's terminal velocity $v_{\infty}$.  
In order to provide a consistency check, we also decided to re-observe two of the stars analyzed in Papers I and II in
order to see how reproducible our answers are given slight differences in instrumentation, rectification, etc.
We describe our data and reduction procedures in \S\ref{Sec-data}.  We present our analysis in \S\ref{Sec-analysis}.
In \S\ref{Sec-results} we give our results, and compare our work with those of others, and incorporate the new
sample into our effective temperature scales.

\section{Observations and Reductions}
\label{Sec-data}

We list in Table~\ref{tab:sample} the 26 stars selected for study.  
These stars were chosen primarily as they were of
O-type, and because they (nominally) had UV spectra in the {\it HST} archives.  We included some stars from previous
studies, including two stars from our own work as a consistency check. Not all 
stars were successfully modeled, for reasons we will describe below.

The spectra used in this study were obtained by 
PM, NIM, and KDE on UT 2004 November 27-29 using the Boller and Chivens Spectrograph on the Clay 6.5-m (Magellan II) telescope on Las Campanas.
The instrument uses a 2048 x 515 Marconi CCD with 13.5 $\mu$m pixels. 
On the first two nights observations
were made in the blue, with a 1200 line mm$^{-1}$ grating blazed at 4000 \AA.  The wavelength coverage extended from 3410-5040 \AA.  
We used a 1" slit (350 $\mu$m) to achieve a spectral resolution of 2.4 \AA\ (3.0 pixels).  The resolution
degraded significantly short-wards of 3500 \AA,  and we do not use those data. On the third night observations were 
made in the red, with a 1200 line mm$^{-1}$ grating blazed at 7500 \AA.  The wavelength coverage was 5315-6950 \AA,
with similar resolution.  No blocking filter was used with either grating.

For the blue exposures, which include most of the weak absorption lines we used in our modeling,
we took a series of three consecutive exposures for each target, aiming for a total S/N of 600 per spectral resolution element at 4500 \AA\ (i.e., each of the three exposures had a S/N of 350.)
To achieve this, we adjusted our exposure times so that we observed a $B=13.0$ mag star for a total of 600 s (i.e., 3x200 s exposures).   For the red, we were
satisfied with a slightly lower S/N (since we were primarily interested in H$\alpha$) and aimed for a total S/N of 400 per spectral resolution element
at 6500 \AA, which we achieved in an 3x200s at $V=13.0$. 
Each observation was obtained at the parallactic angle, as the instrument was used without an atmospheric dispersion corrector.

The CCD data were reduced using IRAF\footnote{IRAF is  
distributed by the National Optical Astronomy Observatory, which
is operated by the Association of Universities for Research in Astronomy, Inc., under
cooperative agreement with the National Science Foundation (NSF). We are grateful to the on-going support of IRAF and the help ``desk" maintained by the volunteers at
http://www.iraf.net.}.  The images
 were trimmed and a bias level subtracted using the overscan.  Nine bias exposures (zero-second exposures) were averaged,
and were used to remove any residual two-dimension structure, although none was evident in our examination of the images.  Nine flat-field exposures 
were obtained each night using the projector flat, with a count level below saturation but whose total number of counts was such that the SNR
of our program data was not degraded.  Twilight sky exposures were used to correct for the slight difference in the illumination function along the slit
between the program data and the dome flats.   A series of long and short flat field exposures were used to construct a bad pixel map for the
CCD, and the data linearly interpolated over bad columns, etc.

Next, the stellar spectra were extracted (following Massey et al.\ 1992) using variance weighting across the spatial direction.  The 
``clean" function was used in order to recognize and remove cosmic rays at this point.  
 A series of long comparison (HeNeAr) arc exposures obtained at the beginning of
 the night was used for the wavelength correction; tests showed that
there was very little flexure within the instrument throughout the night as the instrument rotated about the optical axis at the Nasmyth focus.

Getting the rectification (normalization) ``right" is important for O stars, particularly for weak lines and
for the wings of the Balmer lines. Each individual spectrum was normalized before combining, with the blue data
first clipped below 3800 \AA\ to remove the
effect of the Balmer jump on the normalization.  The normalization was
done interactively using a seventh order cubic spline, with care taken to examine
the regions around the principle classification/modeling lines.  The three normalized
spectra obtained in each wavelength region for each star were then averaged using
an algorithm that rejected deviant pixels.

\section{Analysis}
\label{Sec-analysis}

\subsection{Spectral Classification}

We began by first classifying the stars as
to spectral type and luminosity class.  The spectral types were  determined 
qualitatively
by comparing the spectra with standards from Walborn \& Fitzpatrick (1990).  A quantitative determination 
of the types was also obtained  by measuring the EWs (by direct integration)
of He I $\lambda 4471$ and He II $\lambda 4542$, and calculating 
$\log W'=\log \frac {W ({\rm He I} \lambda 4471)}{W ({\rm He II} \lambda 4542)}$.  
This $\log W'$ value was compared against the calibration of Conti (1988).  

As we discussed in Paper I, we 
expect our ability to detect and measure faint lines to be limited by our spectral resolution $r$ and
the SNR: $\sigma_{\rm EW}=r/{\rm SNR}$.  For $r=$2.4~\AA, and a SNR of 600, we expect the
uncertainty to be about 4~m\AA.  Thus, we should be able to detect lines at the 3$\sigma$ level
of 12~m\AA.  In practice, our measurement of the EWs (which we use only as a secondly check on 
the spectral classification) is limited more by the exact rectification, and is typically 20~m\AA, although in some cases is better than this.

For Milky Way stars, the luminosity class is determined primarily by the ``f" characteristics in the spectra:
a luminosity class ``V" star (dwarf) will have NIII $\lambda \lambda 4634, 42$ in weak emission, but 
He II $\lambda 4686$ in strong absorption, designated by the spectroscopic notation ``((f))".
  A luminosity class ``III" star (giant) will have the NIII lines in somewhat
stronger emission and
the He II $\lambda 4686$ line partially filled in by emission, denoted by the spectroscopic notation
``(f)".   A luminosity class ``I" star (supergiant) will have
both 
the NIII $\lambda \lambda 4634, 42$ lines and He II $\lambda 4686$ line in strong emission, denoted by
the spectroscopic notation ``f".   Thus these classes
would be designated somewhat redundantly as ``V((f))", ``III(f)", and ``If".    The situation is a little more
complicated than that for the later O stars, as the ``((f))" characteristic doesn't manifest itself until luminosity class III,
and by O9 none of these are found in emission and instead one uses the relative strengths of Si IV $\lambda \lambda 4116$
to He I $\lambda 4121$ assign luminosity classes.

The difficulty with extending this to lower metallicity is the following.  As first noted Mihalas et al.\ (1972),
 the NIII $\lambda \lambda 4634, 42$ lines  go into emission
as a result of dielectric recombination, while the He II $\lambda 4686$  emission is formed in the stellar wind.  Thus, at lower
metallicity we might expect that the NIII $\lambda \lambda 4634, 42$ emission formation will take place at about the same
physical conditions, but that the He II $\lambda 4686$ emission will not---at a given effective temperature a higher luminosity
will be needed for the same stellar wind strength and He II $\lambda 4686$ emission.  Thus, in the SMC, it is {\it expected}
on simple theoretical grounds to have ``I(f)" supergiants while in the Milky Way stars with the same basic properties (effective
temperatures, surface gravity, and luminosity) would result in a spectral appearance of ``If".   This was certainly what we found
in Papers I and II.  Additionally, subtle  luminosity/spectral type criteria, such as the presence of Si IV $\lambda \lambda 4089, 16$
(denoted by some by the notation ``f+") go into emission due to selective emission mechanisms, and their behavior may also
differ with spectral type from what we are used to from Milky Way examples.
Therefore in assigning luminosity classes, we first classified the star as if it were a Galactic star, but then paid attention to any large
difference between the absolute visual magnitude we derived, and the absolute visual magnitude we expected for the spectral
type and luminosity class.  Of course, if the visual luminosity was greater than what we expect, this could also be an indication
of binarity, and we carefully examined our spectra for tell-tale signs, such as the inability to obtain good fits of both the He I and
He II lines with a single set of parameters.  The classification of each star is given
in Table~\ref{tab:sample}, and will be discussed individually below in \S~\ref{Sec-types}.
When all is said and done we expect our spectral types are good to one classification step; i.e., an O8~V star could be
reasonably classified as O7.5~V or O8.5~V.

\subsection{Input: $M_V$ and $v_{\infty}$}
In order to model the spectra of the stars with FASTWIND, we need two additional
constraints.
The first of these is 
the absolute visual magnitude $M_V$, which is needed to calculate the stellar radius for
a given set of inputs, and is found from existing photometry and a knowledge of the
intrinsic colors, which in turn comes from the spectral type.
The second of these is the terminal velocity of the wind $v_\infty$, which
we can best measure using the resonance lines in the ultraviolet part of the spectra.

In order to derive an accurate $M_V$, we began by adopting the intrinsic colors corresponding
to the spectral types, using Table 3 of Massey (1998b).  We then calculated $M_V$ from
$$M_V=V-3.1E(B-V)+{\rm DM},$$
where the distance modulus DM is assumed to be 18.9 for the SMC and 18.5 for the
LMC, following van den Bergh (2000). 
This value was compared to that expected for
the assigned luminosity class (Conti et al.\ 1983).  If there was a significant discrepancy,
we reassigned a more appropriate luminosity class as discussed above.  
We then recalculated the final $M_V$ using the more appropriate intrinsic color. (In practice,
this made very little difference, as the differences in intrinsic colors of O-type stars change
by 0.03~mag in $B-V$ as a function of luminosity class.)  The adopted $A_V$
and $M_V$ values are included in Table~\ref{tab:sample}.  We also include
the $A_V$ determined from the $U-B$ colors as a reality check on the $A_V$
values determined from $E(B-V)$. In general the agreement is good, with 
an average difference $A_V (U-B) - A_V (B-V) = 0.15$, and a scatter of 0.13. 
We adopt the $A_V$ determined from $E(B-V)$ as the $B-V$ values are generally
more
accurately determined than $U-B$.  The slight systematic offset may also represent how
poorly we still know the intrinsic colors of O stars at differing metallicities (see, for
example, the nice work by  Martins \& Plez 2006).

The other ingredient for modeling the optical spectra is knowledge of the
terminal velocity ($v_\infty$) of the wind.  Many of our stars had well measured values in the
literature, while for others we measured these ourselves using the methods described in Paper I.  For one of the stars, LMC054383, we obtained new UV data.  We list the measured values,
and sources, in Table~\ref{tab:vinf}.   Two stars require additional comments.
 At the time we planned our Magellan observations, we thought NGC 346-682 had a
 STIS spectrum, but Heap et al.\ (2006) note that the acquisition had failed.
 A second star, NGC 346-487, has only a lower limit on the terminal velocity.
 We have decided to retain both of these stars in our analysis anyway, and, 
 following Kudritzki \& Puls (2000),  we adopt
a value of $v_\infty=2.6v_{\rm esc}$, where the {\it effective} escape velocity $v_{\rm esc}$ is calculated from the stellar radius $R$ and surface gravity $g$,
explictly including the Eddington factor $\Gamma$, i.e.,
$$v_{\rm esc}=\sqrt{1.392\times 10^{11}\ (R/R_\odot)\ g\ (1-\Gamma)},$$
$$\Gamma=0.4\ N_e\  Y_{\rm He}\ g\ [ (1+4Y_{\rm He})\ 1.8913\times 10^{-15}\ T_{\rm eff}^4]^{-1}$$
where $N_e$ is the number of free electrons per He atom (2 for $T_{\rm eff}>28000$, 1 otherwise for OB stars) and $Y_{\rm He}$ is the number ratio of
He to H.  The surface gravity $g$ is in cgs units (i.e., for the sun $\log g=4.438$).   We expect that the factor of
proportionality 2.6 will be fairly insensitive to the metallicity, but one needs to keep in mind that this number is good only
to 20\%, and that this scaling relationship is approximate in any event (Kudritzki \& Puls 2000), so we 
have
used rounded values for these two stars. Given the spectral types, 
neither of these stars is expected to have strong winds. For comparison
we include the values of $2.6v_{\rm esc}$,  drawing on the results of the model fitting presented below,
 for the other stars, where we see that in general there was very good agreement.

\subsection{Methodology}

The spectra were fit by varying the trial surface gravity, effective temperatures and mass loss rates 
in the models, and comparing the synthetic spectra to the observed
spectrum.  The  $\beta$ parameter (related to the
acceleration of the stellar wind) and the He to H ratio were altered, if needed.  Values for the metallicity $Z/Z_\odot$ of 0.2 (SMC) and 0.5 (LMC) were adopted
as mentioned earlier, as we did in Papers I and II. We have scaled the solar abundance of 
Asplund (2003) by these factors, rather than the (older) solar abundances used in Paper I and II, although the differences are in
fact imperceptible.  As we note in Paper I, the actual relative abundances of the interstellar medium in the SMC, LMC, {\it and} the nearby
regions of the Milky Way are non-solar, as discussed by Westerlund (1997), so this is not really
the right thing to do, but lacking detailed abundance analysis of these stars (which is beyond the scope of this paper), this is the
best we can do for now.

In general, certain lines were primarily affected by varying a single stellar parameter.  For instance, the strengths of the model He I and He II lines were sensitive primarily to a
change in $T_{\rm eff}$.  The stars in our sample covered the range of effective temperatures from 29,250 to 53,000 K.  
At the lowest values for $T_{\rm eff}$ the strengths of the Si III and Si IV lines were also useful.  In Paper I and II we argue that
the uncertainty in our effective temperatures was 1000 K, but in many cases our temperatures are
more precisely determined, as  the  models are sensitive to temperature differences of 250-500 K.

The effective surface gravity $g_{\rm eff}$ (surface gravity reduced by the effects of centrifugal acceleration
due to the rotation of the star) was fit by comparing  the wings of H$\gamma$ lines to 
those of the
models, with wider lines 
(more pressure-broadened) indicative of  a larger value.
Values for $\log g_{\rm eff}$ in our sample ranged from 2.9 for the most extreme supergiant to 4.2 for the most compact dwarf.  We could detect differences in the 
line profile of H$\gamma$ by varying $\log g$  by 0.1 dex.
 
The determination of the mass loss rates ($\dot{M}$) depended very strongly on the line profile of H$\alpha$ and to a lesser extent, H$\beta$ and He II $\lambda 4686$.  
The mass loss rates measured ranged from $0.05 \times 10^{-6}$  to $4.5 \times10^{-6}  M_\odot$ yr $^{-1}$.  Models with lower mass loss rates were generally  sensitive to adjustments of 0.1-0.2 $ \times 10 ^{-6} M_\odot$ yr $^{-1}$, while the stars with the largest mass loss rates were
sensitive to adjustments of about 1 $ \times 10 ^{-6}$ M$_{\sun}$ yr $^{-1}$.  The H$\alpha$ profile becomes insensitive to mass-loss rates below about $0.1 \times 10^{-6} M_\odot$ yr$^{-1}$,
and the details of the numerical approach can dominate the results.  Models with the highest mass loss rates had both $H \alpha$ and He II $\lambda$4686 in emission. Typically, lines in emission allowed a determination of  $\beta$, the wind acceleration parameter,  which otherwise was set at default to 0.8.   We note that
in accordance with Papers I and II, the mass-loss rate we determine is with the
assumption of homogeneous stellar winds; the actual mass loss will be related to this by
$\sqrt{1/f}$, where $f$ is the clumping factor.  This factor is likely to be of order 6-10 or less,
if the clumping properties of O stars in the Clouds are similar to those in the 
Milky Way (see, for example, the recent conference workshop by Hamann et al.\ 2008).

We assumed a standard He to H ratio of 0.1, and this sufficed for most of our fitting.  In a few cases we found we had to increase this value in order to obtain sufficiently
strong He lines.  We discuss these individual cases below.

In comparing the star's line profiles to that of the models, we of course had to correct for the radial velocity of the star, and to broaden the model's lines to match the
rotational velocity of the star.   We used the width of several He II lines as a gauge of the rotational velocity.  Strictly speaking, much of this broadening was due to
the instrumental resolution, which created a floor in our measurements of 140 km s$^{-1}$.  Occasionally we had to adjust either the radial velocity or $v\sin i$ values from
our preliminary measurements to obtain better agreement.  

In Table~\ref{tab:results} we list the final adopted models for each star.  We include several derived parameters.  The {\it true} surface gravity is the {\it effective} surface
gravity corrected for the effects of centrifugal acceleration due to the rotation of the star.  This correction can be approximated (in a statistical sense, assuming randomly distributed rotational axis orientations) as simply
the square of the projected rotational velocity divided by the stellar radius $R$:
$$g_{\rm true}=g_{\rm eff}+\frac {(v \sin i)^2} {6.96 R}$$
following Repolust et al.\ (2004).  As discussed above, we expect that low ``rotational velocities" are actually dominated by the instrumental resolution, but these
have a negligible effect.  Even the highest rotational velocities in our sample (240 km s$^{-1}$) contribute only 0.05~dex to $\log g_{\rm true}$, as seen for BI~208 in Table~\ref{tab:results}.
We also include an estimate of the spectroscopically derived mass of the star, based upon our determination of $\log g_{\rm true}$ and the stellar radius 
(since $M_{\rm spec} = g_{\rm true}/g_\odot R^2$, where the mass and radius are in solar units); we will
compare these ``spectroscopic" masses to those derived from evolutionary tracks using the star's effective temperature and luminosity in the next section.

We
 expect some significant number of stars in our sample of 26 stars to have no satisfactory fits, as some fraction will be
actually be composites of two (or more!) stars.  At least 7 out of the 40 stars analyzed in Papers I and II admitted no good solution.  In this way our method has some
significant advantages over that of the ``automatic fitting method" of Mokiem et al.\ (2005, 2006, 2007) which obtains a ``best answer" for every star (see discussion following
de Koter 2008), but does not reject cases where the model is an unlikely match to the data (see discussion in Press et al.\ 1992).

Several of our stars have been previously analyzed, and we included these in our program as such overlap provides a crucial test of the methods (and data) of ourselves and others.  We include in Tables~\ref{tab:compA} and Table~\ref{tab:compB} a comparison of the principle physical properties found by ourselves and others. 
In comparing these values, one should keep in mind that the the effective
temperature ($T_{\rm eff}$) and surface gravities ($\log g_{\rm eff}$) are
not quite independent---in general, a {\it model}
with 2000~K cooler temperature
will require (roughly) a surface gravity that is 0.1~dex lower to preserve the
shape of the Balmer line wings.

Throughout the following section note that the scale of the figures showing the line profiles vary from line to line, and star to star,
as indicated by the labels.  In a few very early-type O stars (i.e., LH64-16 and Sk$-67^\circ$166) 
the He I $\lambda 4471$ line is the only He~I we detect even with our $3\sigma$ limit
of 12 m\AA, and hence is the only He I line we used in the fitting, but 
we still show the comparison between the model and spectra
for He I $\lambda  4387$ and He I $\lambda 4922$ lines for consistency. 
We also call attention to the fact that with FASTWIND
the He I triplet $\lambda 4471$ model line is often too weak for middle and late O-type
giants and supergiants (Repolust et al.\ 2004).  
This so-called ``generalized dilution effect" has been previously
seen by others (e.g., Voels et al.\  1989) and is not well understood.  In this spectral
range we have placed more reliance on the strong singlet He I $\lambda 4387$ line.
In our experience this line is modeled quite well by FASTWIND.  Najarro et al.\ (2006)
call attention to a problem that codes which treat line blanketing exactly (such as
CMFGEN and TLUSTY) have with He I $\lambda 4387$ and other singlets, due probably
to some uncertainties in the atomic data of Fe IV lines overlapping with the He I resonance lines around
584 \AA.  With its approximate treatment of line-blanketing and blocking,  FASTWIND should be less
sensitive to this issue, but future comparisons with other codes are needed to
fully resolve this issue.
 
\subsection{Individual Stars}
\label{Sec-types}

\subsubsection{SMC}

\begin{figure}
\epsscale{0.6}
\plotone{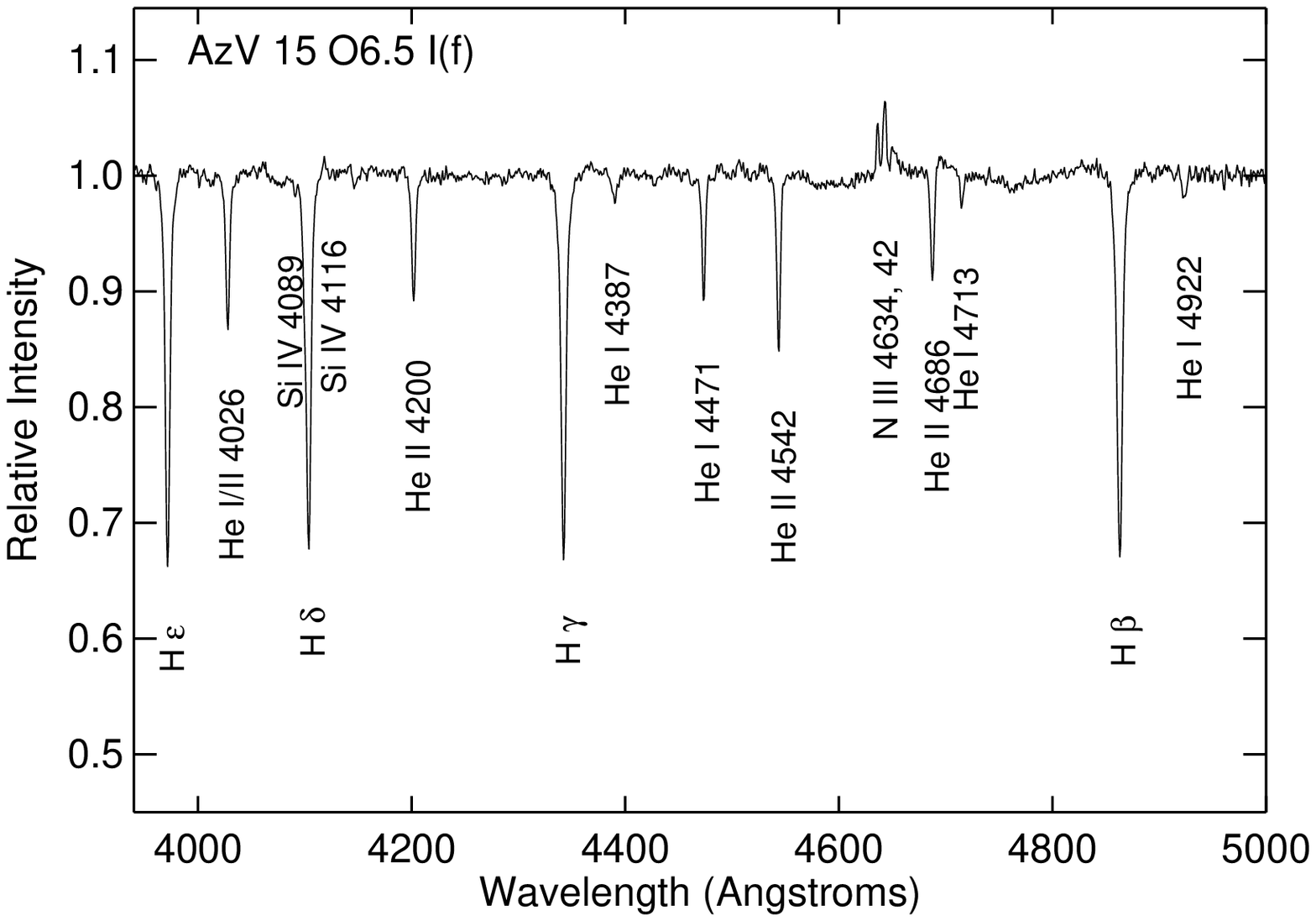}
\vskip 50pt
\plotone{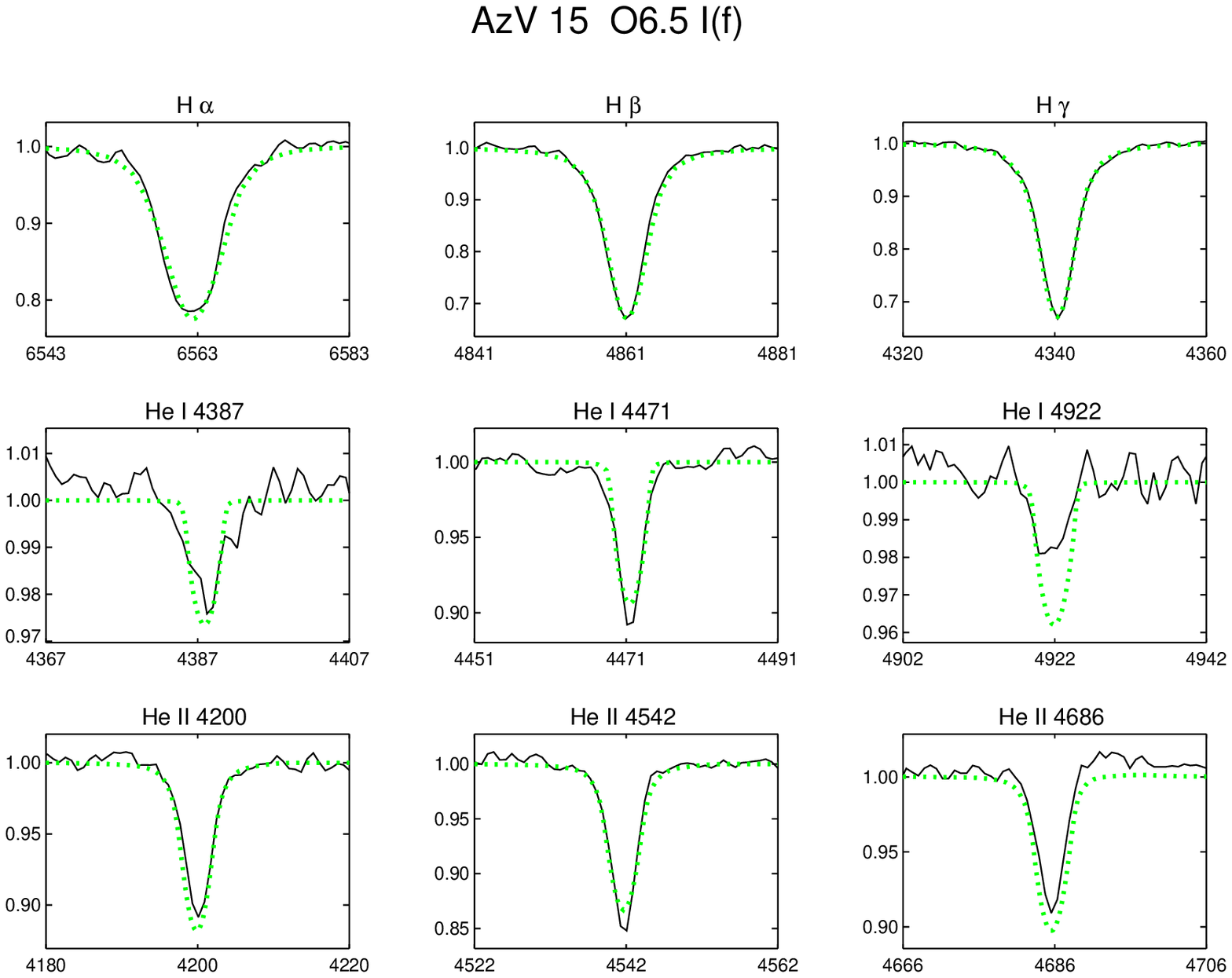}
\caption{\label{fig:AzV15} AzV 15.  The upper figure shows a section of the blue spectrum of this star,
with the prominent lines identified.  The lower figure shows the fits (dotted) for the principle diagnostic lines.}
\end{figure}

{\it AzV 15}.  Visually we would classify this star as O6.5, given the relative strengths of the He~I and He~II lines (Fig.~\ref{fig:AzV15}, upper).  The measured EWs
of He~I $\lambda 4471$ is 430 m\AA, and that of He~II $\lambda 4542$ is 660 m\AA, leading to a $\log W'=-0.19$, also indicative of an O6.5 type.
NIII $\lambda \lambda 4634, 42$ is strongly in emission, but He~II $\lambda 4686$ is in absorption.  This combination would lead to a ``(f)" designation, which 
(at Milky Way metallicity) is associated with the star being of luminosity class III for an O6.5 star. However, the corresponding absolute magnitude is $-6.15$, more
 in keeping with it being a supergiant.  This is similar to what we found for a number
of SMC stars in Papers I and II---the weaker stellar wind (due to the lower metallicity of the SMC) results in He~II $\lambda 4686$ being in absorption for
stars of a given $M_V$ and spectral type that, had they been found in the Milky Way, would have had He~II $\lambda 4686$ in emission.  At our
high SNR we detect very weak Si IV $\lambda 4116$ emission ($-50$ m\AA\ EW), although Si IV $\lambda 4089$ is weakly in absorption (40 m\AA\ EW).  Strong Si IV $\lambda \lambda 4089, 4116$ emission is associated with earlier types in Milky Way stars, but our high SNR apparently allows detection in later types,
as discussed above.  It may also be that at lower metallicity it will be in emission at somewhat later types. We note in passing that the current models do not correctly reproduce the behavior of these lines, not even for extreme Galactic examples where the
lines are in strong emission.
 
 We describe
this star as O6.5 I(f), a designation that would be an oxymoron were it to be applied to a Galactic star.  
The star had previously been called an O7 II by Garmany et al.\ (1987) and as O6.5 II(f) by Heap et al.\ (2006), both in substantial agreement with our designation. 
Of course, an alternative possibility would be that the star is composite; however, the fits for this star are very good (Fig~\ref{fig:AzV15}, lower).  
We emphasize that fitting individual spectral lines
is a very sensitive way to detect composite spectra---only if both components have very similar effective temperatures and luminosities would we likely be able to find
a satisfactory fit both the He~I and He~II lines at the same time. 
The star was previously modeled (Table \ref{tab:compA}) 
by Mokiem et al.\ (2006), who obtained an effective temperature 900 K higher than ours, and by Heap et al.\ (2006) who obtained an effective temperature 1500 K lower
than ours.

\begin{figure}
\epsscale{0.6}
\plotone{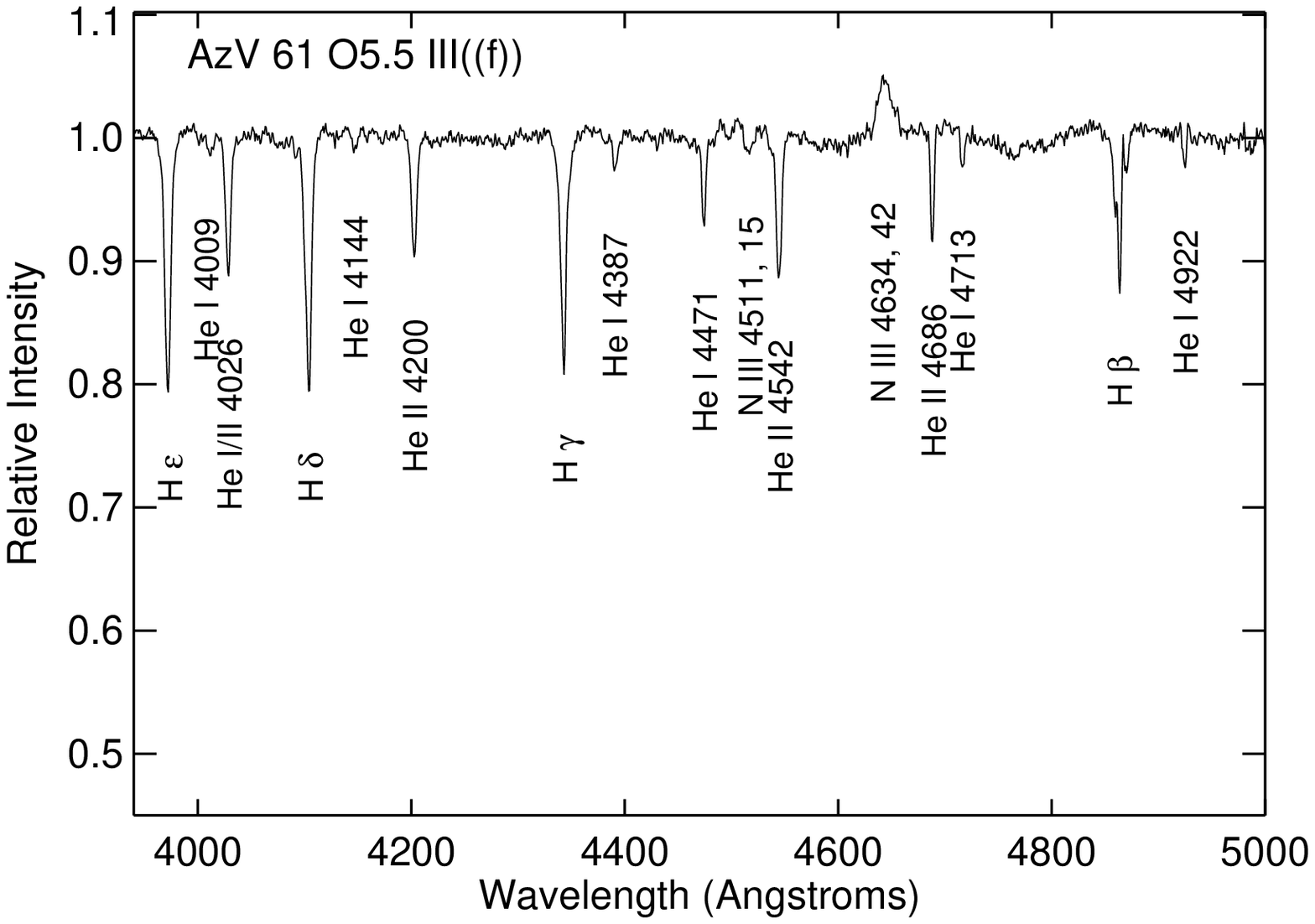}
\vskip 50pt
\plotone{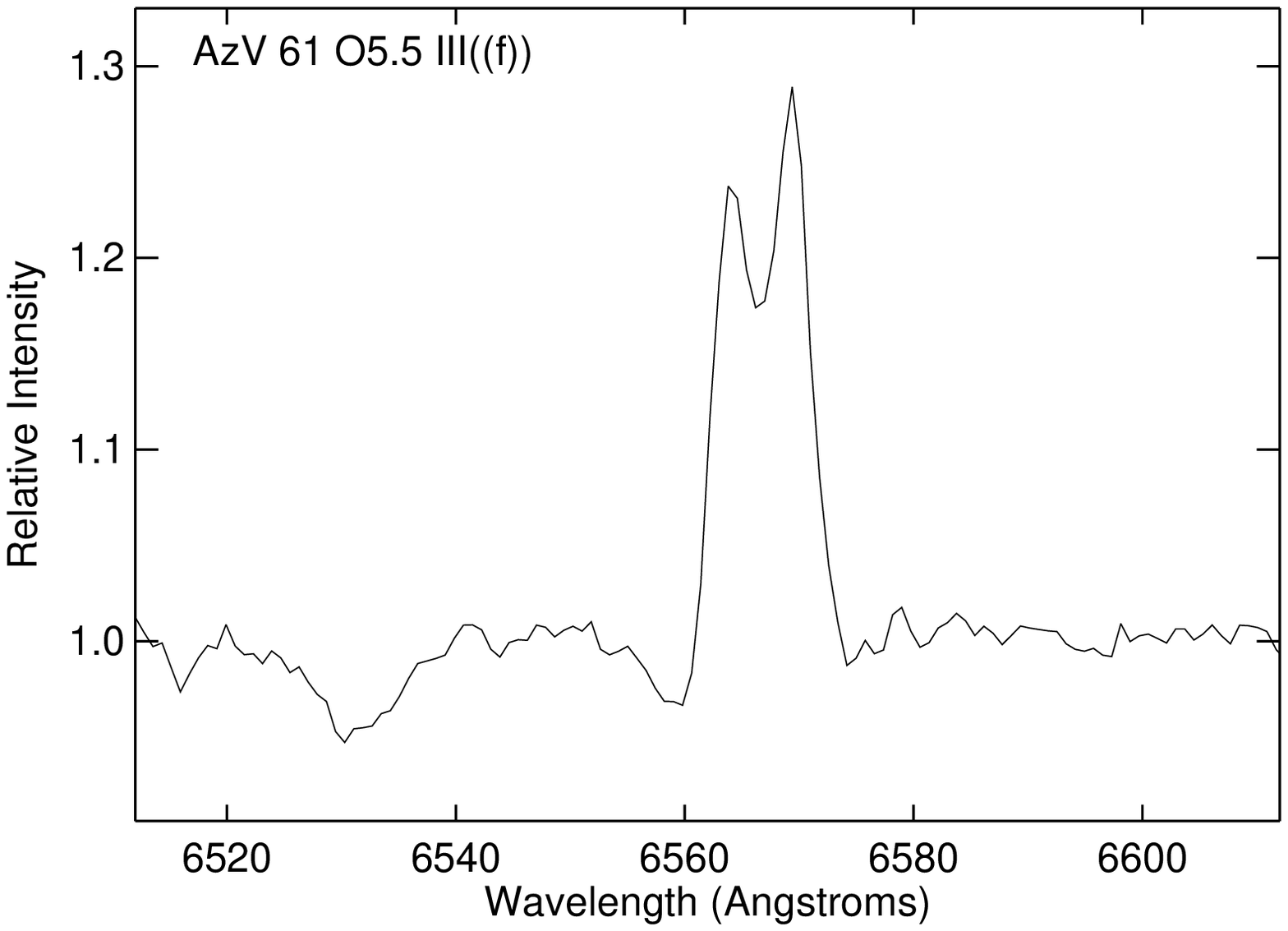}
\caption{\label{fig:AzV61} AzV 61.  The upper figure shows a section of the blue spectrum of this star,
with the prominent lines identified. The lower figure shows the H$\alpha$ profile.}
\end{figure}

{\it AzV 61}.  The spectral subtype is O5.5, with EWs of 300 m\AA\ (He I $\lambda 4471$) and 750 m\AA\ (He II $\lambda 4542$), leading to
$\log W'=-0.40$.  The strength of N III $\lambda \lambda 4634, 42$ emission would lead us to conclude this star is a giant, but again we find that He~II $\lambda 4686$
is strongly in absorption, more consistent with a dwarf. The absolute magnitude $M_V$ we derive is -5.8, also consistent with the star being a giant,
and we thus refer to this star as an O5.5 III((f)).  Previously the star had been called an O5 V by Crampton \& Greasley (1982). The blue spectrum in shown
in Fig.~\ref{fig:AzV61} (upper).

We were unable to obtain a satisfactory fit to this star, due primarily to the fact that  H$\alpha$ is strongly in emission and double-peaked (Fig.~\ref{fig:AzV61}, lower), 
while He II $\lambda 4686$
is strongly in absorption, as shown in the upper panel.
  H$\beta$ is also filled in with emission.
Either this star is a binary, or it is surrounded by a circumstellar disk, as we suspect is the case for two stars discussed below.

\begin{figure}
\epsscale{0.6}
\plotone{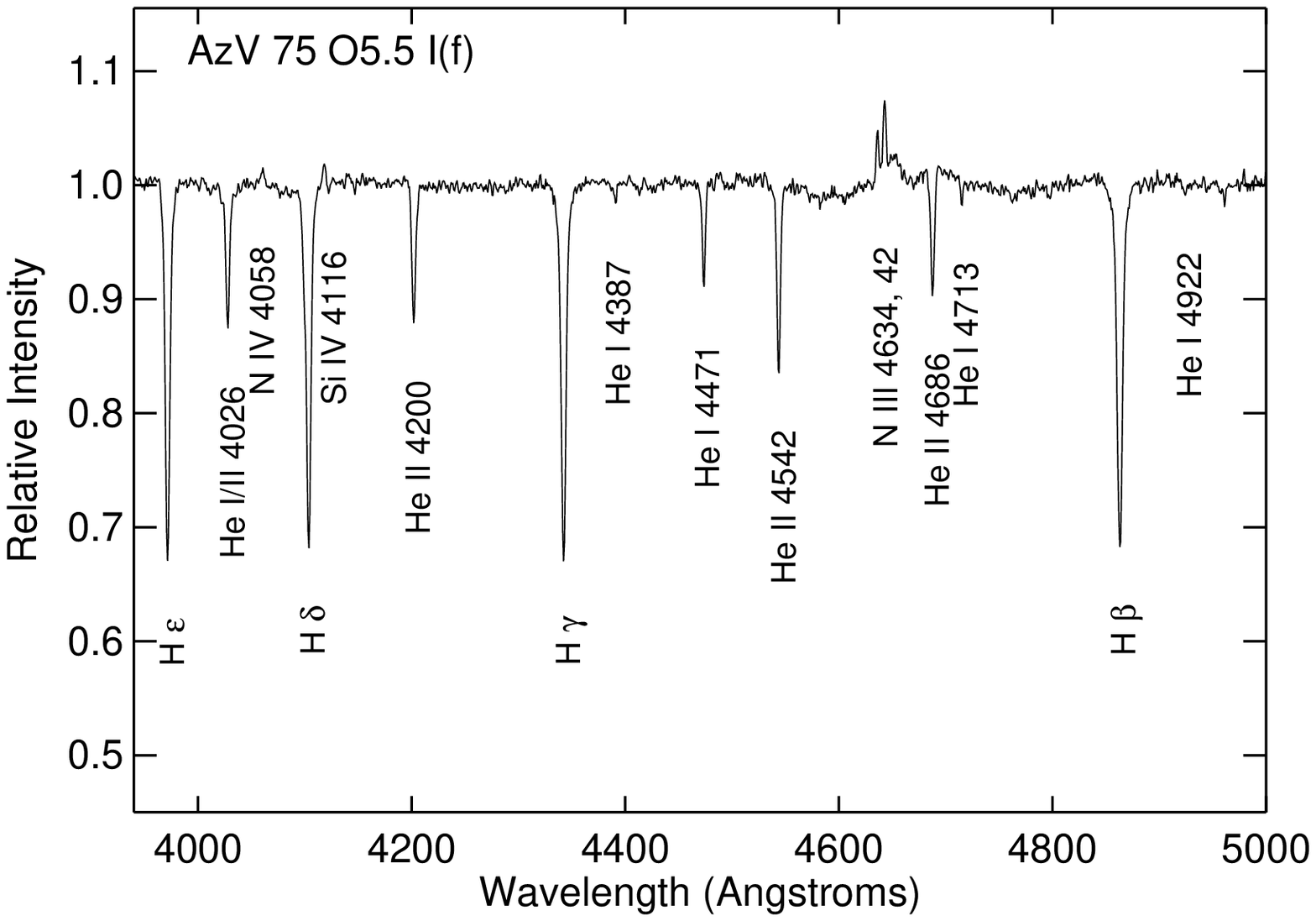}
\vskip 50pt
\plotone{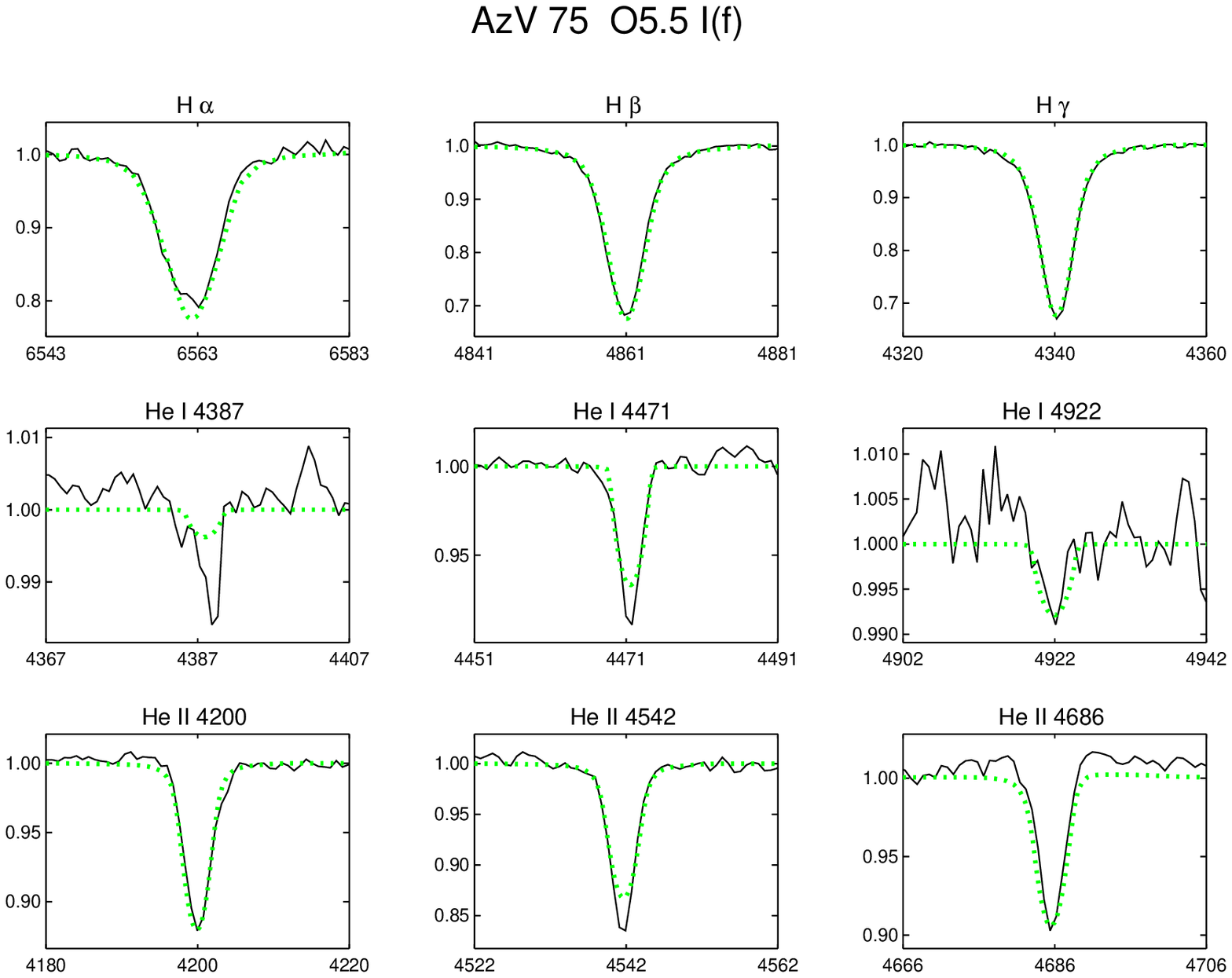}
\caption{\label{fig:AzV75} AzV 75. The upper figure shows a section of the blue spectrum of this star,
with the prominent lines identified.  The lower figure shows the fits (dotted) for the principle diagnostic lines.}
\end{figure}

{\it AzV 75}.  The spectrum in shown in Fig.~\ref{fig:AzV75} (upper). 
The spectral subtype is O5.5, based both upon the visual appearance and the measurements of 
EWs (335 m\AA\ for He~I $\lambda 4471$ and 680 m\AA\ for He~II $\lambda 4542$, leading to $\log W'=-0.31$).  
N~III $\lambda \lambda 4634, 42$ is strongly in emission, consistent with the star being a supergiant.
He II $\lambda 4686$ is in (weak) absorption, which would lead to a giant luminosity class.  Again we find that the absolute magnitude ($M_V=-6.7$) is consistent with
the star being a supergiant, so we again call this an O5.5 I(f). 

The star was included in Paper I, where we independently called it an O5.5 I(f).  However, here we note 
very weak N IV $\lambda 4058$ emission, and possible Si IV $\lambda 4116$ emission.  Heap et al.\ (2006) refer to this star as O5.5 III(f+), with the ``+"
indicating that they recognized the Si IV emission as well.  
The presence of N IV and Si IV emission could be indication of a composite spectrum, as these are usually associated with earlier spectral types, but we discuss above,
the emission mechanisms involved may result in these features not showing the same connection with spectral types as in the Milky Way.

The fits we obtained for this star were reasonably good (Fig.~\ref{fig:AzV75}, lower).  
The He I $\lambda 4471$ model line is a little weak, but that is consistent with the fact that FASTWIND (and
likely other models) do not seem to produce a strong enough He I $\lambda 4471$ line for supergiants and giants of this type and later.  The model He I $\lambda 4922$ is a
good match, but He I $\lambda 4387$ is a bit weak.  We found that if we lowered
the effective temperature by 500-1000 K the agreement with He I $\lambda 4387$ improved, but that He I $\lambda 4922$ and He I $\lambda 4713$ (not shown) both got too strong in the models, so we adopted the hotter fit.   Our model fits were done without reference to Paper I.  Nevertheless, the agreement is excellent, as
shown in Table~\ref{tab:compA}.  Heap et al.\ (2006) also analyzed this star, and derived an effective temperature 1000 K cooler than what we do here.

\begin{figure}
\epsscale{0.6}
\plotone{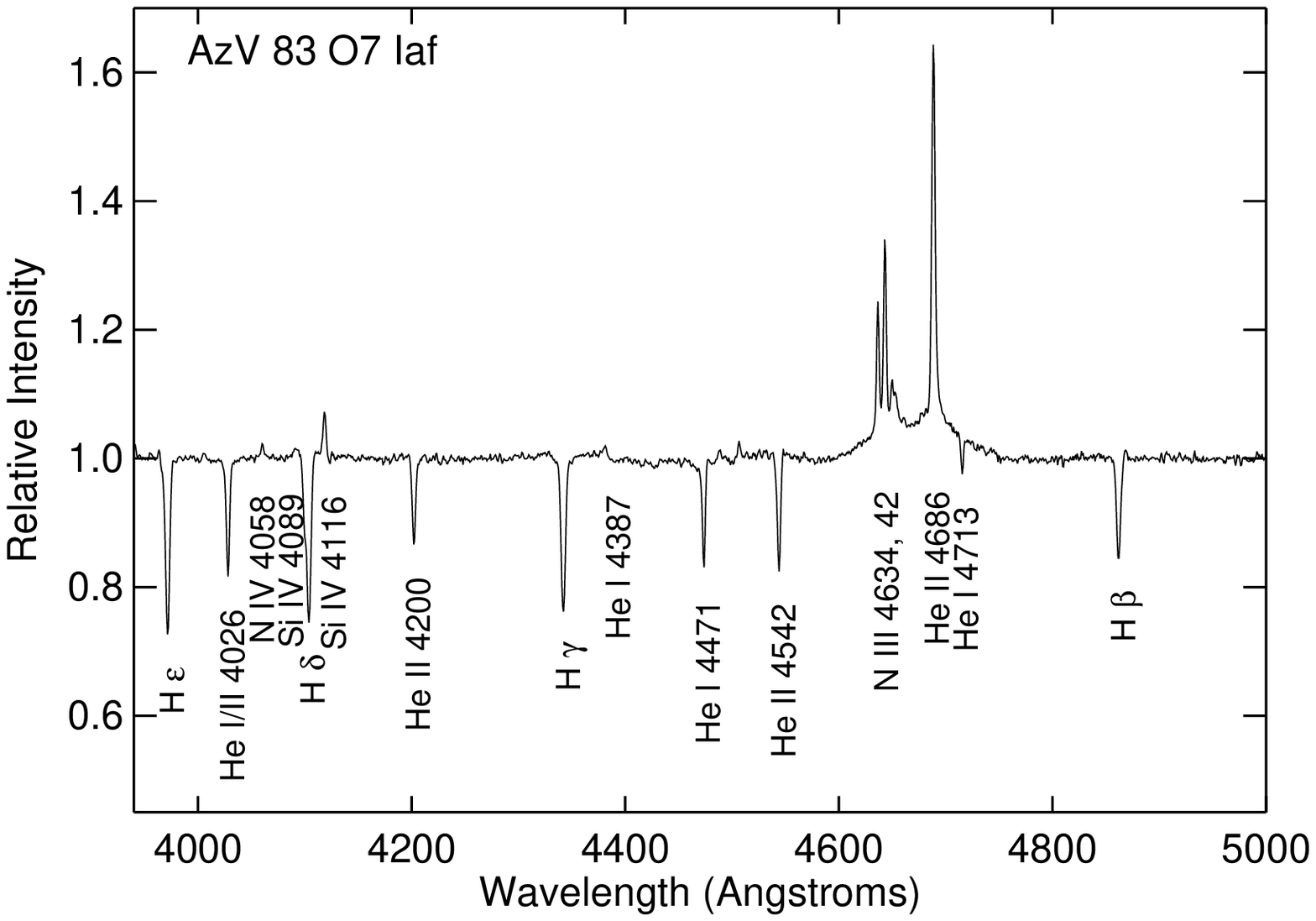}
\caption{\label{fig:AzV83} AzV 83. The figure shows a section of the blue spectrum of this star,
with the prominent lines identified. }
\end{figure}

{\it AzV 83}.  The spectrum is shown in Fig.~\ref{fig:AzV83}. The spectral subtype for this star is O7, based both on its visual appearance and the EW measurements (660 m\AA\ for He I $\lambda 4471$, and 680 m\AA\ for
He II $\lambda 4542$, leading to a $\log W'=-0.01$).  Both N III $\lambda \lambda 4638, 42$ and He II $\lambda 4686$ are strongly in emission, leading to an ``If" luminosity
class, or even an ``Iaf" class, given the strength of emission.  Yet  $M_V=-6.18$ is intermediate between a giant and a supergiant.  
N IV $\lambda 4058$ and Si IV $\lambda \lambda 4089, 4116$ are in  weak emission, similar to that of AzV75, although Si IV is more strongly in emission than in AzV75.
The star was analyzed by Hillier et al.\ (2003), who also call it an O7 Iaf+, although this extreme luminosity classification is somewhat at variance with the absolute
visual magnitude.  Si IV emission is present in the Hillier et al.\ (2003) spectra, and is also present (weakly) in the final adopted model.  N IV $\lambda 4058$ is shown weakly in
emission in their spectra, although their final adopted model has this line weakly in absorption.   The star was also analyzed by Heap et al. (2006).

Despite having examined over 80 models, we were unable to obtain a satisfactory fit to the spectrum of this star.  
Our best solution required an effective temperature in the neighborhood of 36,000-38,000~K in order to match
the weakness of the He~I $\lambda 4387$ line, and we also required a He/H value of 0.2 to obtain sufficiently strong He II absorption in a reasonable temperature regime.  Yet such
a high temperature is inconsistent with the Si IV UV P Cygni profile (Hillier, private communication).  As shown in Table~\ref{tab:compA},
Hillier et al.\ (2003) derived a much lower temperature, 32,800 K, while
Heap et al.\ (2006) found an intermediate temperature (35,000 K).   As Heap et al.\ (2006) emphasized, the temperatures derived for this star depended in large
part upon which He I lines are given the stronger weight in the fit, as the results are inconsistent.   We note the following: (1) The absolute visual magnitude, $M_V=-6.18$, is
considerably fainter than we would expect for such an ``extreme" Of star; it is more
in keeping with a normal supergiant. (2) The O7 Iaf star AzV232 (discussed below) has {\it weaker} luminosity indicators and yet has $M_V=-6.96$.
This is hard to understand if both of these are single stars.  
However, if the NIII $\lambda \lambda 4634, 42$ and He II $\lambda 4686$ emission
was primarily from a hotter component in AzV~83,
this would make sense, although the sum of the stars' luminosities would still have
to be no larger than that of a single supergiant.   Also, the terminal velocity determined from the UV lines is  consistent with a
supergiant, and there is no evidence of the faster wind we would associate with a dwarf---any composite explanation would
have to satisfy both the optical and UV spectra.  Still,
although we do not see a significant radial velocity difference ($\Delta v_r < 30$ km s$^{-1}$) between the adjacent nights on which
our spectra (blue and red) were obtained, we believe that a radial velocity study might be useful.
 
\begin{figure}
\epsscale{0.6}
\plotone{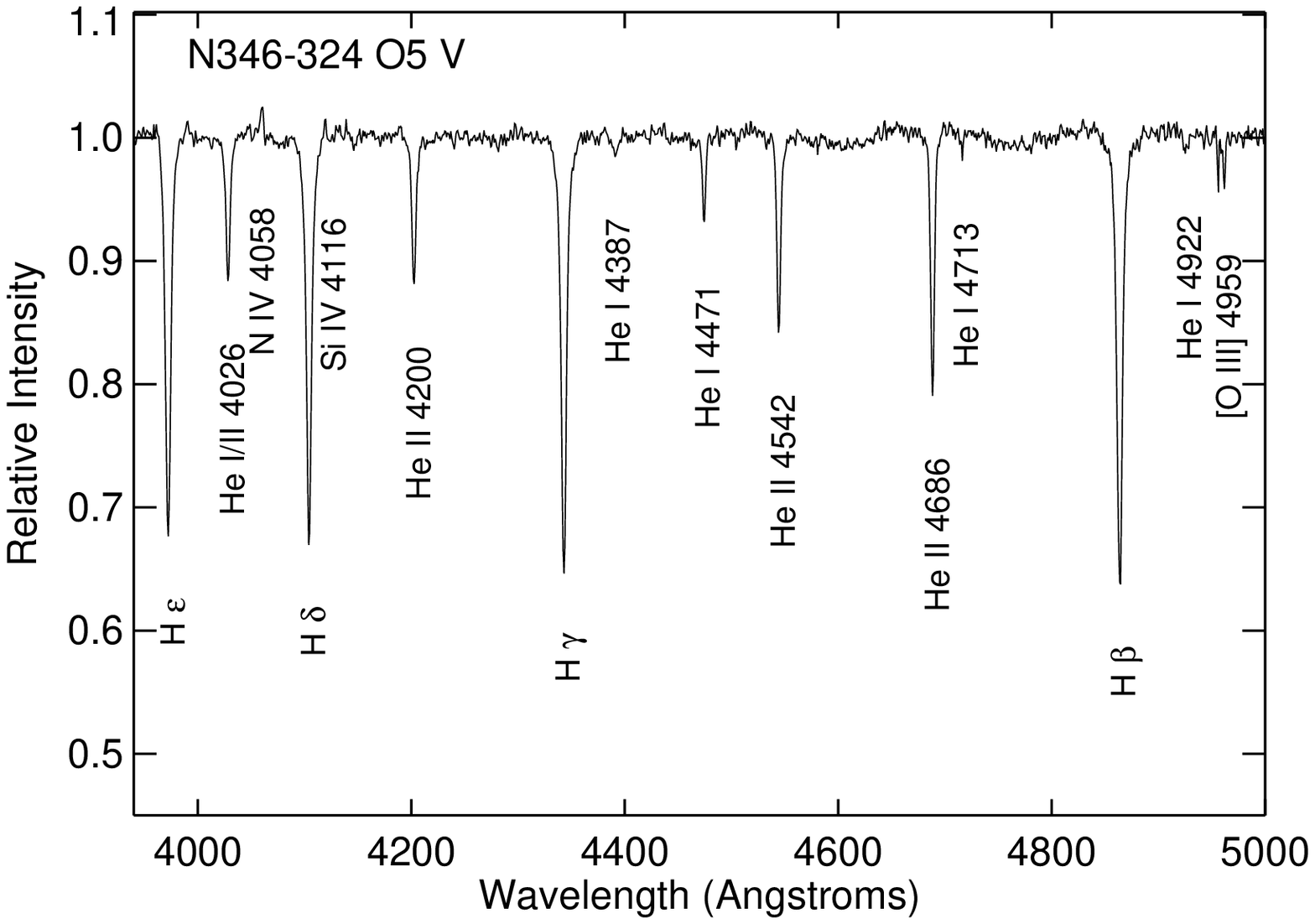}
\vskip 50pt
\plotone{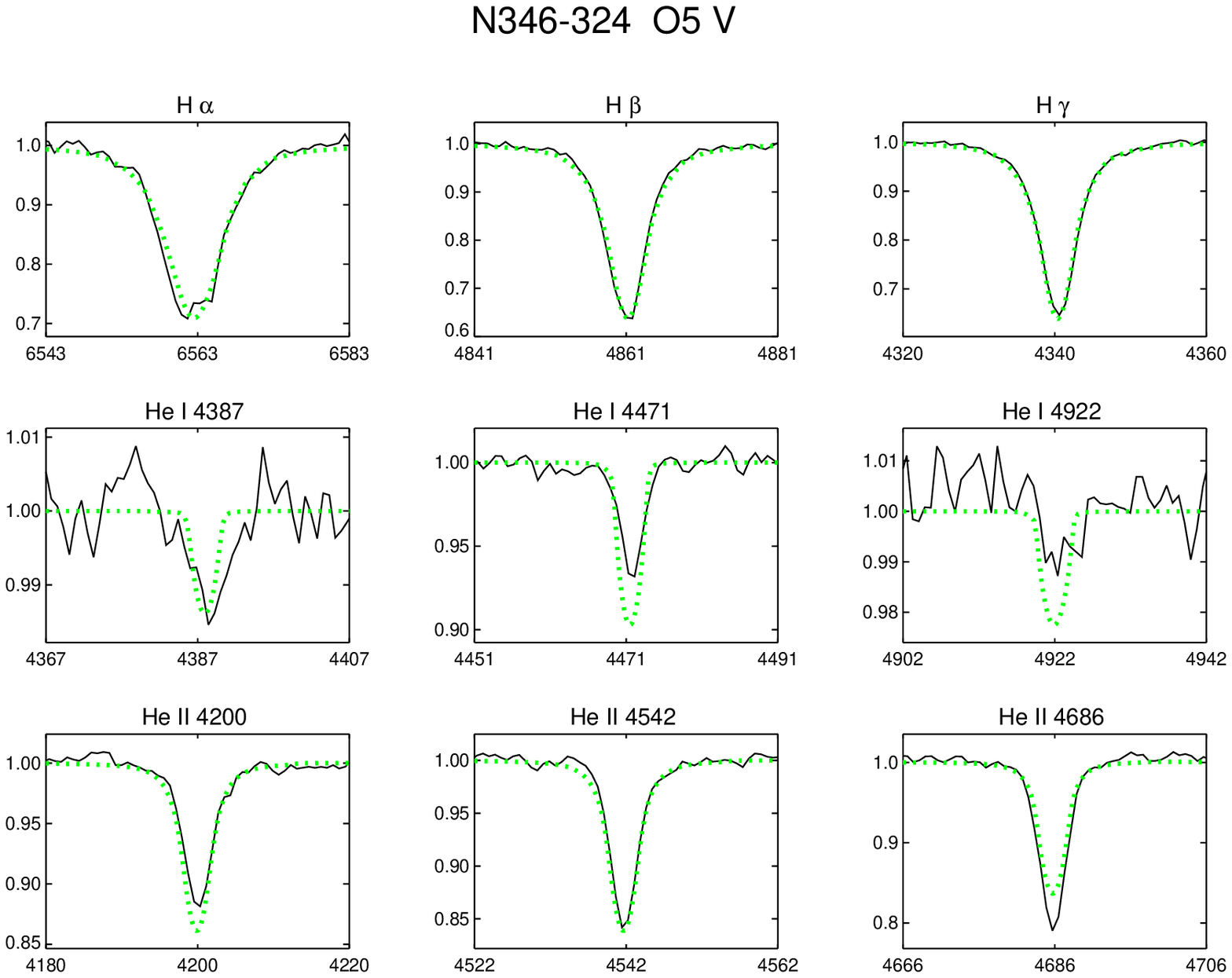}
\caption{\label{fig:N346324} NGC 346-324. The upper figure shows a section of the blue spectrum of this star,
with the prominent lines identified.  The lower figure shows the fits (dotted) for the principle diagnostic lines.
}
\end{figure}

{\it NGC346-324}.  The spectrum is shown in Fig.~\ref{fig:N346324} (upper).  We assign a spectral subtype of O5 for this star, both based on its overall visual appearance and the
measured EWs of He I $\lambda 4471$ and He II $\lambda 4542$ (250 m\AA\ and 760 m\AA, respectively), leading to a $\log W'=-0.48$.  There is no N III $\lambda \lambda 4634,42$
or He II $\lambda 4686$ emission, which would suggest a ``V" luminosity class.  The absolute visual magnitude is -5.13, also quite consistent with the ``V" class.  The star
has been classified as O4 V((f)) by both Massey et al.\ (1989) and Walborn et al.\ (2000), but we are satisfied with our slightly later classification. Niemela et al.\ (1986) called
the star an O4-5 V.  The spectra has some nebular
contamination, as judged by the over-subtraction of [OIII] $\lambda 4959$ and $\lambda 5007$.  Weak NIV $\lambda 4058$ (EW=$-100$ m\AA) is also present, as is very
weak SiIV $\lambda 4116$ emission (EW=$-25$ m\AA).  Again, such features would go undetected at lesser SNRs.  

The fits we obtained were good (Fig.~\ref{fig:N346324}, lower): He I $\lambda 4471$ is a little too strong in our adopted model, as is He II $\lambda 4200$.  Other lines show excellent agreement.   The star
has been previously analyzed by Bouret et al.\ (2003) and Puls et al.\ (1996); the agreement in Table~\ref{tab:compA} is quite good, despite the differences in code and data used
in these studies.

The star was included in the ``automated" fitting program of Mokiem et al.\ (2006), who obtained nearly identical results to ours
(Table~\ref{tab:compA}). They add a footnote to their Table 1 suggesting that this star is a binary, based, apparently, on variable
radial velocities although Mokiem et al.\ (2006) cryptically notes that none of their three detected binaries ``appear to have massive companions", and hence conclude that their derived fitting parameters are valid. They do not quote any radial velocities.  The star has also
been analyzed by Heap et al.\ (2006), again with very similar results\footnote{Mokiem et al.\ (2006) state that their effective temperature
for this star differs significantly from that derived by Heap et al.\ (2006), but quote an effective temperature that corresponds to a different
NGC 346 star studied by Heap et al.\ (2006).  The agreement for this star is actually reasonably good.}.

\begin{figure}
\epsscale{0.6}
\plotone{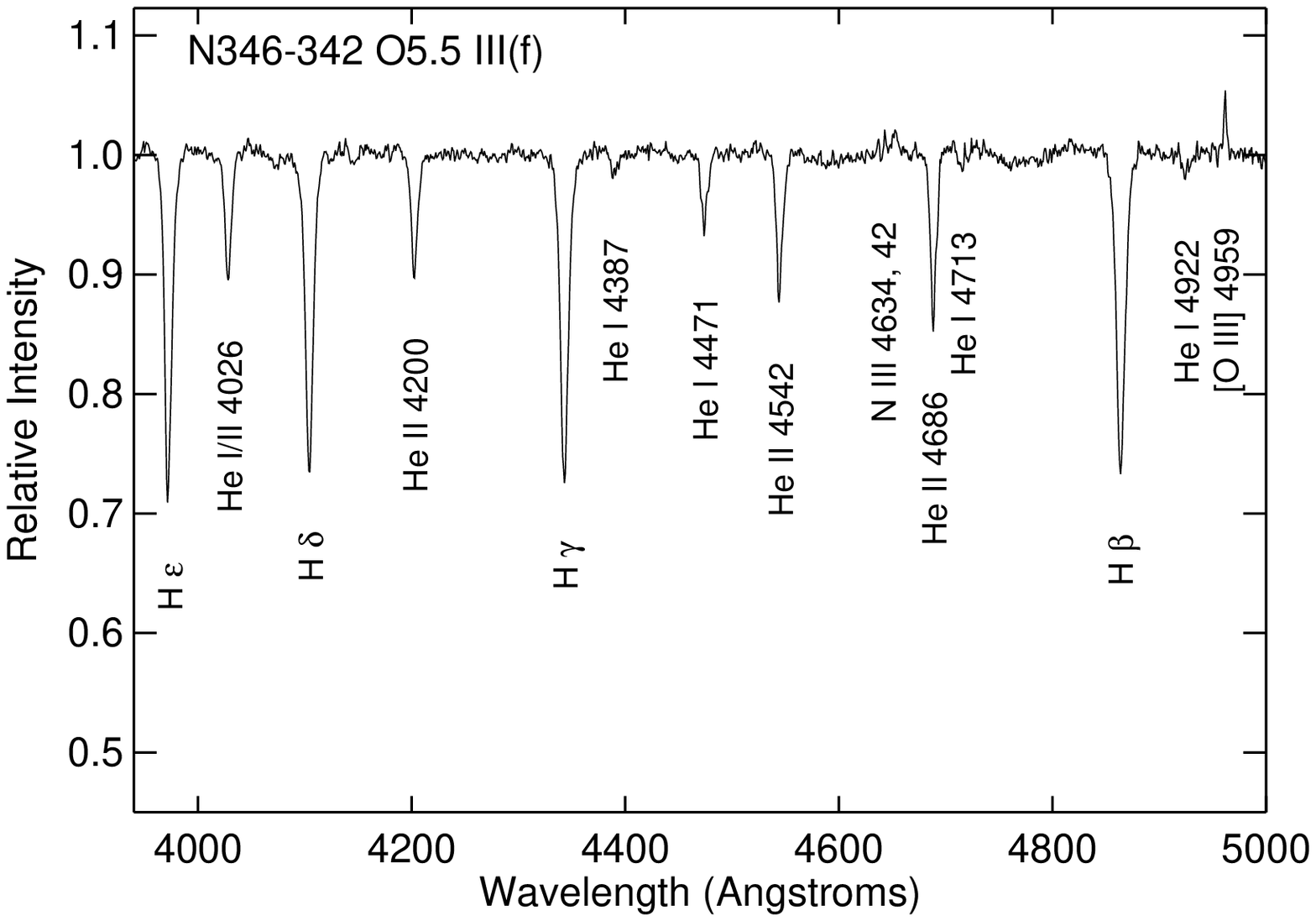}
\caption{\label{fig:N346342} NGC 346-342. The figure shows a section of the blue spectrum of this star,
with the prominent lines identified. 
}
\end{figure}

{\it NGC346-342}.  The spectrum is shown in Fig.~\ref{fig:N346342}.  We obtain a spectral subtype of O5.5 for this star, based on its general appearance and the measured
EWs (410 m\AA\ for He~I $\lambda 4471$ and 850 m\AA\ for He II $\lambda 4542$), leading to $\log W'=-0.32$.  NIII $\lambda \lambda$ 4634, 42 is in emission but 
He~II $\lambda 4686$ is in absorption, and we might call this an O5.5 V ((f)), as did Massey et al.\ (1989), but  the absolute magnitude is $-5.52$, more like that of a giant.  
We would, therefore, call this star an O5.5 III((f)).
 However, we conclude that the star is composite.  First, we were unable to obtain consistent
fits to the He I and He II lines.  Second, close examination of the spectrum revealed an inflection point on the He I lines suggesting incipient double lines.  Third, and most 
convincingly, 
the star
has been found to be an eclipsing binary in data obtained by
several of the co-authors (PM, NIM, KDE)  in connection with another project. These data show
eclipses of 0.25~mag and a period of 2.35 days.  A
follow-up spectrum obtained in January revealed clear double lines, with spectral types of O5~V and O7~V, with some indication of a third component, which has now been
confirmed by additional spectroscopy at Magellan.  A follow-up radial velocity study is in progress.

\begin{figure}
\epsscale{0.6}
\plotone{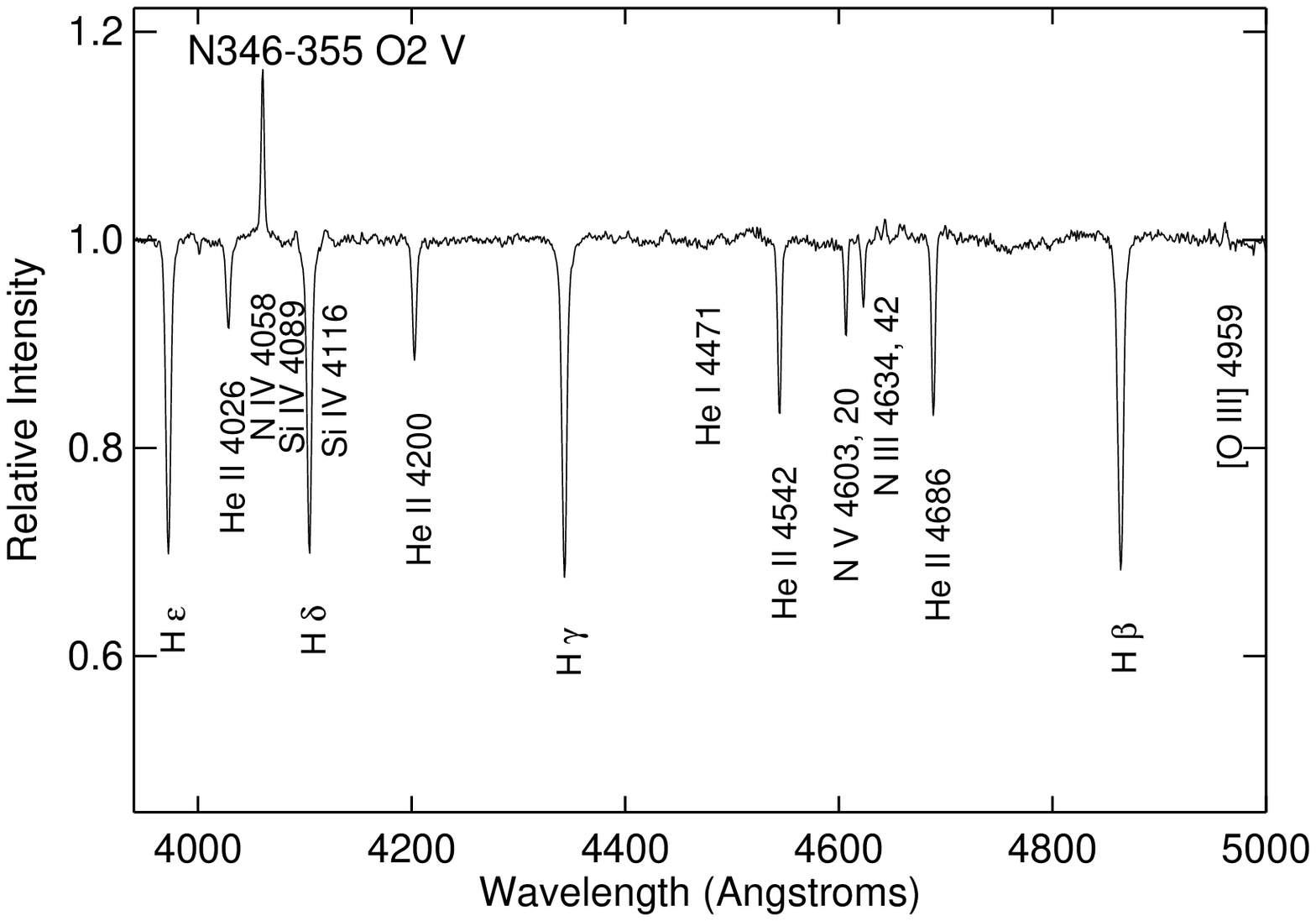}
\vskip 50pt
\plotone{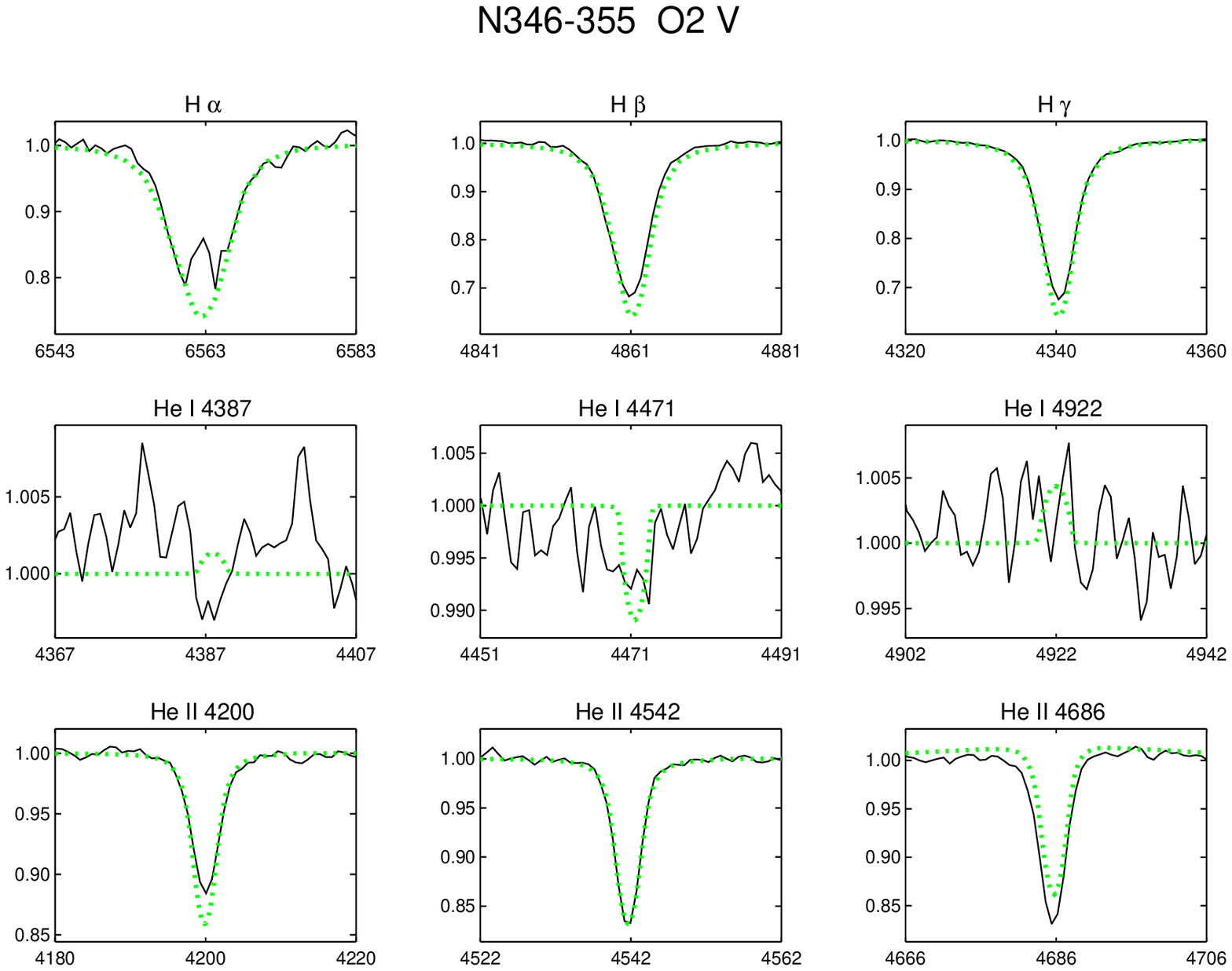}
\caption{\label{fig:N346355} NGC 346-355. The upper figure shows a section of the blue spectrum of this star,
with the prominent lines identified.  The lower figure shows the fits (dotted) for the principle diagnostic lines.
}
\end{figure}

{\it NGC346-355}.  The spectrum is shown in Fig.~\ref{fig:N346355} (upper).  It is the earliest type star in our sample, and indeed is offered by Walborn et al.\ (2002) as
a representative of the earliest defined
spectral class, O2.   Walborn et al.\ (2002) assign
a luminosity class of III(f*).   We measure an EW of $40\pm10$ m\AA\ for He I $\lambda 4471$ in our spectrum, 
consistent with what we found in Papers I and II, and also with the statement by 
Walborn et al.\ (2002) that O2 giants and supergiants show ``no or very weak" He I.    The EW of He II $\lambda 4542$ is 720 m\AA.
For the earliest type stars Walborn et al.\ (2002) propose a classification scheme based on the relative strengths of the NIV and NIII emission
lines, and consistent with that we find that NIV $\lambda$ 4058 is much stronger than NIII $\lambda \lambda$ 4634,42 emission. 

The effective temperature of our  final model (Fig.~\ref{fig:N346355}, lower) is strongly constrained by the weak He~I $\lambda 4471$ we measure, as the strengths of the He~II lines are no longer sensitive
to changes of effective temperatures in this regime.  Over the range 48,500 to 50,500 K
the EW of He I $\lambda$ 4471 changes from 51 to 23m\AA, which may be
compared to the $40\pm10$ m\AA\ we measure. 
Our effective temperature
for this star is  significantly lower than others in the literature
(Table~\ref{tab:compA}), which is necessitated by the 
(very weak) He~I $\lambda 4471$ we detect.  At 52,500 (found by Bouret et al.\ 2003 and Walborn et al. 2004 from the data described by Walborn et al.\ 2000) the expected EW of He I $\lambda 4471$ is only 8 m\AA, inconsistent with what we find with our
higher SNR data.

There is some incomplete nebular 
subtraction of the [OIII] lines, and we see residual emission in the H$\alpha$ profile, which we
attribute to nebulosity despite our careful sky (and nebular) subtraction.

\begin{figure}
\epsscale{0.6}
\plotone{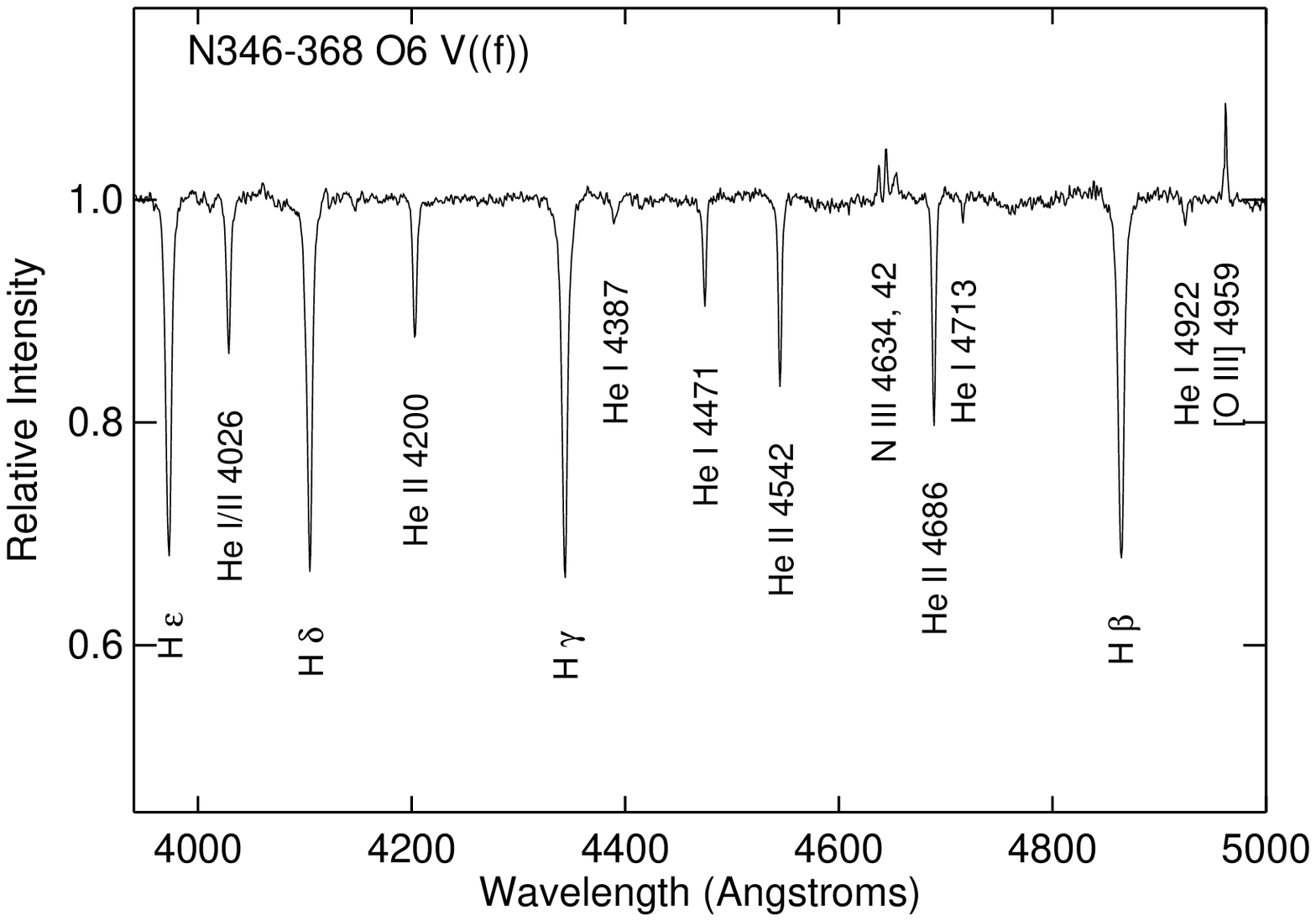}
\caption{\label{fig:N346368} NGC 346-368. The figure shows a section of the blue spectrum of this star,
with the prominent lines identified. 
}
\end{figure}

{\it NGC346-368}. The spectrum is shown in Fig.~\ref{fig:N346368}. It shows a (composite) spectral type of O6, with an 
EW of He~I $\lambda 4471$ of 380 m\AA\ and an EW of He II $\lambda 4542$ of
750 m\AA, yielding $\log W'=-0.30$.  NIII $\lambda \lambda 4634, 42$ is evident in 
emission but He II $\lambda 4686$ is strongly in absorption, suggesting a 
luminosity class of ``V((f))", consistent with the measured absolute magnitude
$M_V=-5.0$.  However, when we measured the radial velocities of the lines we
found a significant difference (50 km s$^{-1}$)
between the values for He~I and He~II.   The star shows no light variations in the
eclipsing binary search alluded to above, but we conclude it is likely a binary.  We attempted
a few fits, but found nothing satisfactory.

The spectrum was modeled by Bouret et al (2003), who call the star an O4-5 V((f)), considerably
earlier than what we find here for the composite type, or for the O6 V((f)) type found by Massey et al.\ (1989).
Heap et al.\ (2006) give further details of this fitting, noting
that the radial velocities from the UV spectrum and optical disagreed by 70 km s$^{-1}$,
and suggesting that the star might be a binary.  Nevertheless, they include their analysis
of this star in their effective temperature compilation. 

\begin{figure}
\epsscale{0.6}
\plotone{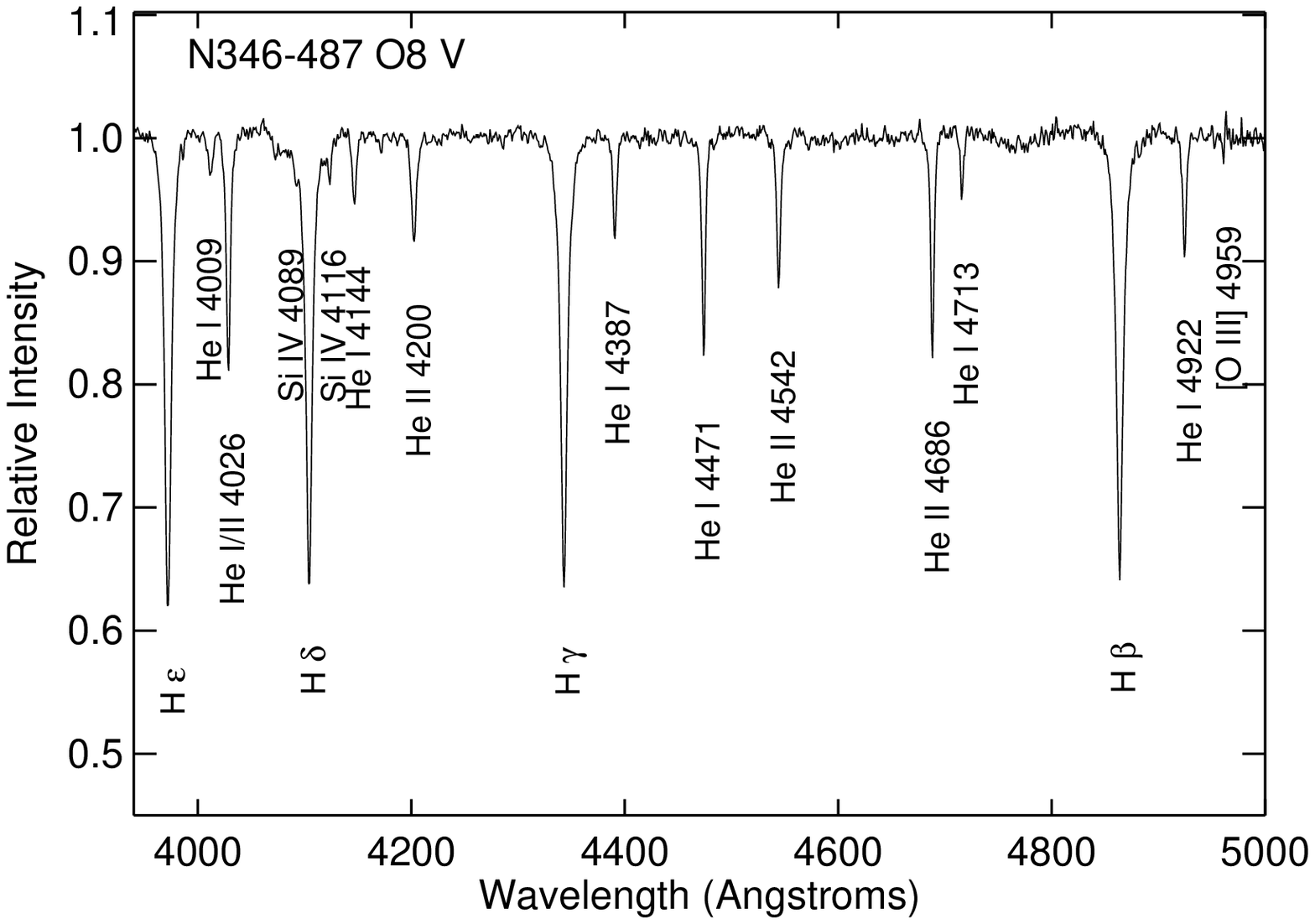}
\vskip 50pt
\plotone{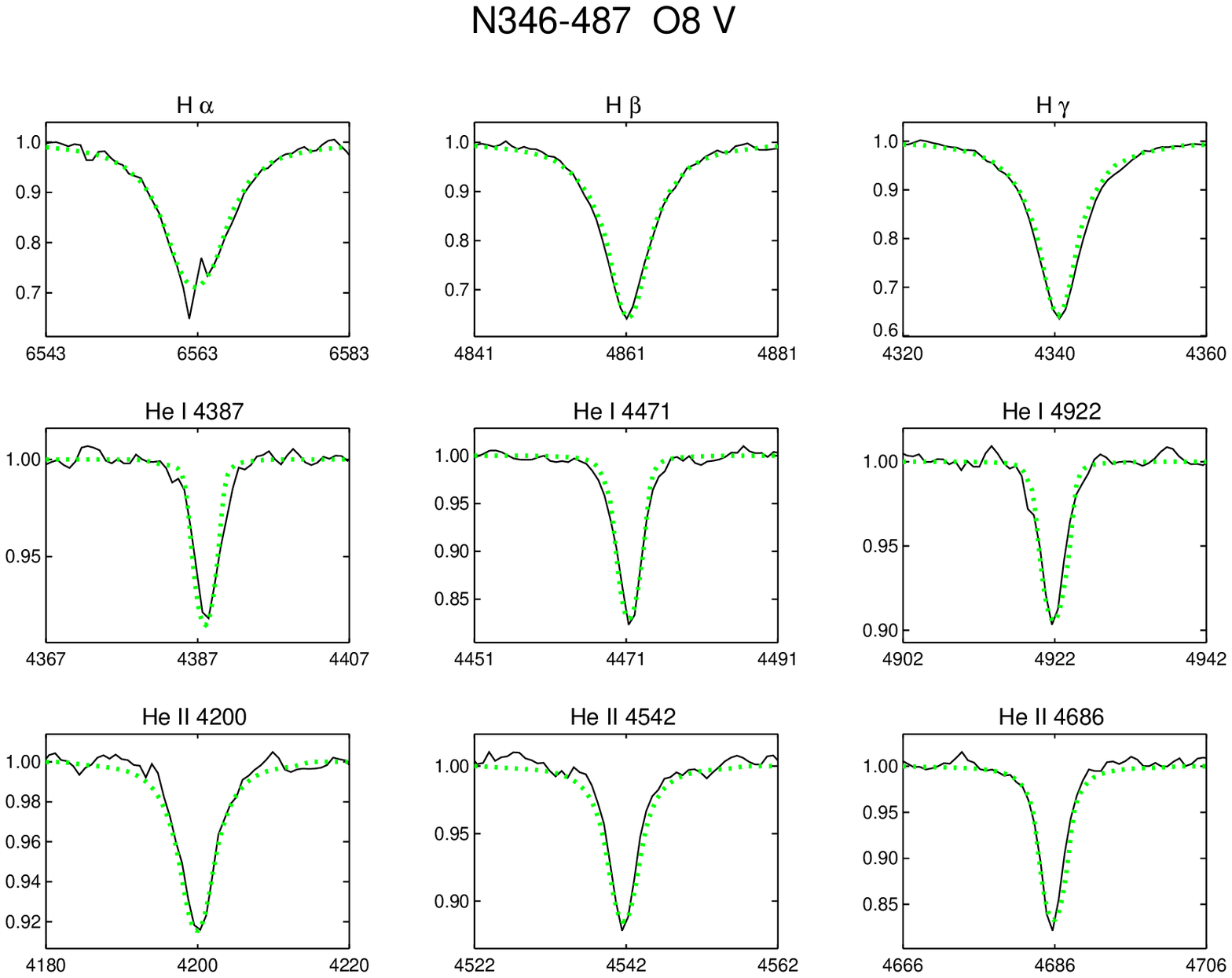}
\caption{\label{fig:N346487} NGC 346-487. The upper figure shows a section of the blue spectrum of this star,
with the prominent lines identified.  The lower figure shows the fits (dotted) for the principle diagnostic lines.
}
\end{figure}

{\it NGC346-487}.  The spectrum of this star is shown in Fig.~\ref{fig:N346487} (upper).  Visually
the spectral subtype is O8, as He~I $\lambda 4471$ is slightly stronger than He~II $\lambda 4542$.  
We measure EWs of 790 m\AA\ and 615 m\AA, respectively, leading to $\log W'=+0.11$,
also indicative of an O8 type.  He II $\lambda 4688$ is strongly in absorption, and there
is no NIII $\lambda \lambda 4638, 42$ emission, and so we call this an O8 V, which is
also in accord with the $M_V=-4.6$.  The star was classified quite a bit earlier, O6.5 V, by Massey
et al.\ (1989).   Our fits, shown in Fig.~\ref{fig:N346487} (lower), were good.

The star was also analyzed by Bouret et al.\ (2003), who derived an effective temperature 3000 K cooler than
our result, along with a significantly (0.2~dex) smaller value for $\log g$ (Table~\ref{tab:compA}).
Their values were derived primarily from the UV, as they were unable to obtain a satisfactory fit to
the optical without assuming an unlikely low value for the metallicity.  They concluded that their optical
spectrum was contaminated by other stars on the slit.   Below, we argue that the problem with their optical
data was moonlight contamination, and compare their data with ours (\S~\ref{Sec-compare}).

\begin{figure}
\epsscale{0.6}
\plotone{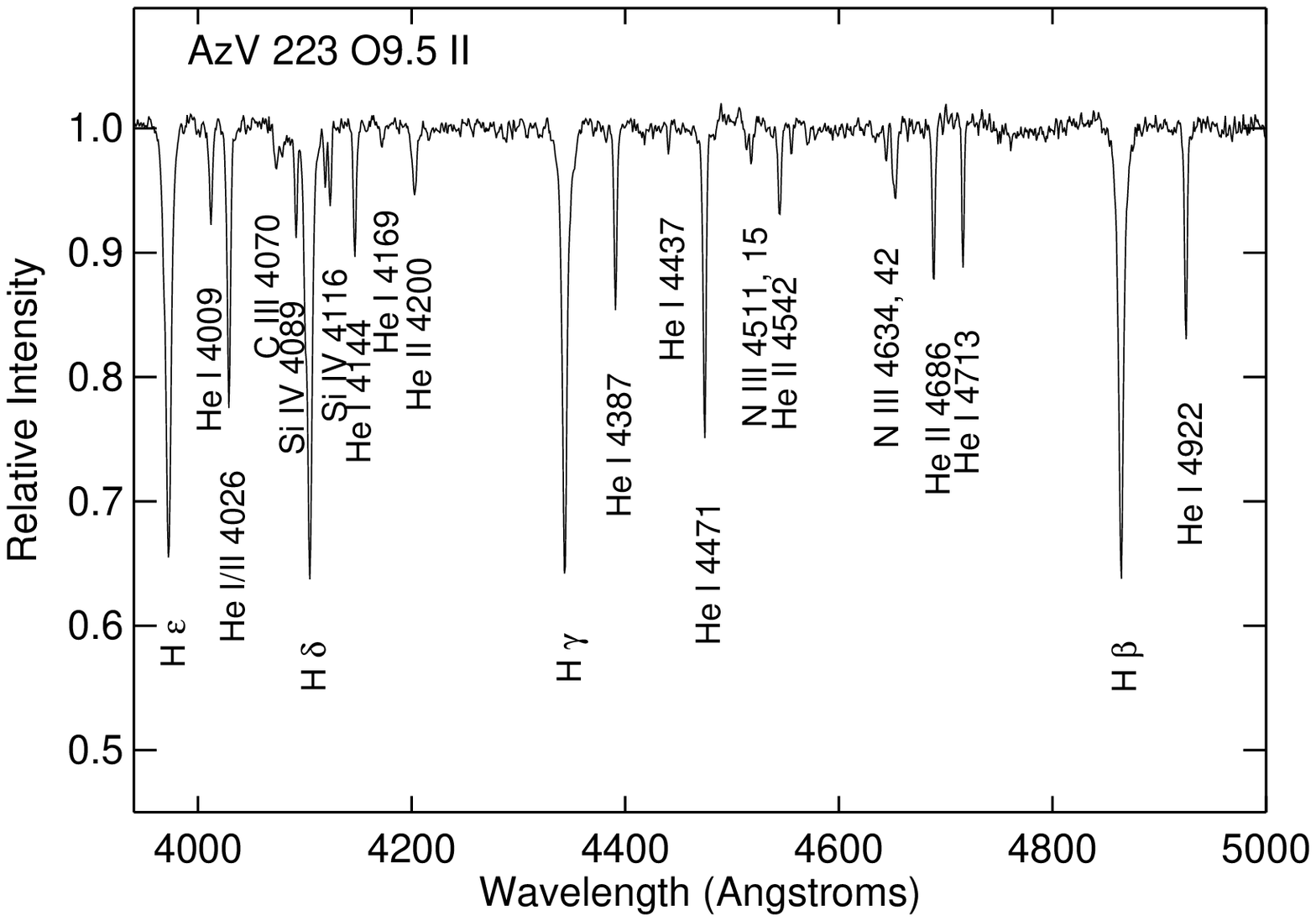}
\vskip 50pt
\plotone{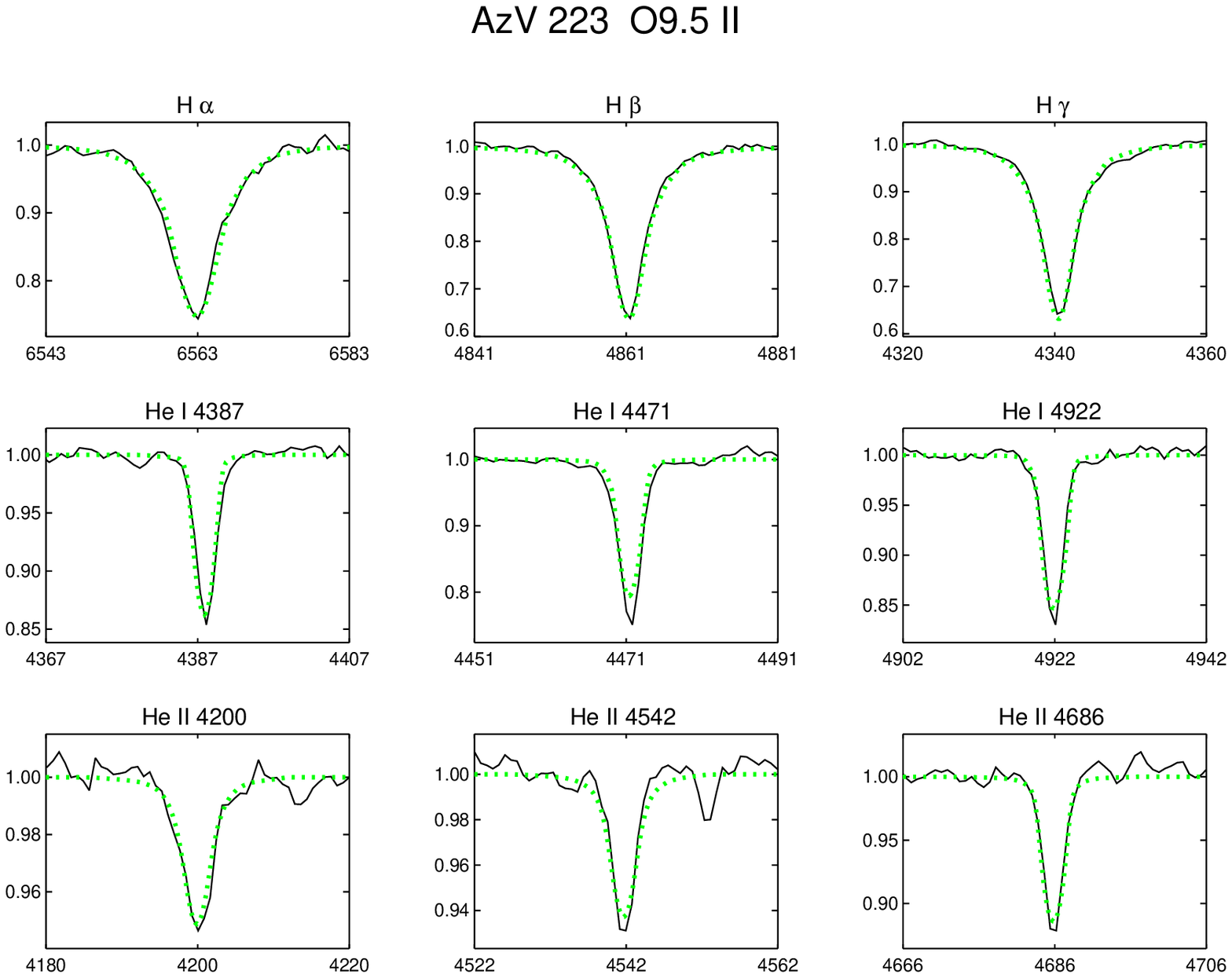}
\caption{\label{fig:AzV223} AzV 223. The upper figure shows a section of the blue spectrum of this star,
with the prominent lines identified.  The lower figure shows the fits (dotted) for the principle diagnostic lines.
}
\end{figure}

{\it AzV 223}.  We show the spectrum of this star in Fig.~\ref{fig:AzV223} (upper).  The He~I $\lambda 4471$ line
(EW of 950 m\AA)
is much stronger than He~II $\lambda 4542$ (EW of 250 m\AA), leading to an O9.5 type ($\log W'=0.58$). 
He~II $\lambda 4200$ is still reasonably strong (EW of 285 m\AA) and so we were not tempted to call this
a B0, although others might call it an O9.7 type.  Were this a Galactic star, we would call the luminosity class
``V" or ``III" based on the strength of He~II $\lambda 4686$ absorption.  However, the absolute magnitude $M_V=-5.6$
is more intermediate between a ``III" and a ``I". Given the expected weakness of the wind lines in the SMC, we adopt
an O9.5 II classification for this star.  Previously the star was classified as O9 III by Garmany et al.\ (1987).
The fits we obtained are quite good, and are shown in Fig.~\ref{fig:AzV223} (lower).

\begin{figure}
\epsscale{0.6}
\plotone{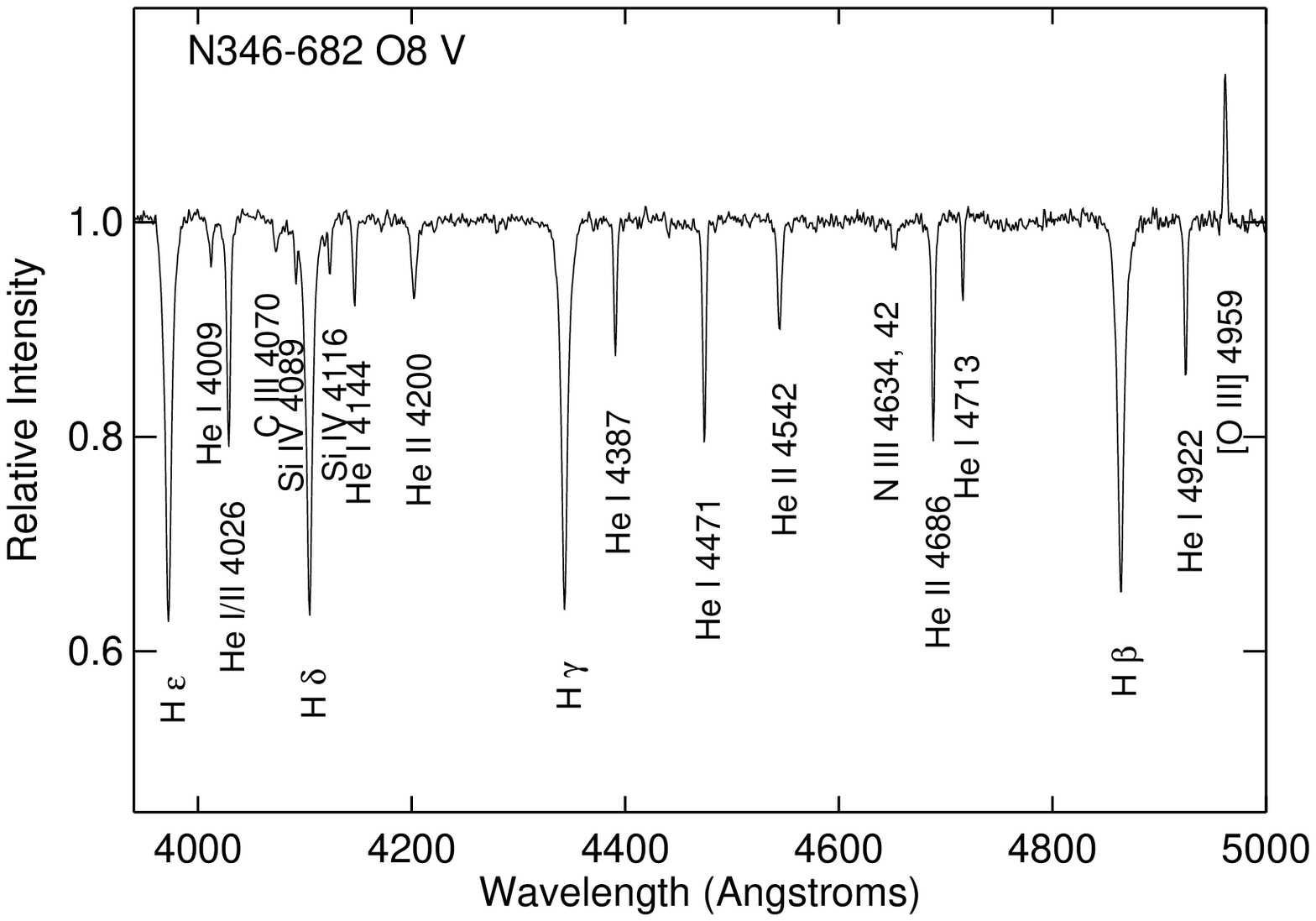}
\vskip 50pt
\plotone{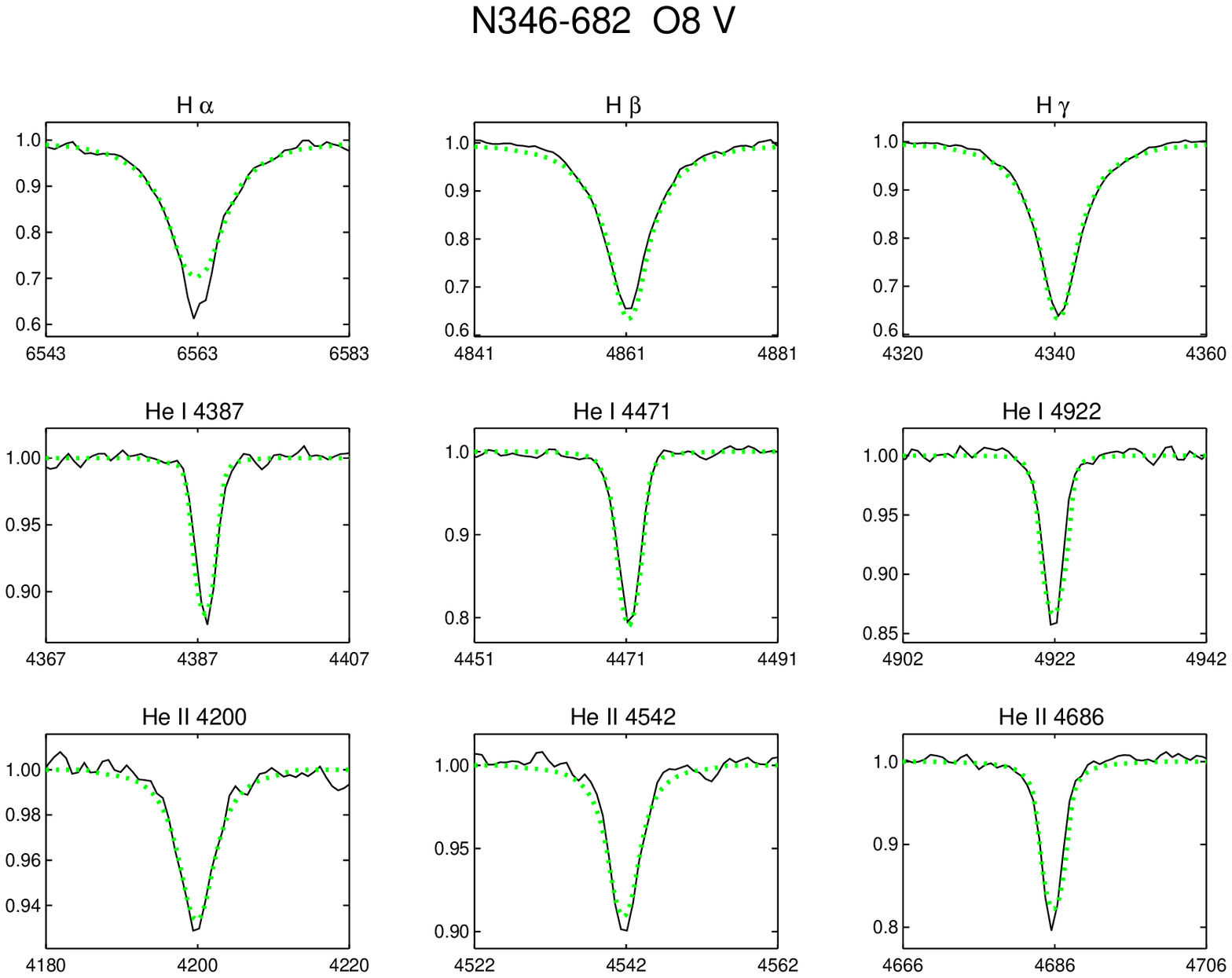}
\caption{\label{fig:N346682} NGC 346-682. The upper figure shows a section of the blue spectrum of this star,
with the prominent lines identified.  The lower figure shows the fits (dotted) for the principle diagnostic lines.
}
\end{figure}

 {\it NGC 346-682}.  The spectrum is shown in Fig.~\ref{fig:N346682} (upper).  We measure EWs of 810 m\AA\ and
 530 m\AA\ for He~I $\lambda 4471$ and He~II $\lambda 4542$, respectively, leading to $\log W'=+0.18$, 
 consistent with the O8 subtype suggested by the visual appearance.  We conclude that the star is a
 dwarf, based upon the strong He~II $\lambda 4686$ absorption and the fact that $M_V=-4.2$.  Our O8 V type 
 is identical to that found by Massey et al.\ (1989).
 
 The model for this star gave good agreement with the observed spectrum (Fig.~\ref{fig:N346682}, lower).  The star had
 been previously modeled both by Mokiem et al.\ (2006) and Heap et al.\ (2006).  As can be seen in Table~\ref{tab:compA}
 the FASTWIND optical results (both ours and Mokiem et al.\ 2006) give somewhat higher temperatures and larger values
 for $\log g$  than the TLUSTY analysis by Heap et al.\ (2006).
   
\begin{figure}
\epsscale{0.6}
\plotone{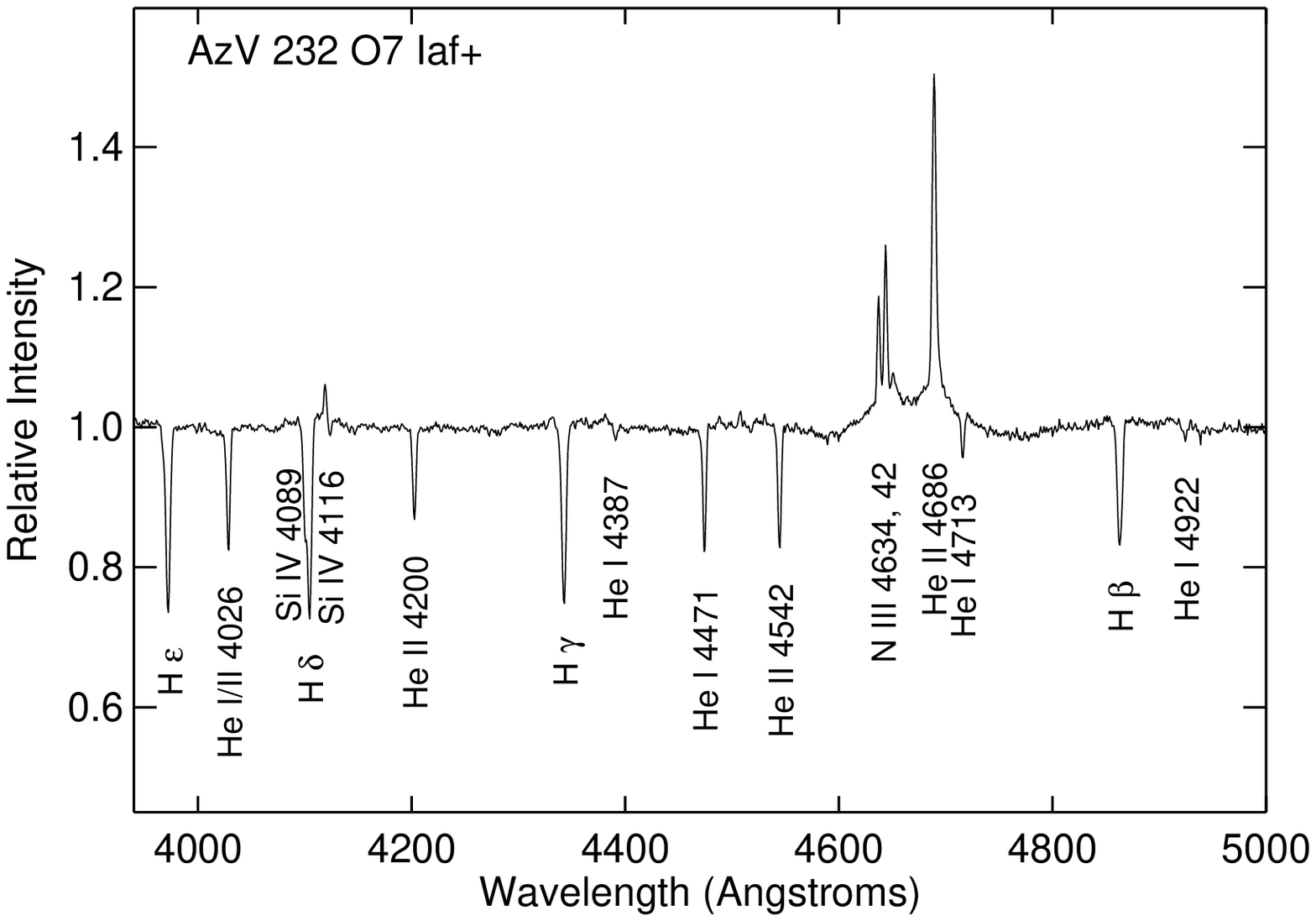}
\caption{\label{fig:AzV232} AzV 232. The figure shows a section of the blue spectrum of this star,
with the prominent lines identified. 
}
\end{figure}

 {\it AzV 232}.  This star, also known as Sk 80,  is a well-known O7 Iaf+ star (see, for example, Walborn \& Fitzpatrick 1990), a classification which we do not dispute.
 Our spectrum is shown in Fig.~\ref{fig:AzV232}.
 We measure EWs of 610 m\AA\ and 680 m\AA\ for He I $\lambda 4471$ and He II $\lambda 4542$, respectively; $\log W'=-0.05$, consistent
 with the O7 designation.  The star has extremely strong NIII $\lambda \lambda 4638, 42$ and He II $\lambda 4686$ emission.  Note that
 the Si IV $\lambda 4116$ emission is considerably stronger than that of Si IV $\lambda 4089$.  We noticed this in a number of our spectra; this 
 inequality is visible in Figure 5 of the Walborn \& Fitzpatrick (1990) atlas but is quite obvious in our higher SNR spectrum.  
 
 Extreme supergiants are often hard to fit, but in this case we were completely stymied after 40 models.
   Even by significantly increasing the fractional He ratio
 we could not obtain sufficiently strong He I or He II lines.  Mokiem et al.\ (2006) describe this star as a binary
 based on radial velocity variations, although their automatic fitting procedure arrived at some ``best fit" properties,
 given in Table~\ref{tab:compA}.  The star was also modeled successfully by Crowther et al.\ (2002), who derive
 cooler temperature (32000 K vs 34100 K) than Mokiem et al.\ (2006) and a somewhat lower surface gravity (3.1 vs. 3.4).
 
\begin{figure}
\epsscale{0.6}
\plotone{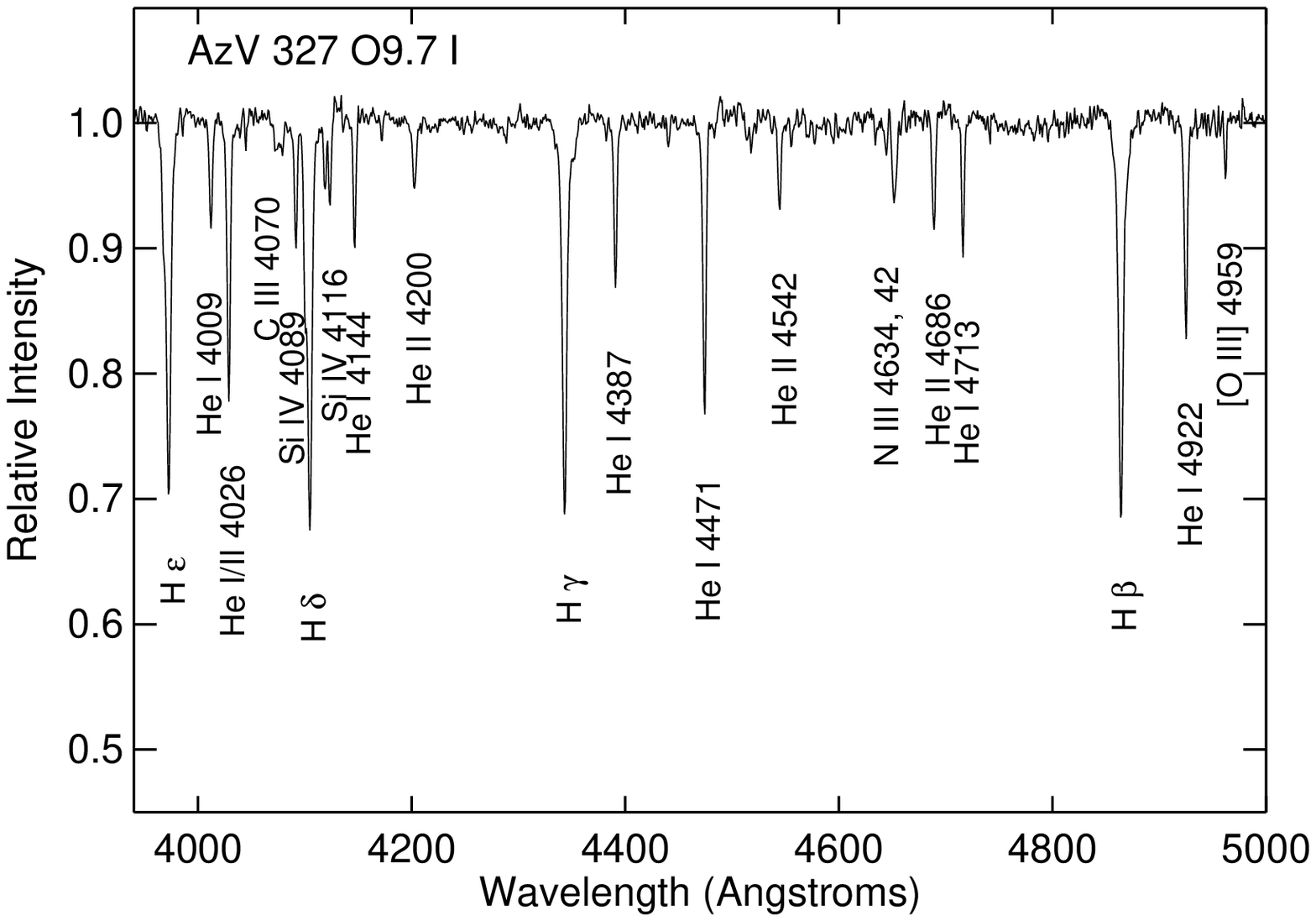}
\vskip 50pt
\plotone{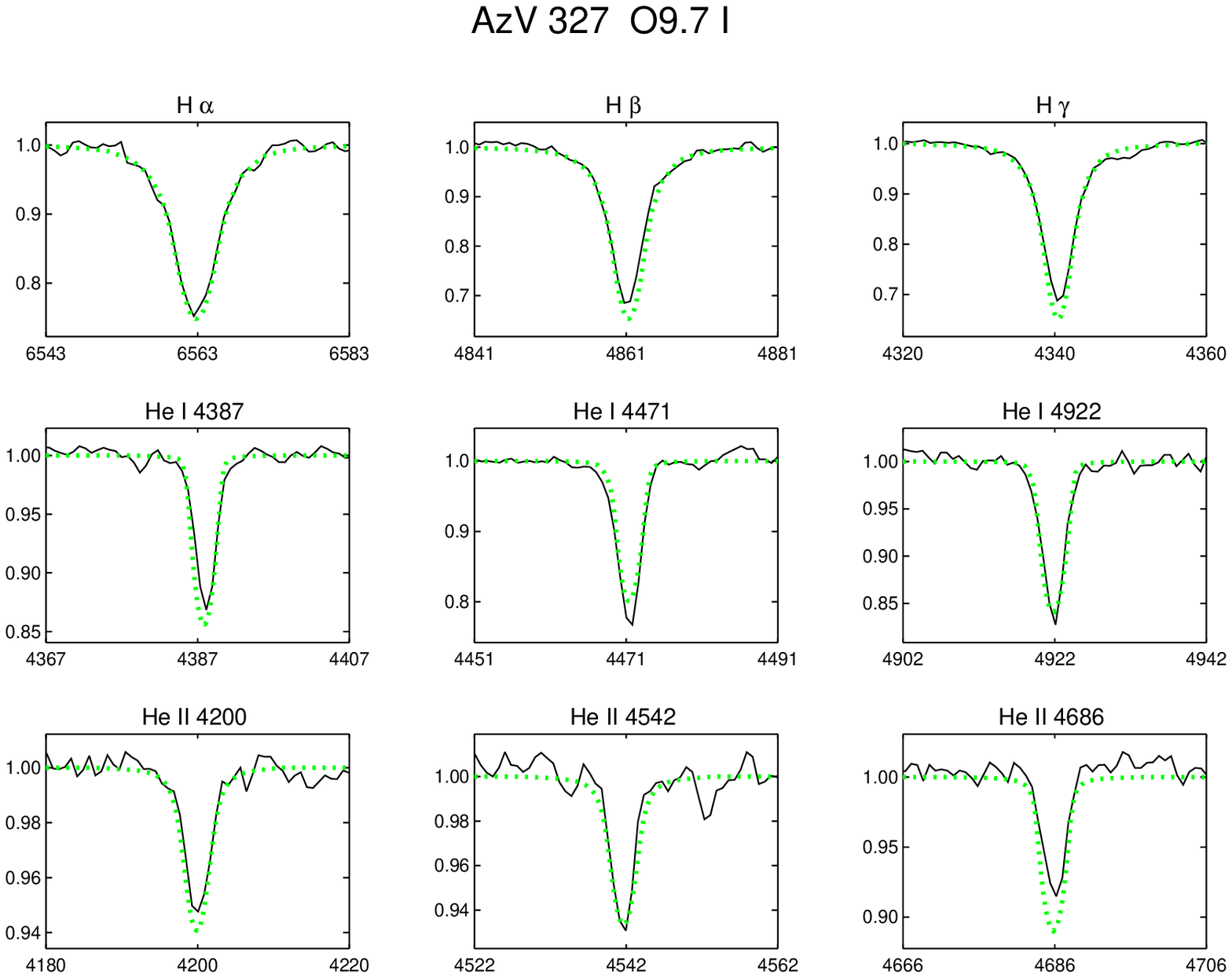}
\caption{\label{fig:AzV327} AzV 327. The upper figure shows a section of the blue spectrum of this star,
with the prominent lines identified.  The lower figure shows the fits (dotted) for the principle diagnostic lines.
}
\end{figure}

 {\it AzV 327}.  This star is a late-type O supergiant, as evidenced by weak He II and strong He I, sharp and deep lines, and strong metal lines, such
 as Si IV and N III (Fig.~\ref{fig:AzV327}, upper).  We measure EWs of 905 and 240 m\AA\ for He~I $\lambda 4471$ and He~II $\lambda 4542$,
 respectively, leading to $\log W'=0.58$, making this at least O9.5 or later.  A  B0 I is precluded by the strength of He II 
 (especially, say, He II $\lambda 4200$), but the strength of the Si IV features led us to use the intermediate spectral type O9.7 I,
 as the spectrum closely resembles that shown for $\mu$ Nor by Walborn \& Fitzpatrick (1990).  The star had been previously classified as
 O9 I by Massey et al.\ (2000), and is listed as  ``O9.5 II-Ibw" by Walborn et al.\ (2000).
 
 We obtained very nice fits for this star (Fig.~\ref{fig:AzV327}, lower), although it was necessary to use a slightly elevated He/H ratio (0.15) in order
 to make the He lines (both He I and He II) strong enough.  The star has been previously modeled by Heap et al.\ (2006), who derived very similar parameters (Table~\ref{tab:compA}). They found that the star had enriched N.  
  
\begin{figure}
\epsscale{0.6}
\plotone{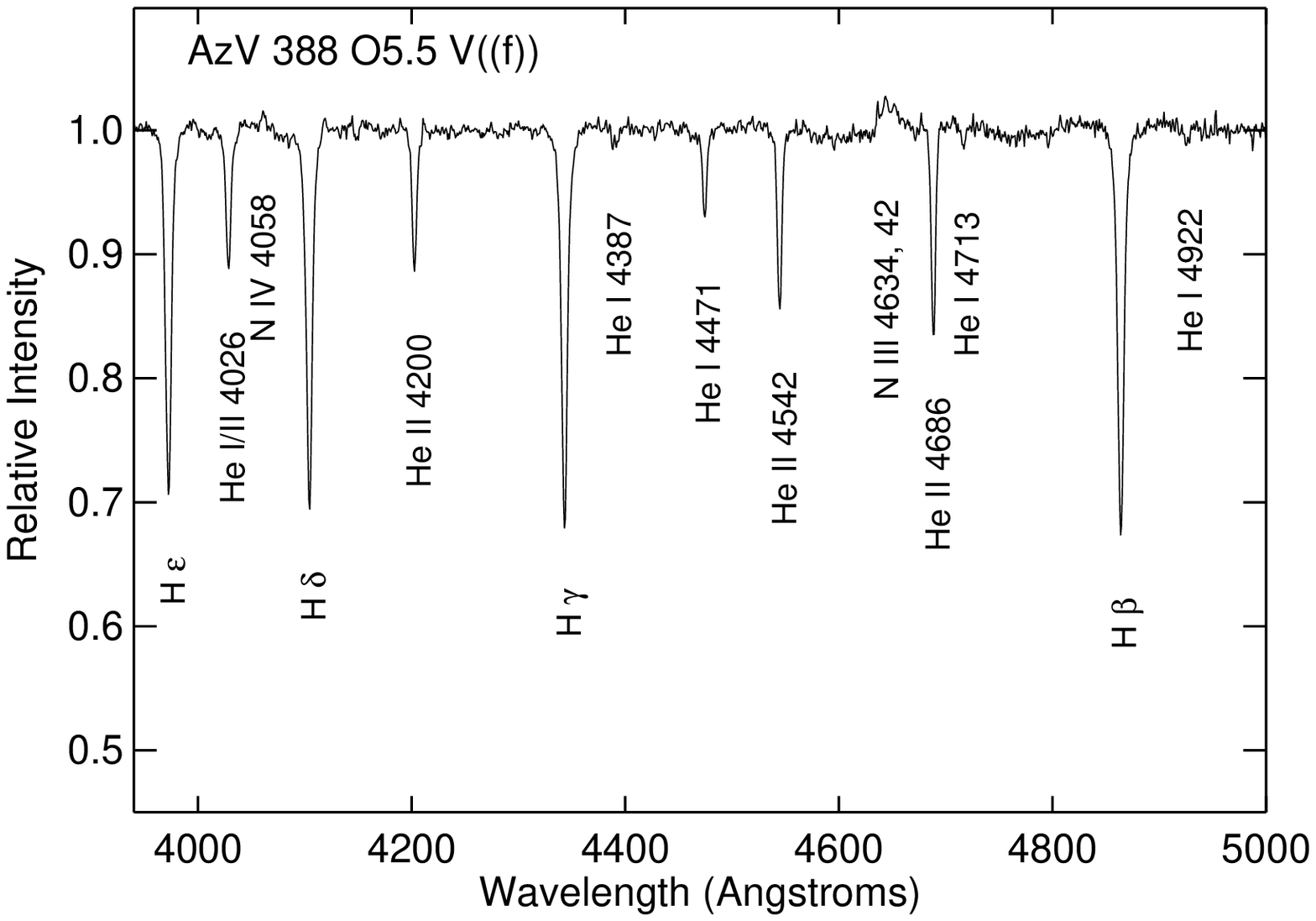}
\vskip 50pt
\plotone{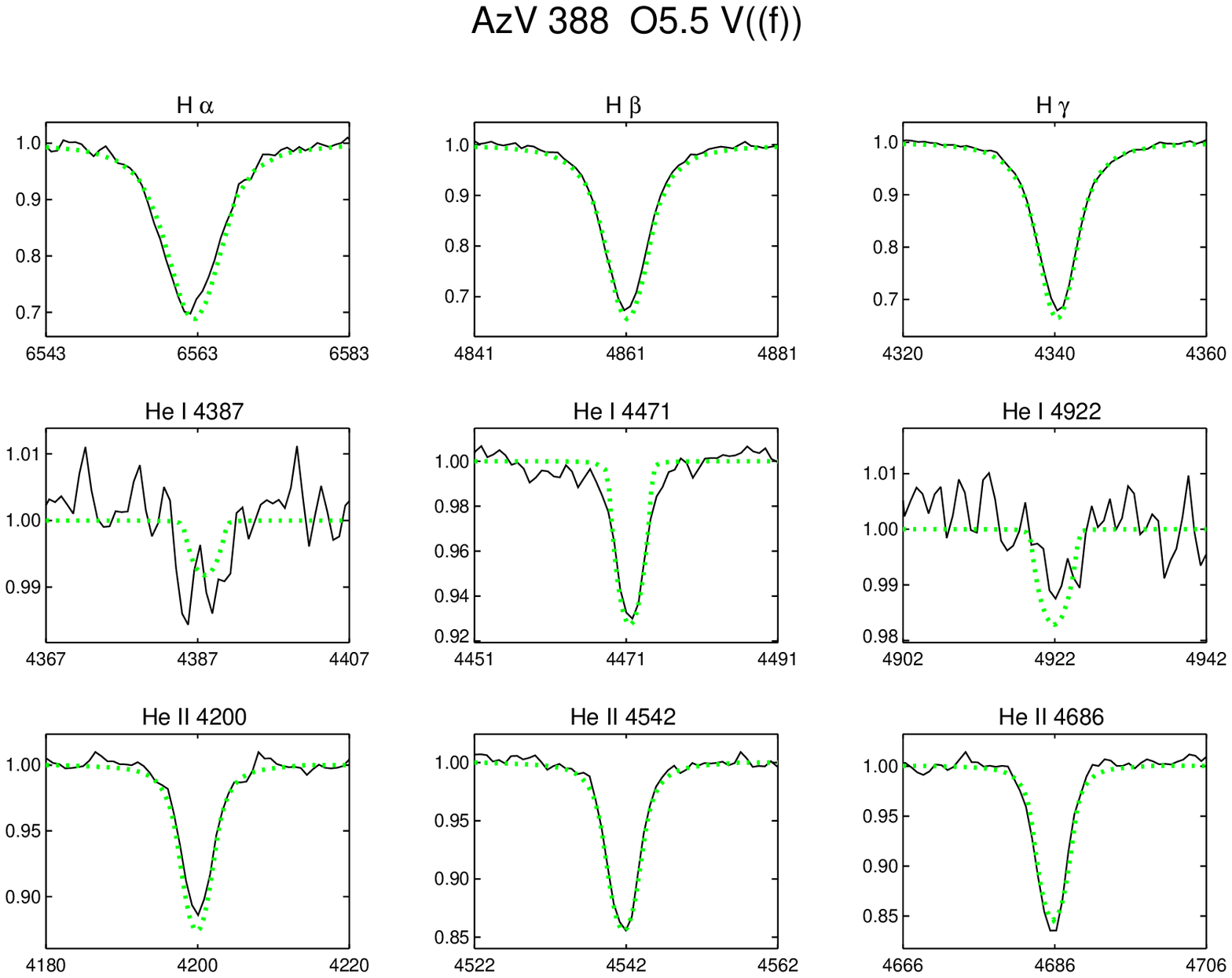}
\caption{\label{fig:AzV388} AzV 388. The upper figure shows a section of the blue spectrum of this star,
with the prominent lines identified.  The lower figure shows the fits (dotted) for the principle diagnostic lines.}
\end{figure}

 {\it AzV 388}.  This star is a mid-early O star, with a visual type of O5.5 (Fig.~\ref{fig:AzV388}, upper).  We measure EWs of 325 m\AA\ and 800 m\AA\ for He~I $\lambda 4471$
 and He~II $\lambda 4542$, respectively, leading to $\log W'=-0.39$, or an O5.5 subtype.  NIII $\lambda \lambda 4634, 42$ shows weak
 emission, but He~II $\lambda 4686$ is strongly in absorption.   The absolute magnitude is $-5.15$ (Table~1).  All of this is consistent with
 a dwarf luminosity class, and we call the star O5.5 V((f)).  The star was previously classified as O4 V by Garmany et al.\ (1987) and Walborn et al.\ (1995).  
 
 The fits are quite good, and are shown in Fig.~\ref{fig:AzV388} (lower).  The star had been previously modeled by Mokiem et al.\ (2006) using 
 a similar version of FASTWIND, with nearly identical results (Table~\ref{tab:compA}).
 
 \subsubsection{LMC}

\begin{figure}
\epsscale{0.6}
\plotone{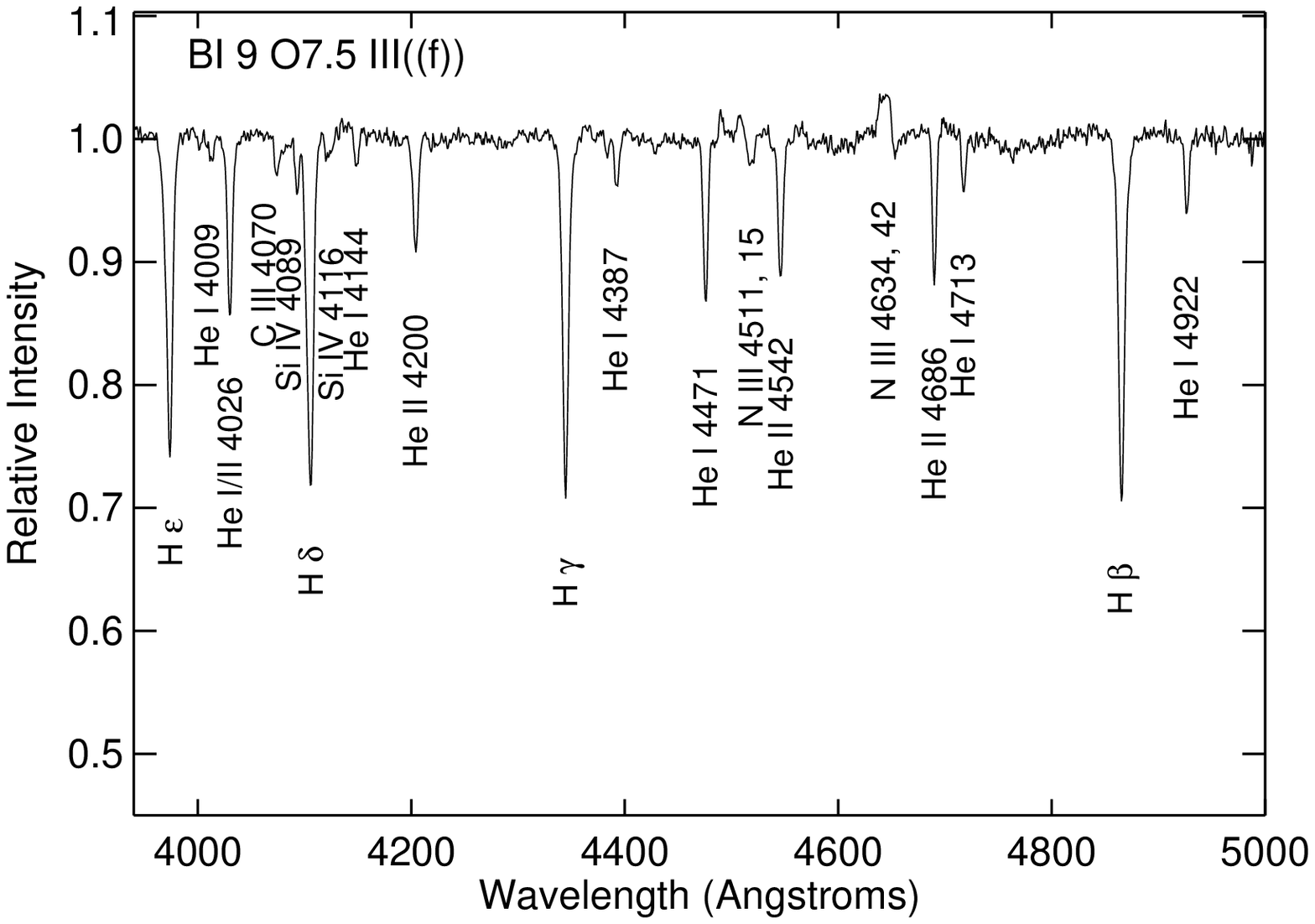}
\caption{\label{fig:BI9} BI 9. The figure shows a section of the blue spectrum of this star,
with the prominent lines identified. 
}
\end{figure}

 {\it BI 9}.  This star appears to be  a mid-O dwarf or giant (Fig.~\ref{fig:BI9}). 
 We measure EWs of 670 m\AA\ and 650 m\AA\ for He I $\lambda 4471$ and He II $\lambda 4542$,
 respectively, leading to $\log W'=+0.01$, or an O7.5 subtype.  NIII $\lambda \lambda 4634,42$ is strongly in emission.  This would suggest
 that the star is a giant, but He II $\lambda 4686$ is 
strongly in absorption.  Were this star in the SMC, with its lower metallicity, we would attribute the behavior of He II $\lambda 4686$ purely to the
weaker stellar winds we expect.  A giant luminosity class is consistent with the absolute magnitude, $-5.8$, but we were distrustful of the
strong He II $\lambda 4686$, which suggests that perhaps this star is a binary. We are forced to call this star an O7.5 III((f)).  It was called
an O8 III by Crampton (1979).

Our best fit models have $\log g=3.5$ and $T_{\rm eff}=$36,000-36,250 K.  However, we noticed that while the He I lines required
a $v \sin i\sim 180$ km s$^{-1}$, the He II lines were substantially broader, with an apparently $v \sin i \sim 250$ km s$^{-1}$.  The
situation is reminiscent of P.M.'s first foray into the study of massive stars, the study of HDE 228766 (Massey \& Conti 1977).  There Walborn (1973) had asserted that the He I lines were broad and the He II lines were sharp.  This led to the first double-lined
spectroscopic orbit for this interesting star.  Here we take the converse situation (He I narrow and He II broad) as an equally valid
indicator that the star is likely a binary.

\begin{figure}
\epsscale{0.6}
\plotone{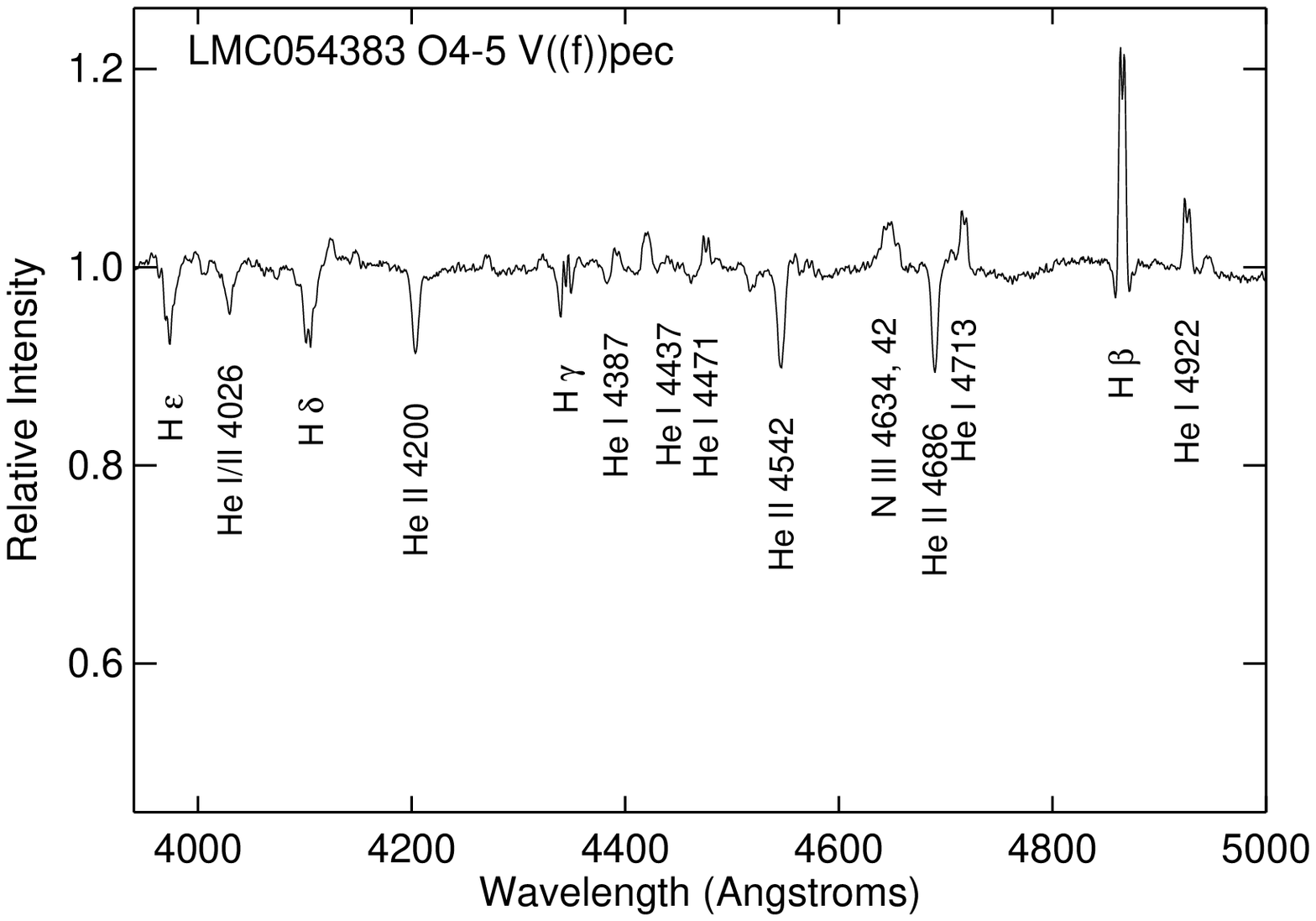}
\vskip 50pt
\plotone{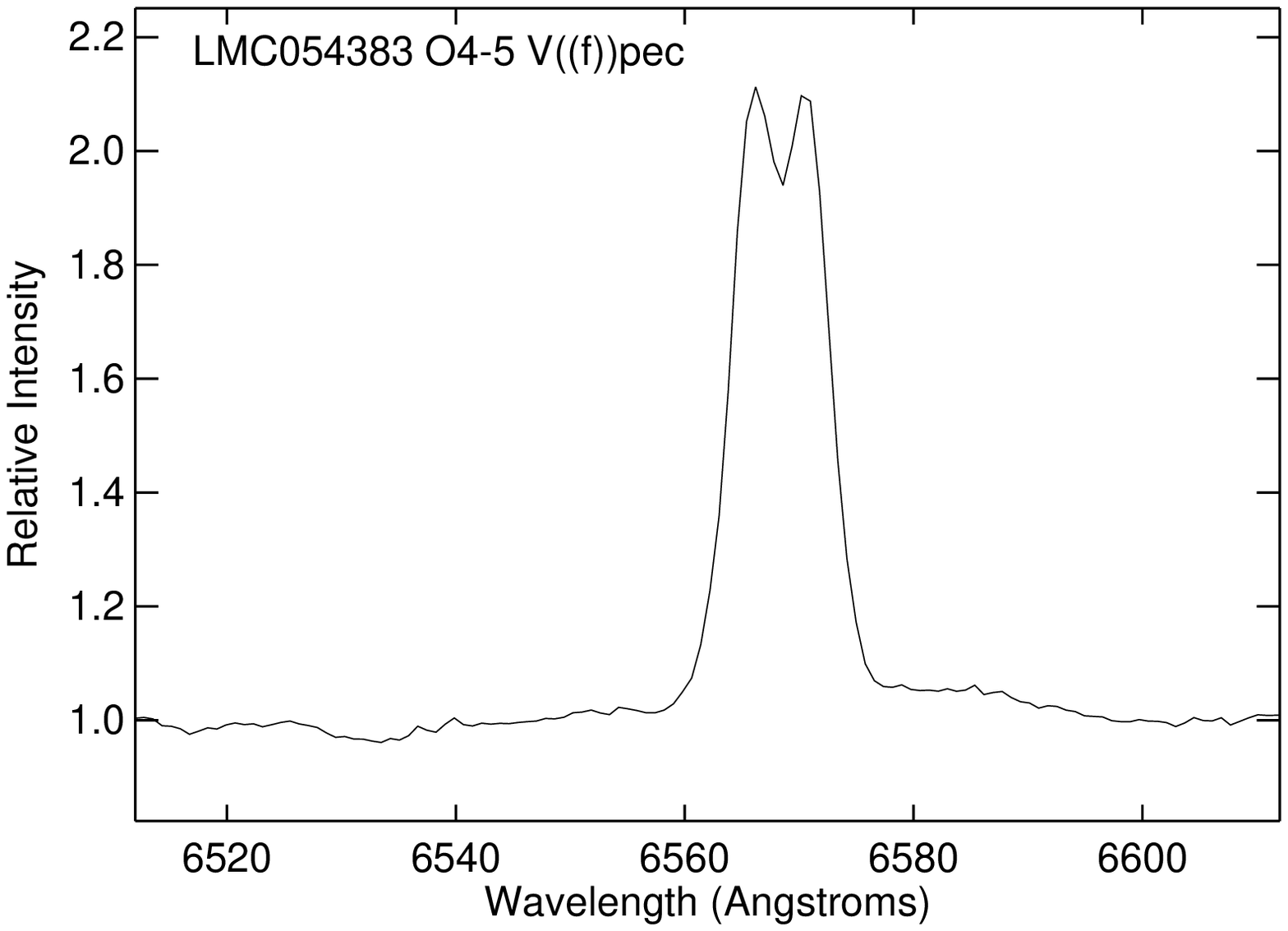}
\caption{\label{fig:LMC054383}
LMC 054383. The upper figure shows a section of the blue spectrum of this star,
with the prominent lines identified.   The lower figure shows the H$\alpha$ profile.
}
\end{figure}

{\it LMC 054383}.  This star was included in our program because an earlier spectrum, obtained with the fiber positioner Hydra, showed very weak hydrogen lines compared to He II lines, and we had described its spectrum to several colleagues as being a H-poor O3 star.
However,  with the higher SNR, long-slit spectrum we obtained here we find that the hydrogen and HeI lines are simply filled in by
emission (Fig.~\ref{fig:LMC054383}, upper).  Furthermore, the emission is double-peaked.  This is most clearly evident in the H$\alpha$ emission
profile (Fig.~\ref{fig:LMC054383}, lower).  The obvious interpretation is that the star is surrounded by a disk of material.  Note that this spectrum is not that of a classical ``Oe"
star
since NIII $\lambda \lambda 4634, 42$ is also in emission; the spectral type is early, possibly an O4-O5 given the amount of He I absorption that we see.  Given its absolute visual magnitude (-4.83) we call this an O4-O5V((f))pec. We made no attempt to model the star's spectrum.

\begin{figure}
\epsscale{0.6}
\plotone{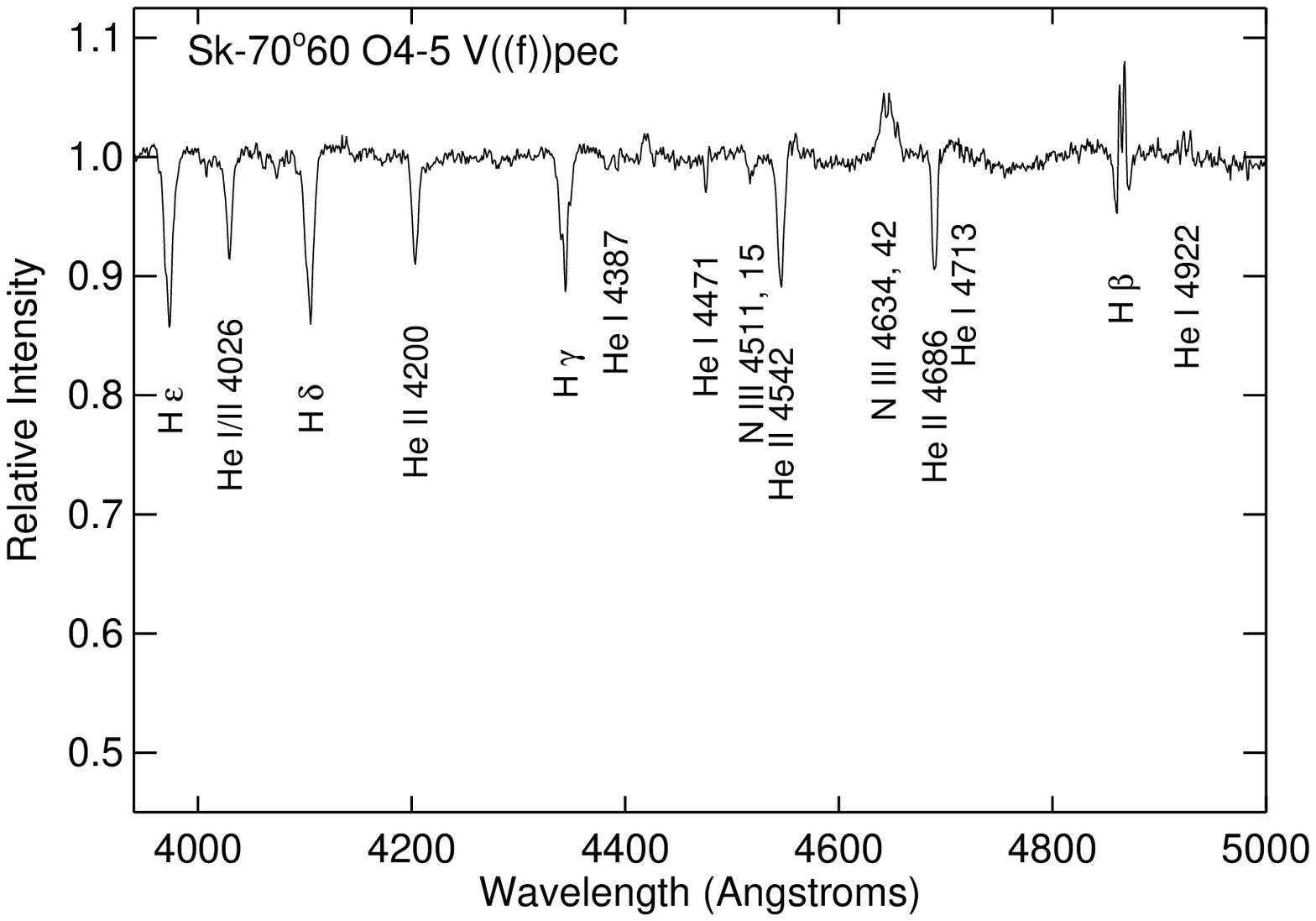}
\vskip 50pt
\plotone{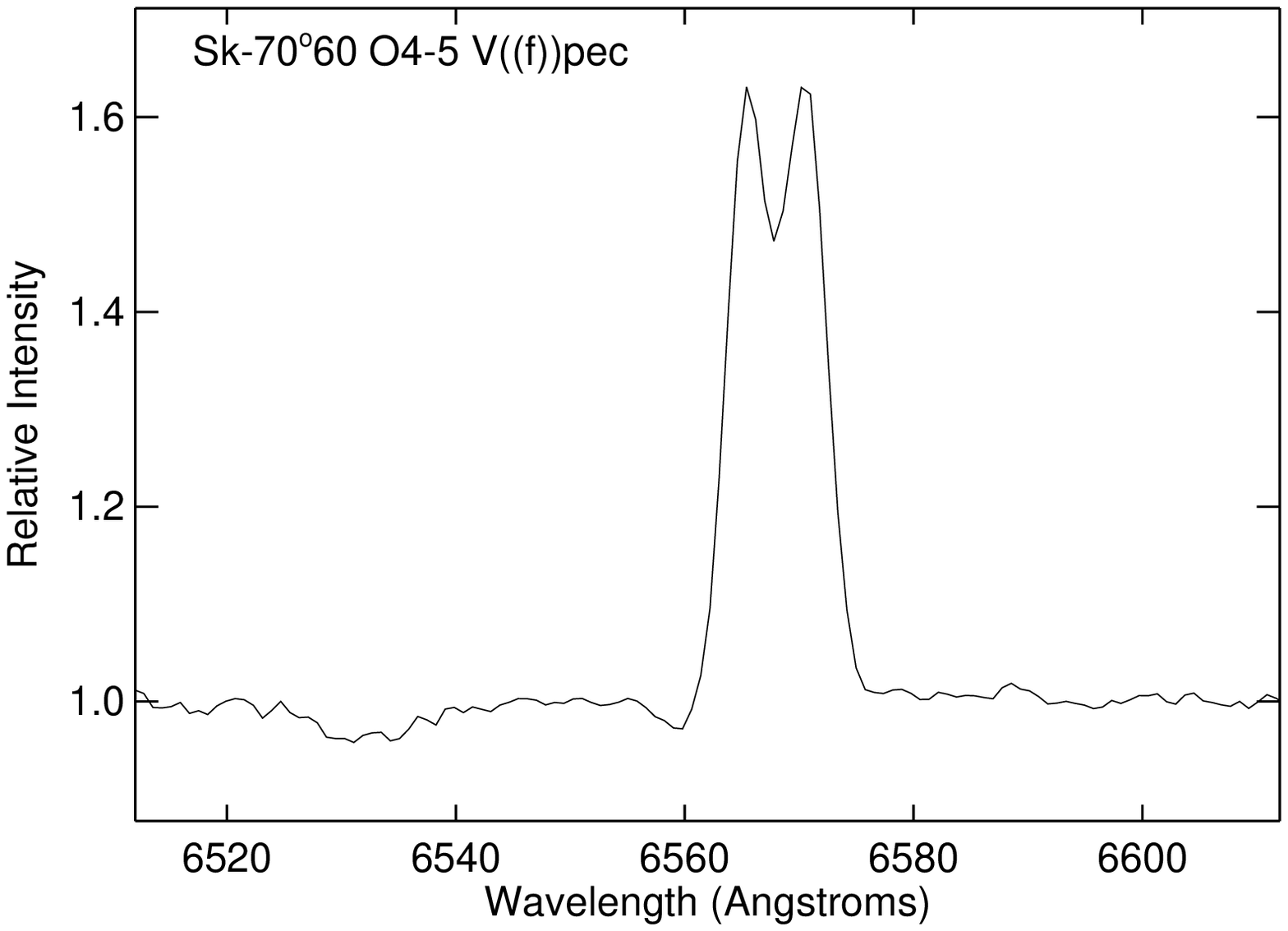}
\caption{\label{fig:Sk7060} Sk$-70^\circ 60$.The upper figure shows a section of the blue spectrum of this star,
with the prominent lines identified.   The lower figure shows the H$\alpha$ profile.
}
\end{figure}

{\it Sk$-70^\circ 60$}.  Oddly, this star is very similar to that of LMC 054383, with both Balmer and He I lines showing double-peaked emission.
The spectral type is early, O4-O5V((f))pec.  The blue spectrum is shown in Fig.~\ref{fig:Sk7060} (upper), and the H$\alpha$ profile in Fig.~\ref{fig:Sk7060} (lower).
 Again no attempt was made to model the
star.

\begin{figure}
\epsscale{0.6}
\plotone{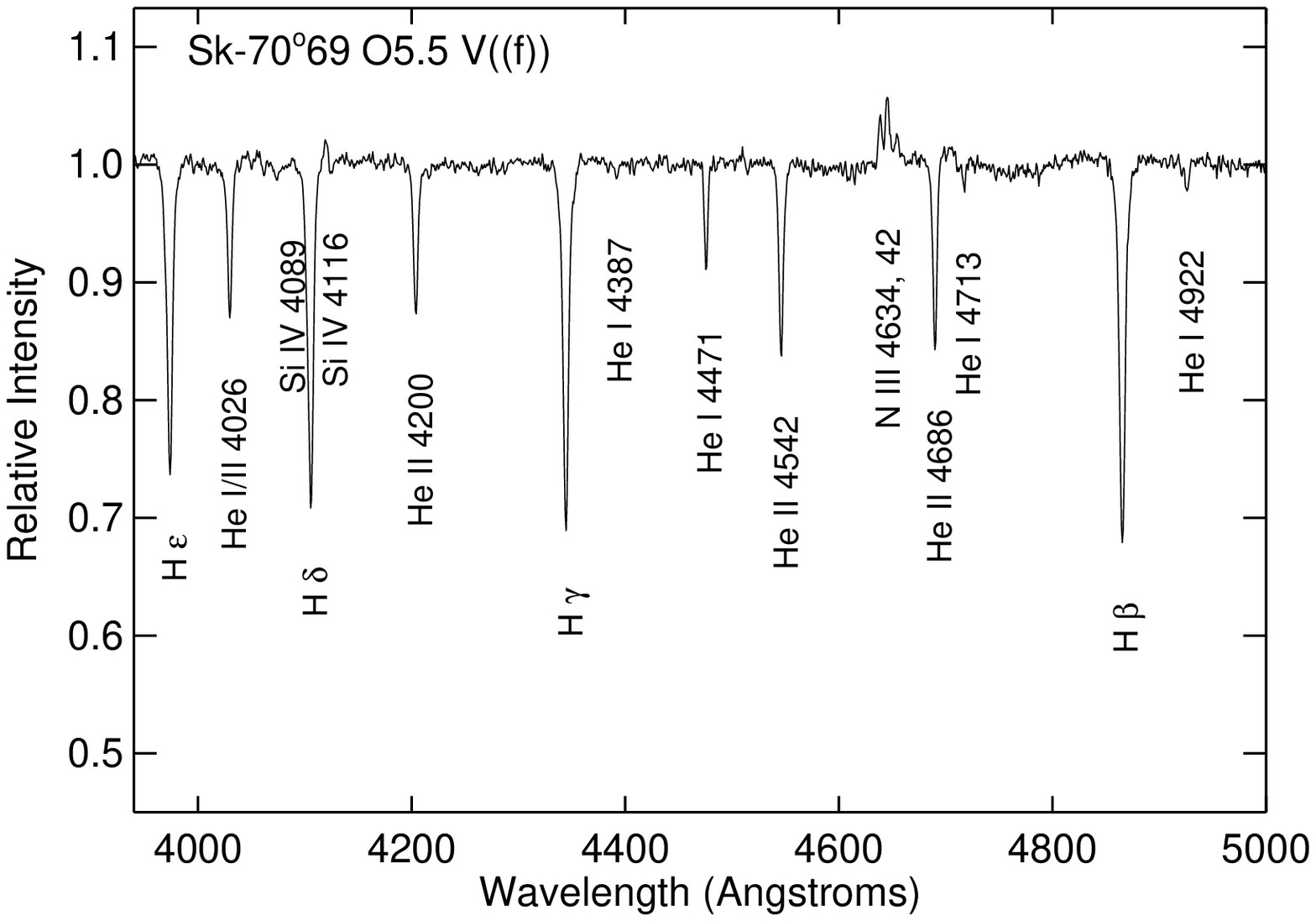}
\vskip 50pt
\plotone{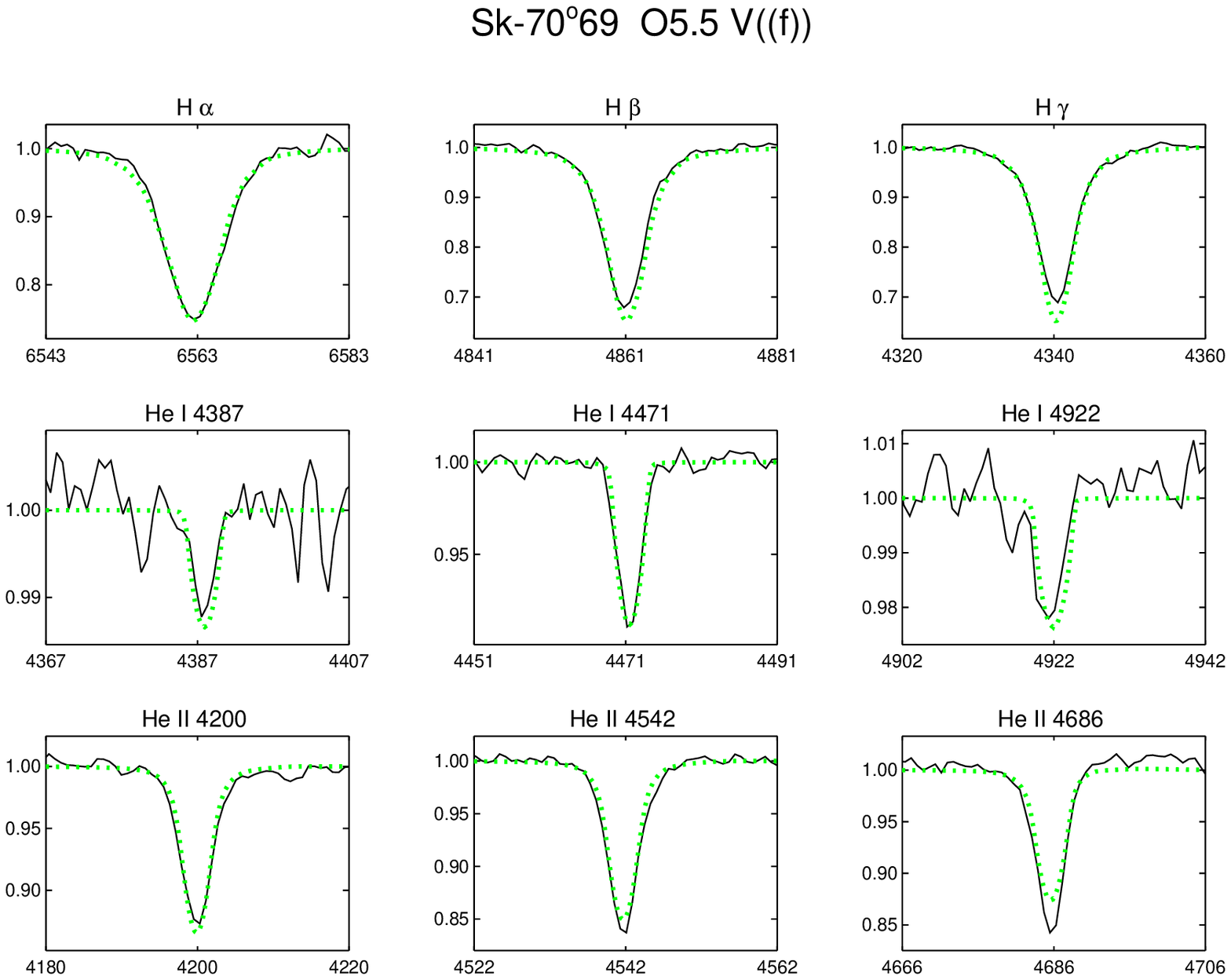}
\caption{\label{fig:Sk7069} Sk$-70^\circ 69$. The upper figure shows a section of the blue spectrum of this star,
with the prominent lines identified.  The lower figure shows the fits (dotted) for the principle diagnostic lines.
}
\end{figure}

{\it Sk$-70^\circ 69$}.  This is an early O dwarf (Fig.~\ref{fig:Sk7069}, upper), and we measure EWs of 340 m\AA\ and 870 m\AA\ for He I $\lambda 4471$ and He II $\lambda 4542$, respectively, leading to $\log W'=-0.41$, an O5.5 subtype.  NIII $\lambda \lambda 4634,42$ emission is expected in even a dwarf at these early
types, and  He II $\lambda 4686$ is strongly in absorption, suggesting a dwarf luminosity class.  The absolute magnitude -4.83 is also consistent
with a dwarf, and so we call this star an O5.5 V((f)).

 The fits we obtained for this star were excellent (Fig.~\ref{fig:Sk7069}, lower).  The star has also been analyzed by Mokiem et al.\ (2007), who found
 a higher effective temperature and somewhat higher values for the surface gravity
 and He/H content  (Table~\ref{tab:compA}).

\begin{figure}
\epsscale{0.6}
\plotone{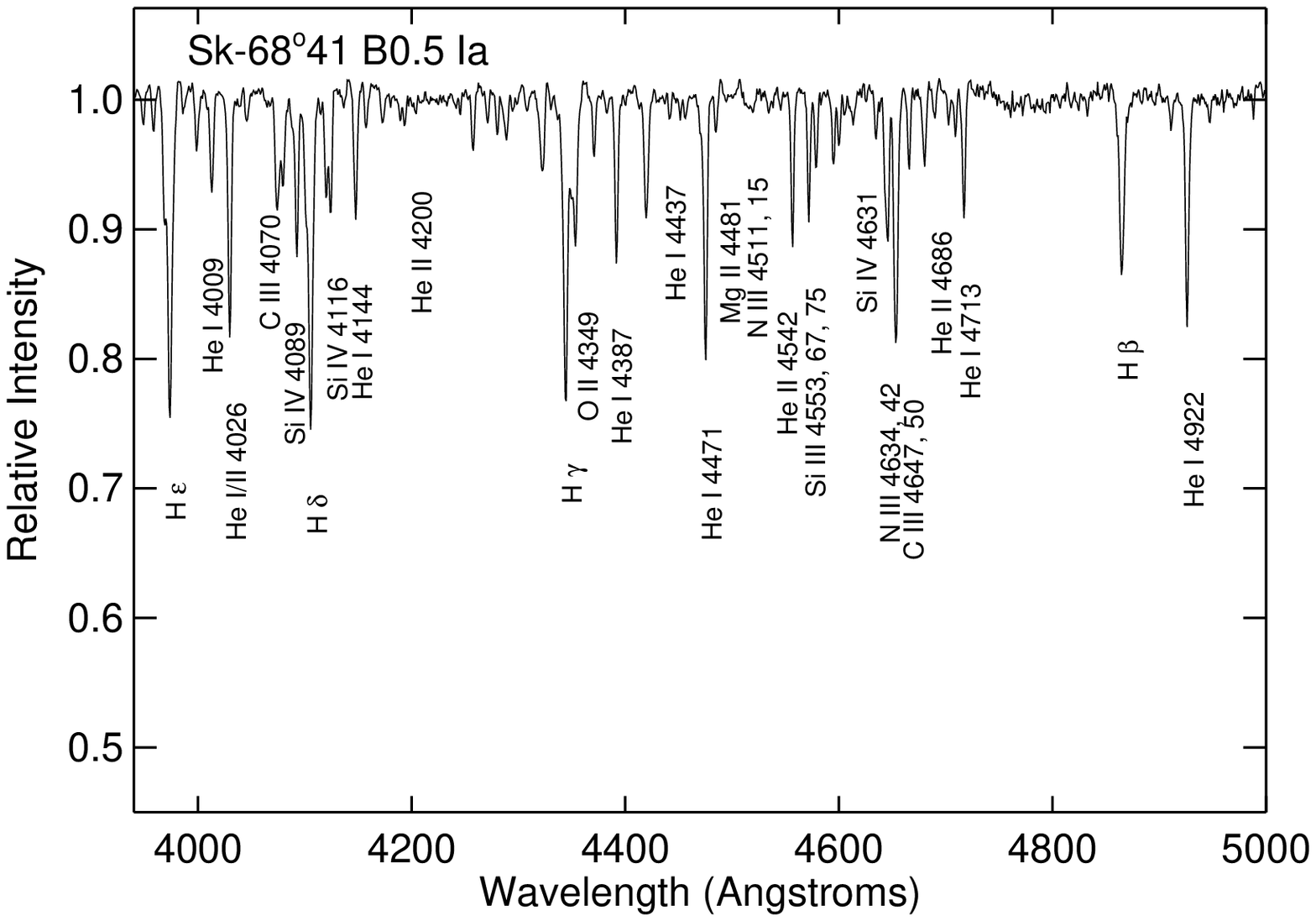}
\vskip 50pt
\plotone{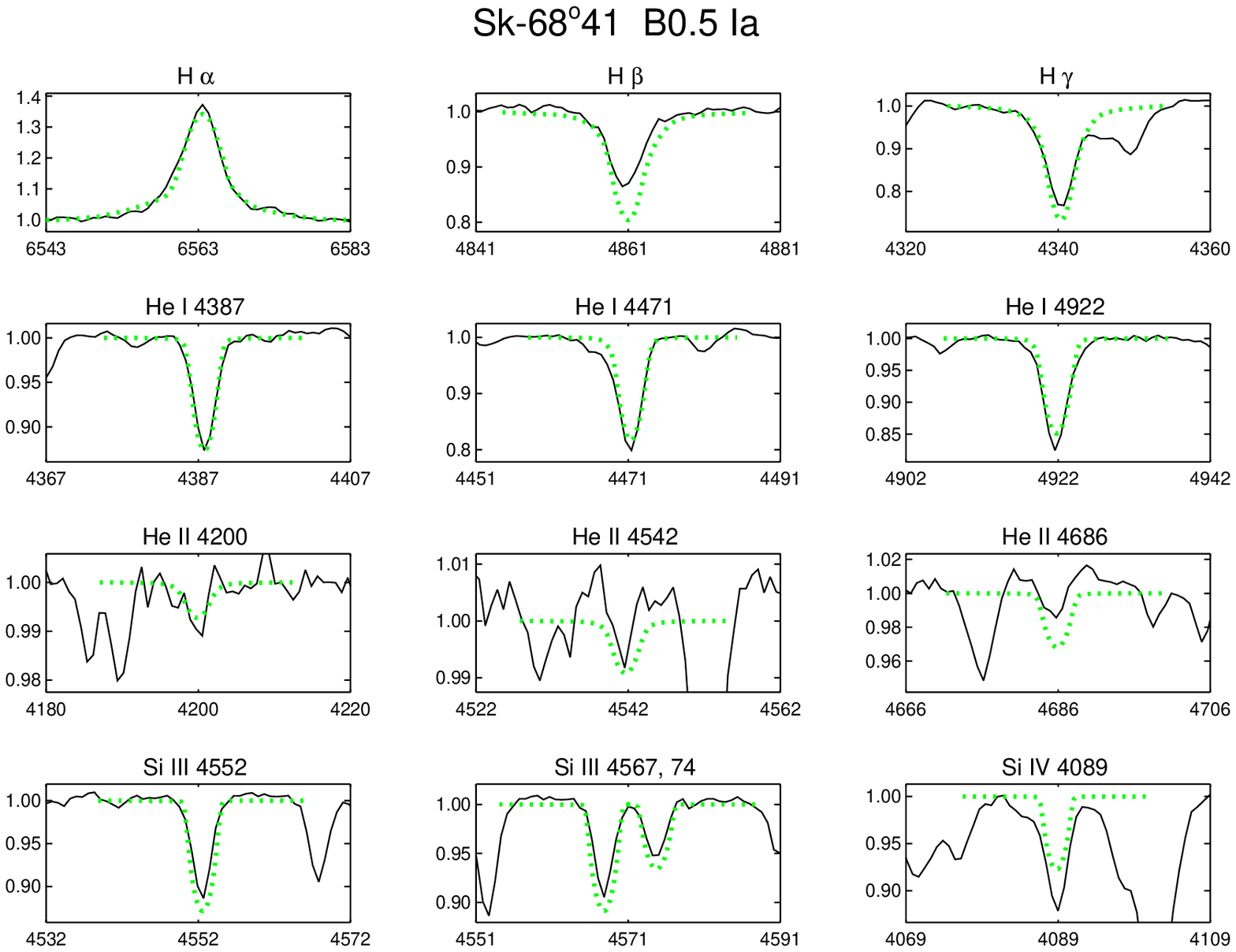}
\caption{\label{fig:Sk6841} Sk$-68^\circ 41$. The upper figure shows a section of the blue spectrum of this star,
with the prominent lines identified.  The lower figure shows the fits (dotted) for the principle diagnostic lines.
}
\end{figure}

 {\it Sk$-68^\circ 41$}.  This is an early-type B supergiant (Fig.~\ref{fig:Sk6841}, upper), the latest star in our sample.  We classify it as B0.5 Ia, based upon visual comparison with the Walborn \& Fitzpatrick (1990) atlas.  Our classification is based on the fact that Si IV $\lambda \lambda 4089, 4116$  absorption is about
 equal in strength to the strongest Si III $\lambda 4552$ line, the (near) lack of He II absorption, the strength of the O II lines, and the weakness
 of Mg II $\lambda 4481$.  The luminosity class is consistent with the absolute visual magnitude of the star, $-6.71$.
 
 At our very high SNR we actually detect weak He II lines; their EWs are 15-25 m\AA,
 not evident in Fig~\ref{fig:Sk6841} (upper), but 
 visible in the rescaled version showing the
 fits in Fig.~\ref{fig:Sk6841} (lower).   However, these provide a powerful way of constraining the
 effective temperature.  For this star we also used the model's Si III $\lambda 4553$ and Si IV $\lambda 4089$ profiles.  A single effective
 temperature gave consistent results for He I, He II, Si III, and Si IV.  We find remarkable consistency, and feel confident that the effective
 temperature here is determined to a precision of about 250 K (1\%), unlike the typical 1000 K uncertainty for the O-type stars.
 H$\alpha$ emission was also well fit, but required a a value of $\beta=2.0$.  H$\beta$ was not well fit in any model with reasonable
 agreement in H$\alpha$.  The fits are shown in Fig.~\ref{fig:Sk6841} (lower).

\begin{figure}
\epsscale{0.6}
\plotone{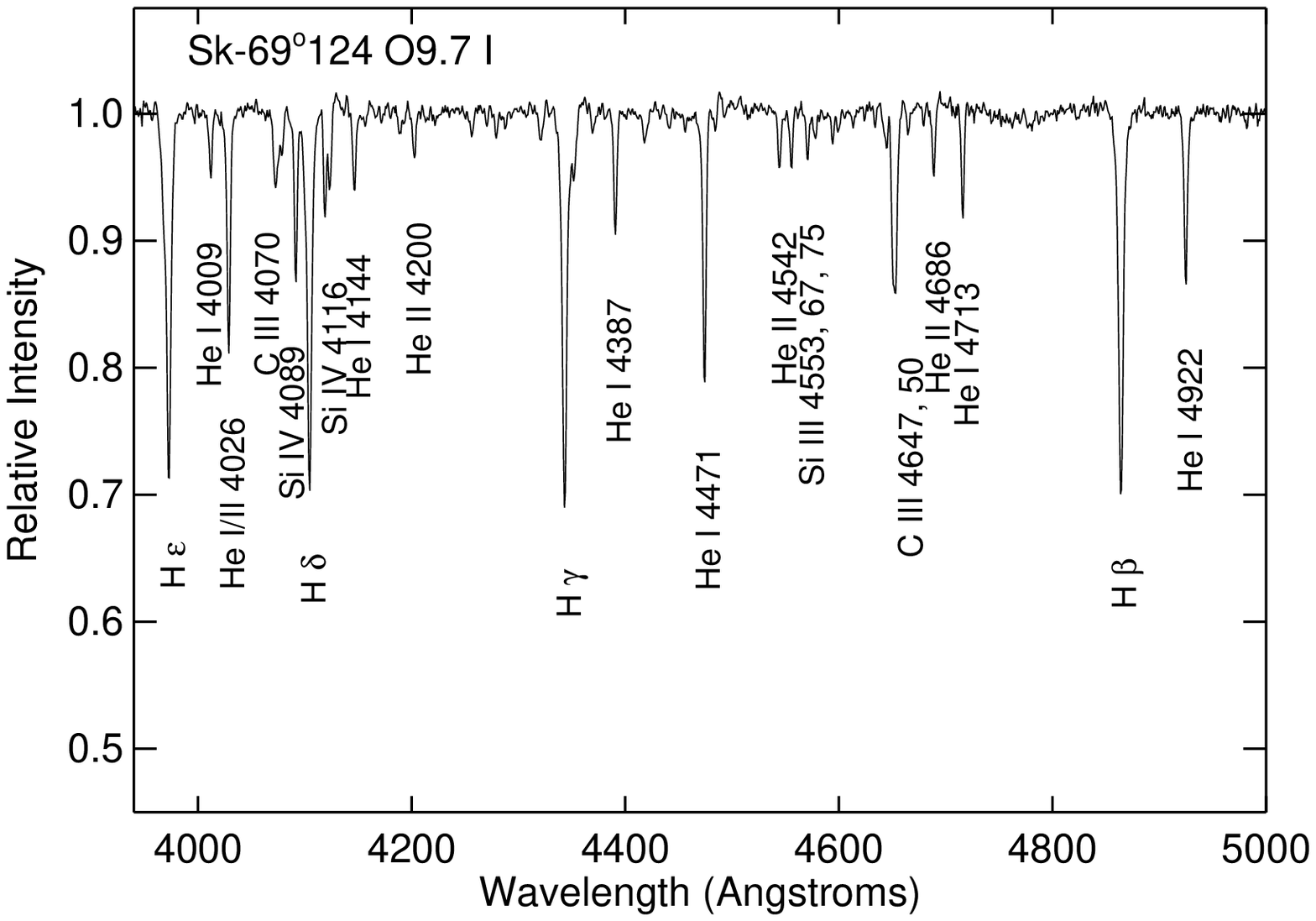}
\vskip 50pt
\plotone{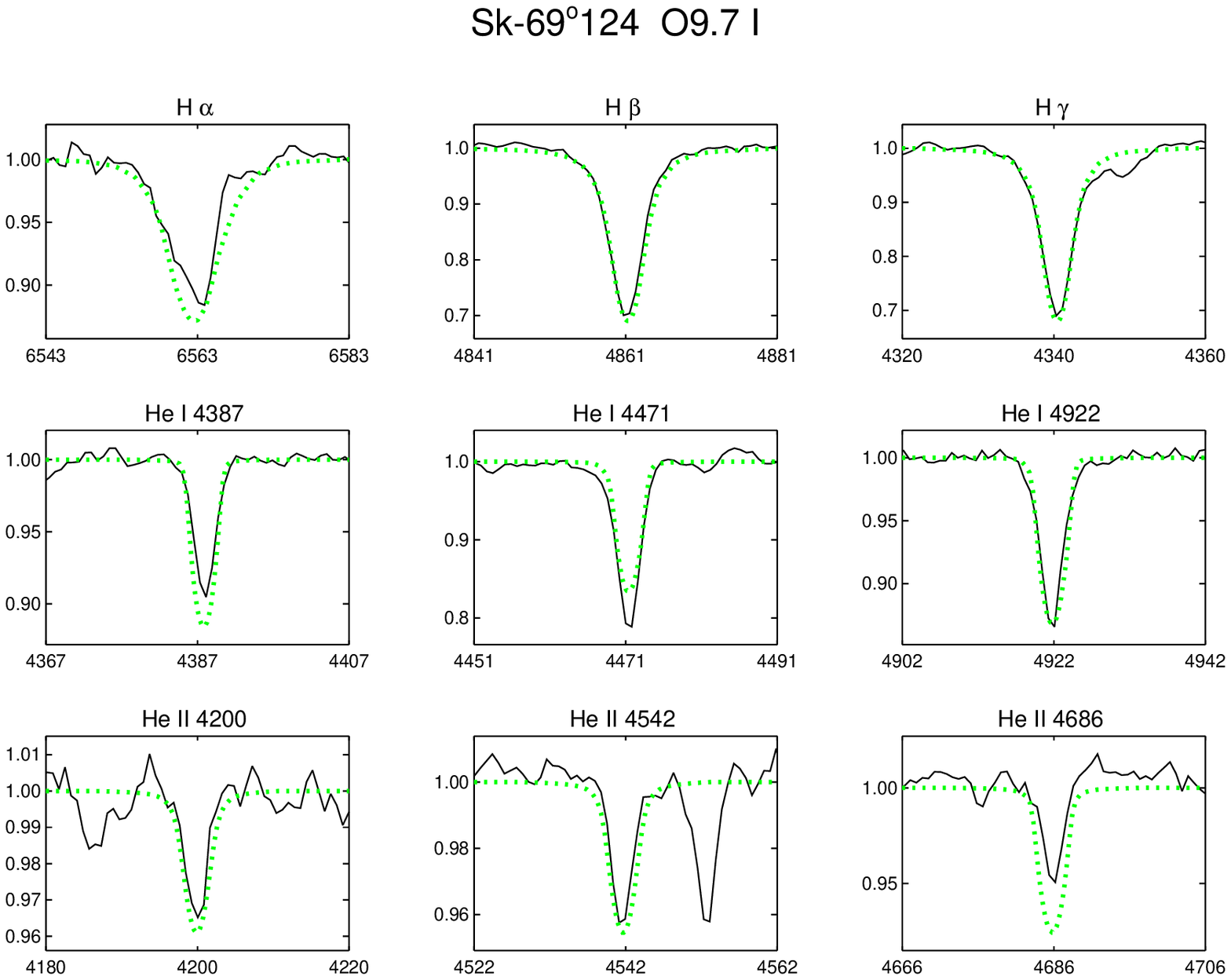}
\caption{\label{fig:Sk69124} Sk$-69^\circ 124$. The upper figure shows a section of the blue spectrum of this star,
with the prominent lines identified.  The lower figure shows the fits (dotted) for the principle diagnostic lines.}
\end{figure}

 {\it Sk$-69^\circ 124$}.  This is a late-type O supergiant (Fig.~\ref{fig:Sk69124}, upper).  We measure EWs of 830 m\AA\ and 145 m\AA, or $\log W'=0.76$.
 This is well beyond the $\log W'=0.45$ for an O9.5 star, and so we call this O9.7.  A B0 subtype is excluded on the basis that He II $\lambda 4200$
 is still reasonably strong (120 m\AA) in our spectrum.  The luminosity class is I, based upon the sharpness and strengths of the features; the
 O9.7 I type is also consistent with the absolute magnitude, $M_V=-6.07$.  The fits are good (Fig.~\ref{fig:Sk69124}, lower).
 
\begin{figure}
\epsscale{0.6}
\plotone{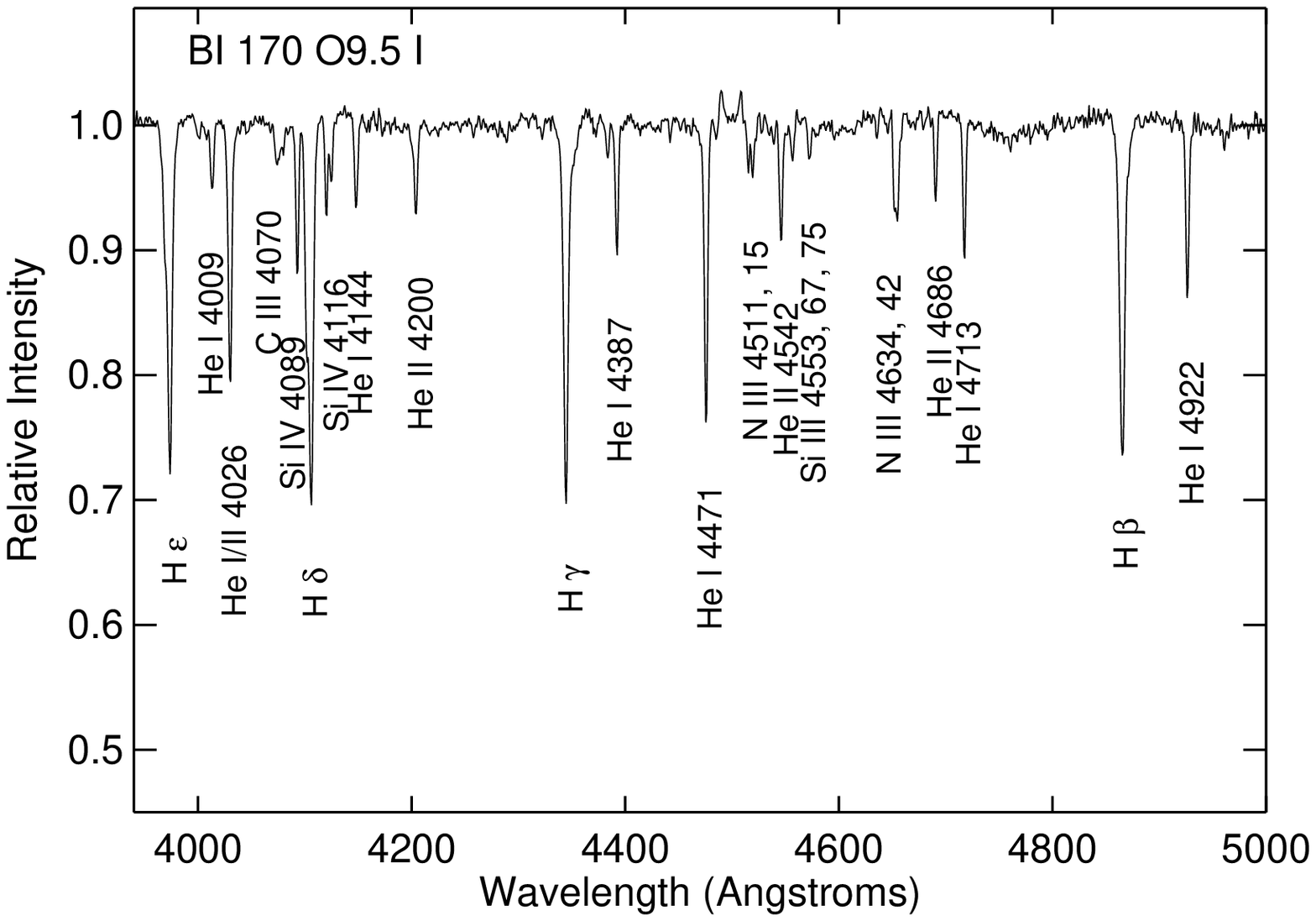}
\vskip 50pt
\plotone{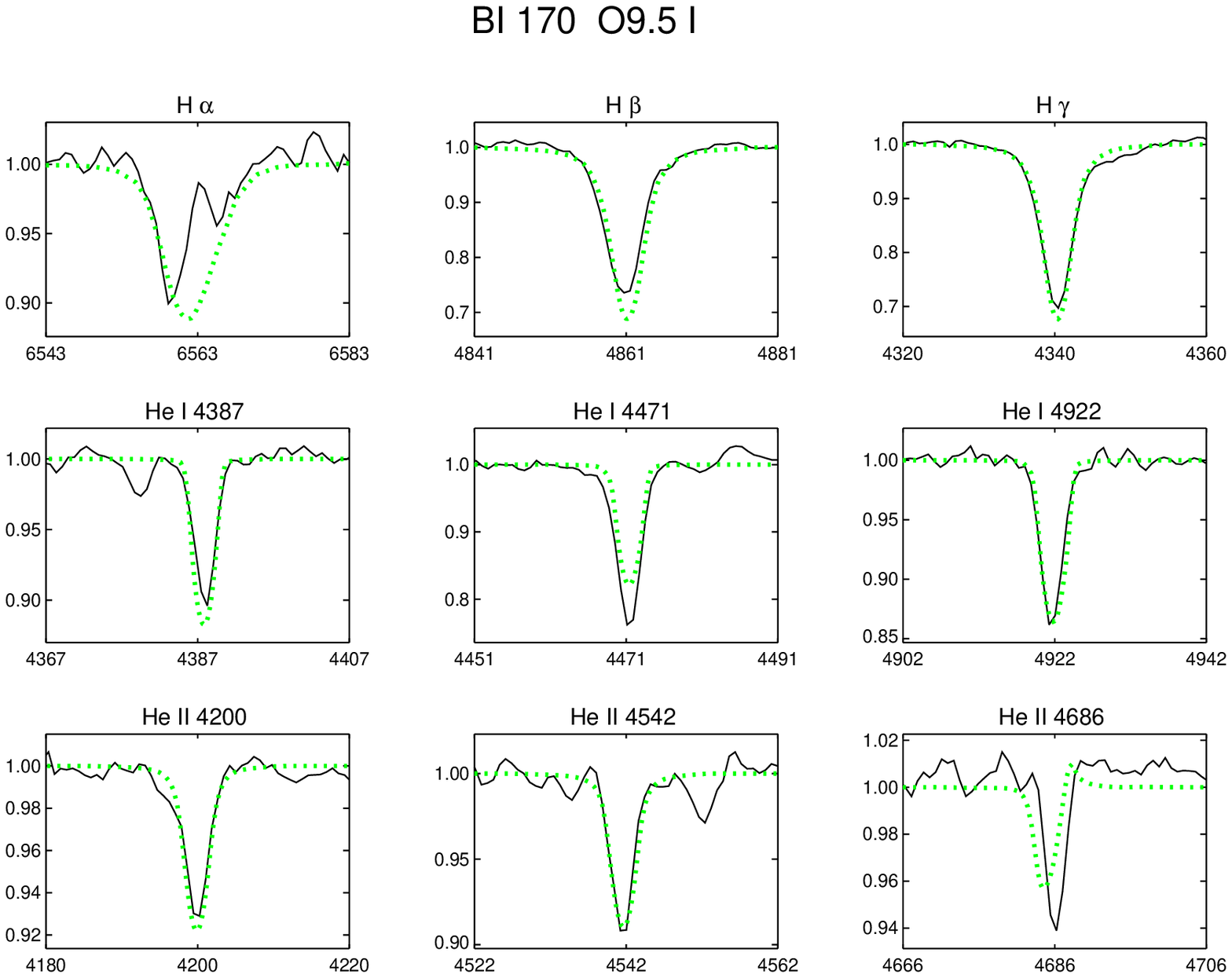}
\caption{\label{fig:BI170} BI 170. The upper figure shows a section of the blue spectrum of this star,
with the prominent lines identified.  The lower figure shows the fits (dotted) for the principle diagnostic lines.}
\end{figure}

 {\it BI 170}.  This too is a late type O supergiant (Fig.~\ref{fig:BI170}, upper), although clearly not quite as late as Sk$-69^\circ$ 124.  We classify this
 as O9.5 I, based both on our visual comparison with the Walborn \& Fitzpatrick (1990) atlas and our EW measurements for He I $\lambda 4471$
 (930 m\AA) and He II $\lambda 4542$ (310 m\AA), which leads to $\log W'=0.48$.   The $M_V=-5.66$ is consistent with the visual appearance
 of the spectrum, which suggests the star is a supergiant.
 
 The model fits to this star were good, except for the wind lines (Fig.~\ref{fig:BI170}, lower).  
 No combination of mass-loss rate and $\beta$ gave us good matches at both
 H$\alpha$ and He II $\lambda 4686$.  
 
\begin{figure}
\epsscale{0.6}
\plotone{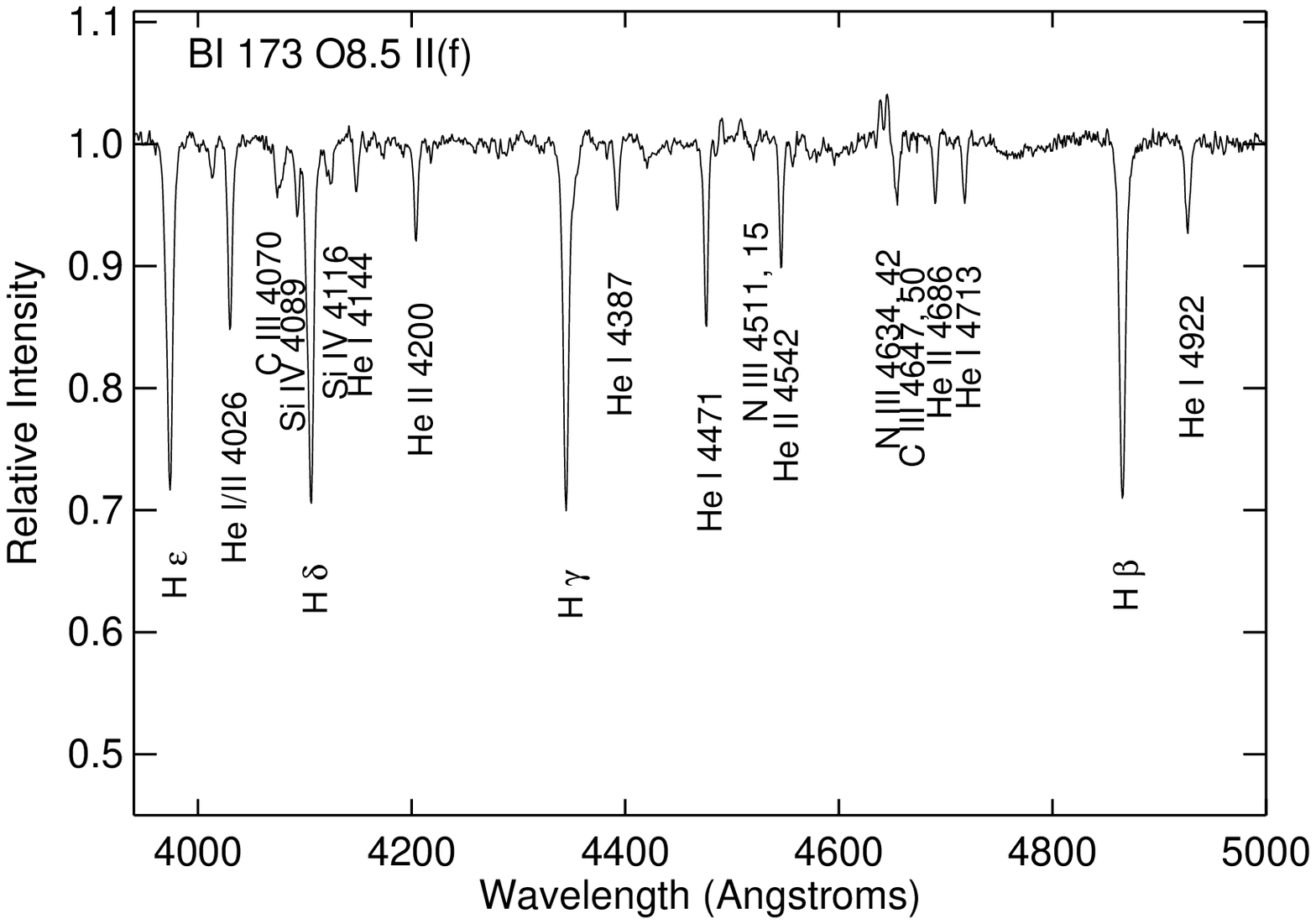}
\vskip 50pt
\plotone{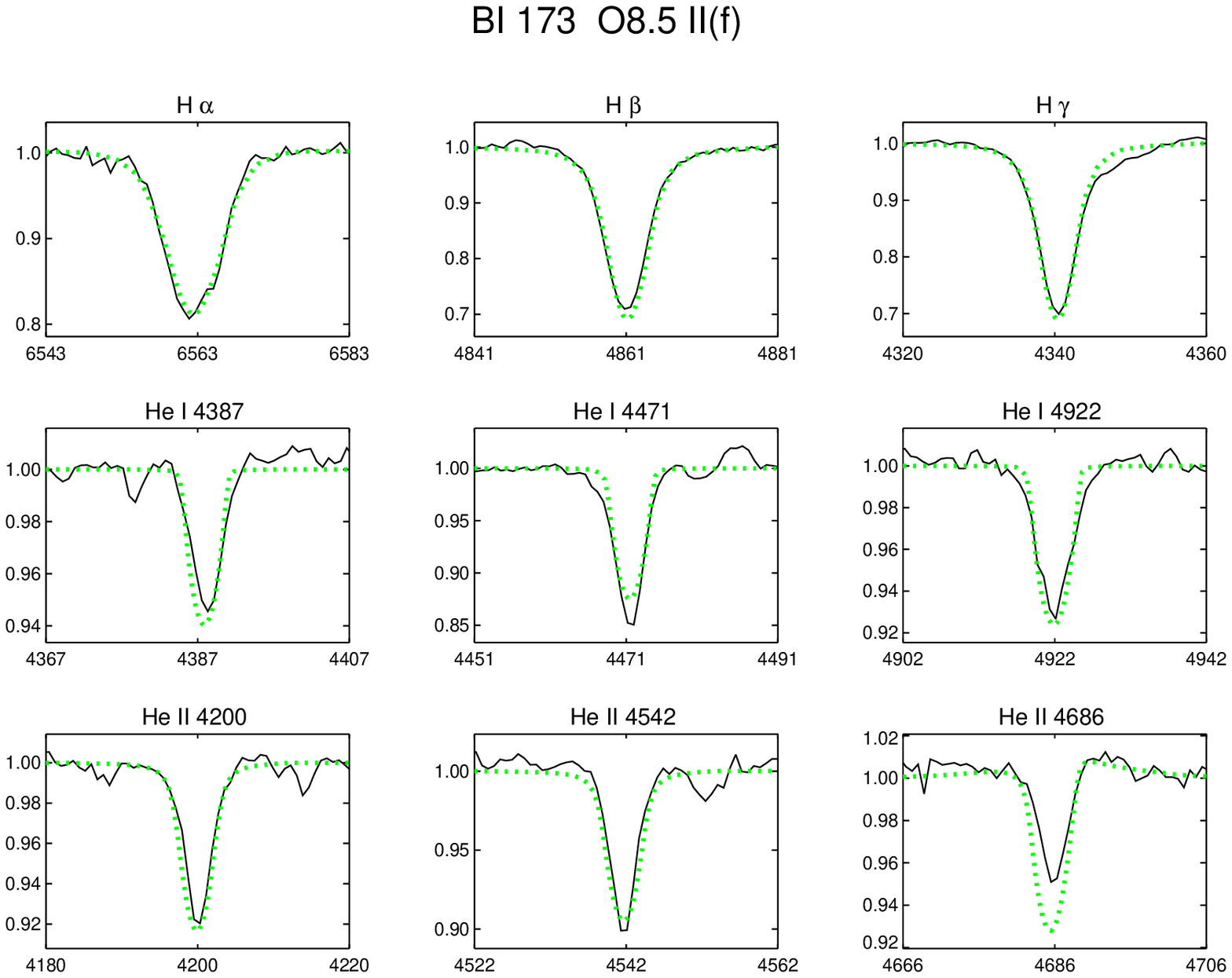}
\caption{\label{fig:BI173} BI 173. The upper figure shows a section of the blue spectrum of this star,
with the prominent lines identified.  The lower figure shows the fits (dotted) for the principle diagnostic lines.}
\end{figure}

 {\it BI 173}. This is a mid-to-late O type giant/supergiant (Fig.~\ref{fig:BI173}, upper).  We measure EWs of 715 m\AA\ and 425 m\AA\ for He I $\lambda 4471$ and 
 He II $\lambda 4542$, respectively, leading to $\log W'=0.23$, and O8.5 subtype.  The luminosity class is intermediate between giant
 and supergiant: NIII $\lambda \lambda 4634, 42$
 is in emission, while He II $\lambda 4686$ is weakly in absorption.  The  absolute visual magnitude, $-5.93$, is closer to a supergiant than
 a giant.  We call this star an O8.5 II(f).
  The fits were good, except perhaps for He II $\lambda 4686$ which is slightly too strong in the model (Fig~\ref{fig:BI173}, lower).  The rotation
  is somewhat  fast, with $v \sin i = 200$ km s$^{-1}$.
 
\begin{figure}
\epsscale{0.6}
\plotone{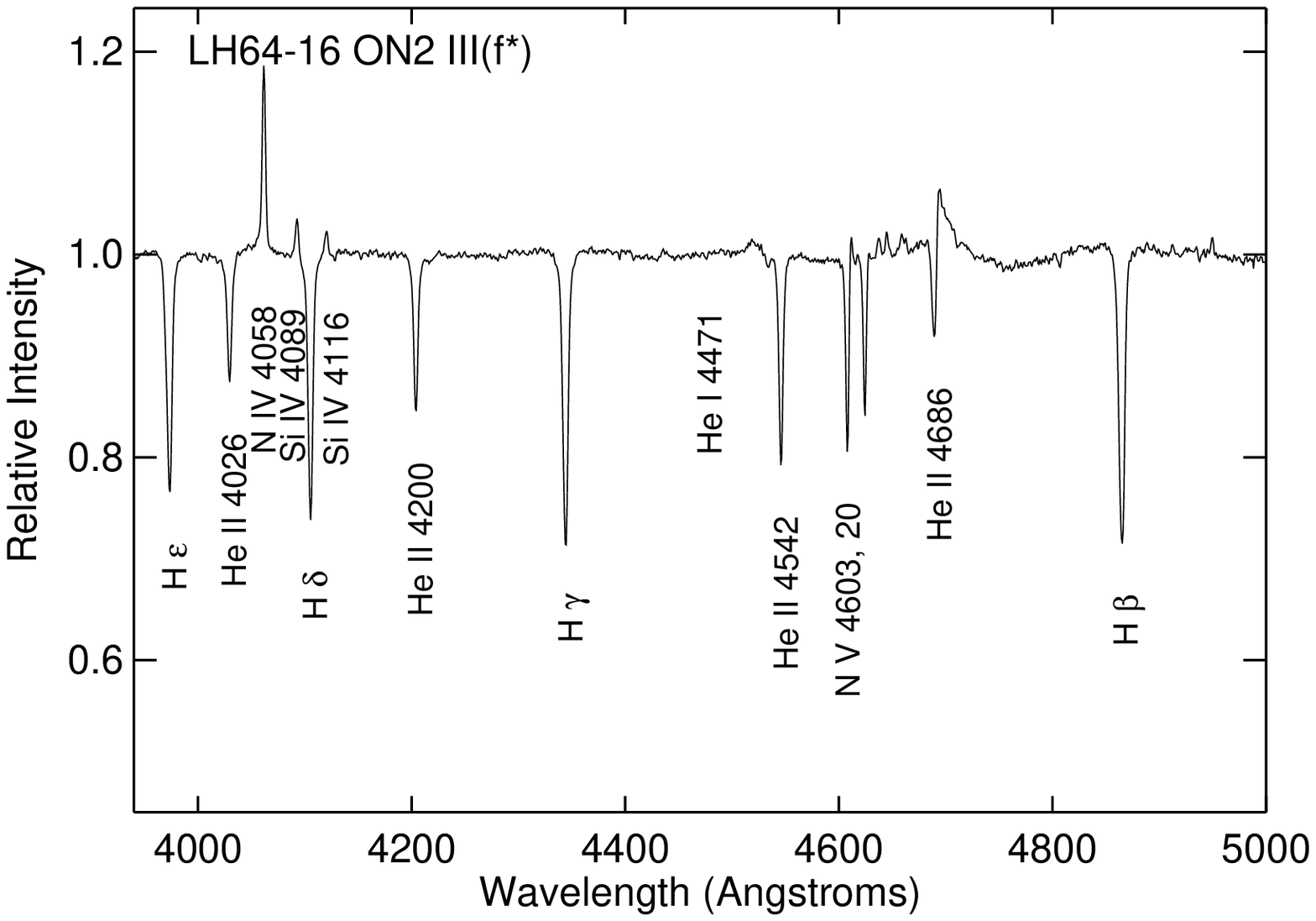}
\vskip 50pt
\plotone{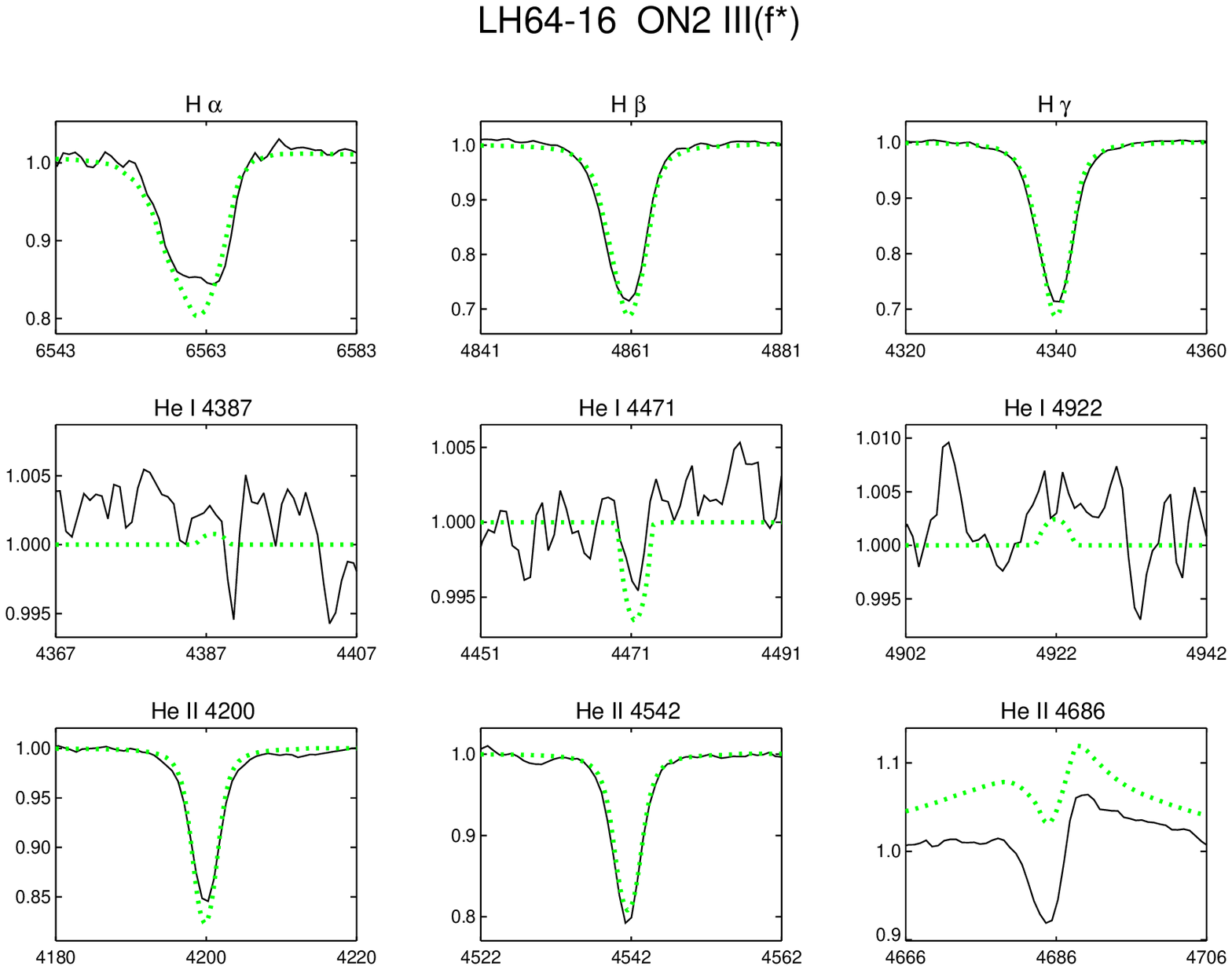}
\caption{\label{fig:LH6416} LH64-16. The upper figure shows a section of the blue spectrum of this star,
with the prominent lines identified.  The lower figure shows the fits (dotted) for the principle diagnostic lines.}
\end{figure}

 {\it LH64-16}.  This  is  a very early type O giant (Fig.~\ref{fig:LH6416}, upper) whose spectrum was first described as ``O3 III: (f*) by  Massey et al.\ (2000).  It formed
 one of the prototypes of the new O2 III class introduced by Walborn et al.\ (2002).  A subsequent analysis by Walborn et al.\ (2004) showed that
 the star was chemically enriched (both He and N) with an extremely high effective temperature, and called it an ON2 III(f*), with the ``N" denoting the enhanced nitrogen abundance.  The star was re-modeled in Paper I.   We 
 included it in our current study because of its extreme properties, and to see if we could independently reproduce its properties with our new
 spectra.  
 
We aimed to obtain a much higher SNR spectra of this star than what we used in Paper I as the He I $\lambda 4471$ line was
only marginally detected in that study\footnote{Note that the EW of He~I $\lambda 4471$ was incorrectly reported in Paper II as 100 m\AA, rather than $\sim$10m\AA.  We have compared our new spectra to that used in Paper I and they are quite similar; the problem was a typographical error in Table 7 of Paper II, and the subsequent use of this number to report an erroneous $\log W'$ in Paper II.}.
We achieved a SNR of 1000 per 2.4 \AA\ spectral resolution element, which
confirms our detection of He I $\lambda 4471$ absorption in this O2 star, again emphasizing the point made
 in Papers I and II that the ``lack of He I" criteria originally applied to O3 stars and subsequently to the O2 class can be a SNR issue, and
 needs to be quantitative in order to be useful.  We
 measure an EW of 12 m\AA\ for He I $\lambda 4471$, and
 an EW of He~II $\lambda 4542$ of 990 m\AA, leading to a $\log W'=-1.9$.
 (Given the much higher SNR of these data, we find that the uncertainty on
the EW of the He I $\lambda 4471$ line is about 5m\AA, again dominated by
uncertainty in the rectification.)
 Our fits for this star are reasonably good (Fig.~\ref{fig:LH6416}, lower).  Note that we do not consider neither He I $\lambda 4387$
 nor He I $\lambda 4922$ to have been detected; we compare the models and the spectra of these two lines in the figure purely
 for consistency with the other stars analyzed here.  The He I $\lambda 4471$ line {\it was} used in the fit, and and the results very similar to what we obtained in Paper I, as shown in Table~\ref{tab:compA}.  A very high He/H ratio was needed to obtain a reasonable fit, as we found in Paper I.   There we
 note that the $7\times$ N enrichment found by Walborn et al.\ (2002) is consistent with any high He/H ratio (0.25-2.0), as it is simply
 the CNO-burning equilibrium ratio.   Note in Table~\ref{tab:results} that the spectroscopic mass is only 24$M_\odot$, while the
 mass inferred from the evolutionary tracks is much higher, $74M_\odot$.  In accord with Paper I, we suggest that
this star is the result of binary evolution other than a normal evolutionary process. 
 
\begin{figure}
\epsscale{0.6}
\plotone{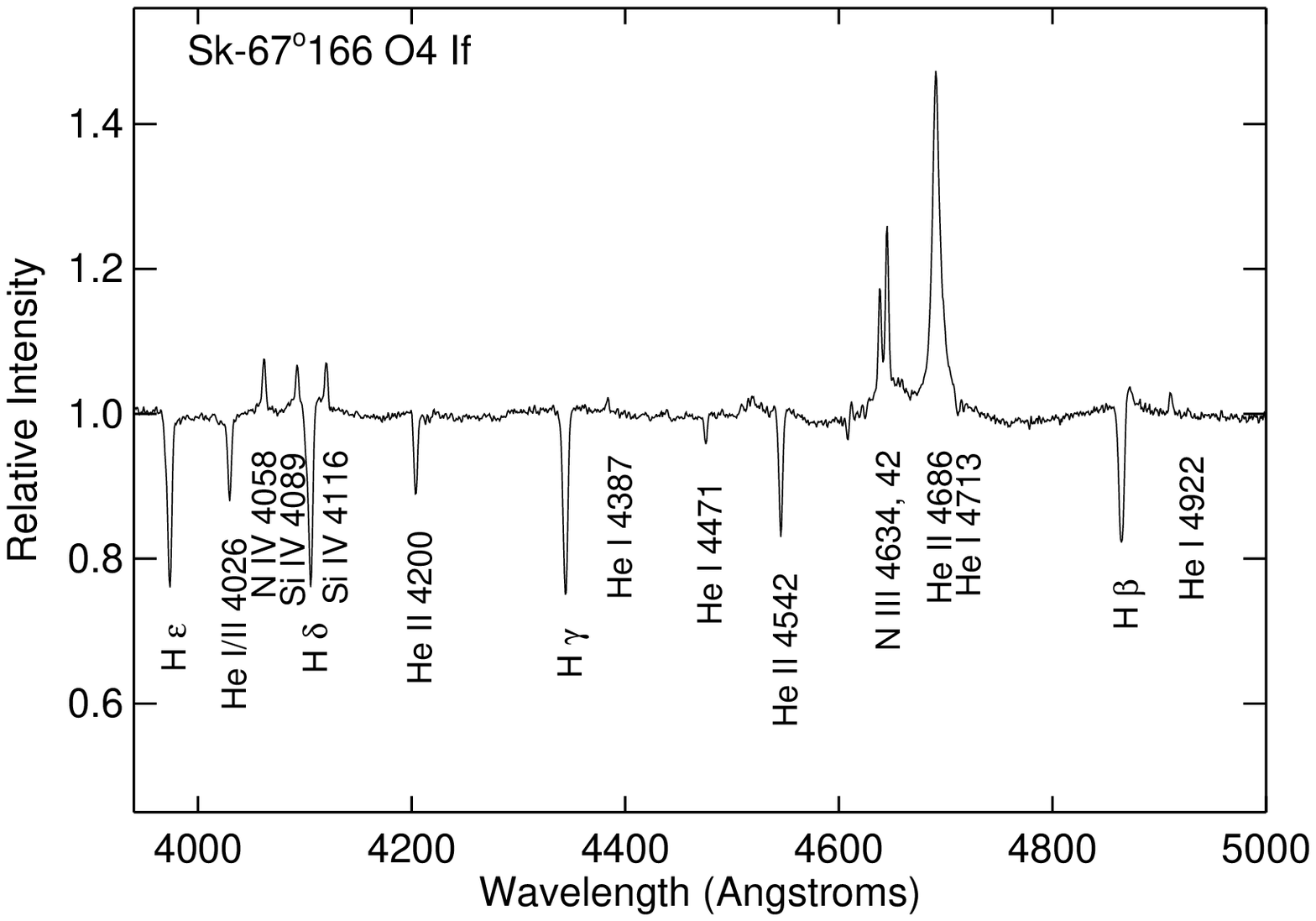}
\vskip 50pt
\plotone{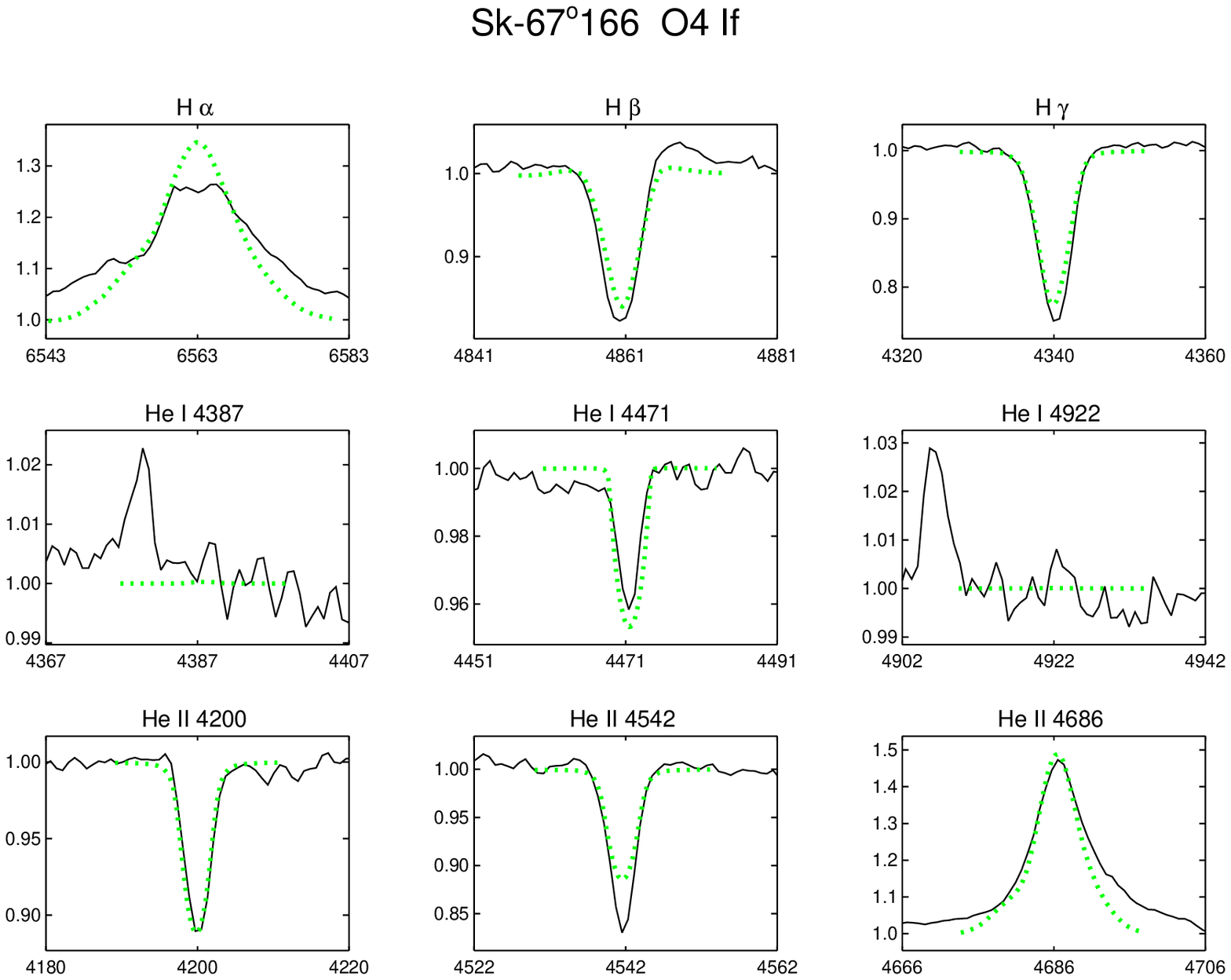}
\caption{\label{fig:Sk67-166} Sk$-67^\circ 166$.The upper figure shows a section of the blue spectrum of this star,
with the prominent lines identified.  The lower figure shows the fits (dotted) for the principle diagnostic lines.}
\end{figure}

 {\it Sk$-67^\circ 166$, HD 269698}.  This star has been previously called an O4 If+ star by Walborn (1977), a classification with which
 we concur (Fig.~\ref{fig:Sk67-166}, upper): we measure EWs of 115 m\AA\ and 670 m\AA\ for He I $\lambda 4471$ and He II $\lambda 4542$, respectively, leading to
 $\log W'=-0.76$.  NIII $\lambda \lambda 4638, 42$ and He II $\lambda 4686$ are strongly in emission, leaving no ambiguity about the
 luminosity class being a supergiant, in agreement with $M_V=-6.5$.  
 
 Our fits for this star were relatively good (Fig.~\ref{fig:Sk67-166}, lower), although we had to increase the He/H ratio to 0.20.   The He~II $\lambda 4542$ model line is weaker
 than that observed, although the He II $\lambda 4200$ line is in good agreement.  For He I we detected only the $\lambda 4471$ line, but include
 the $\lambda 4387$ and $\lambda 4922$ lines in the figure for consistency.  The star had been previously modeled by Puls et al.\ (1996) and
 Mokiem et al.\ (2007).  Our parameters are very similar to those derived by the latter study, which also used the modern version of FASTWIND.

\begin{figure}
\epsscale{0.6}
\plotone{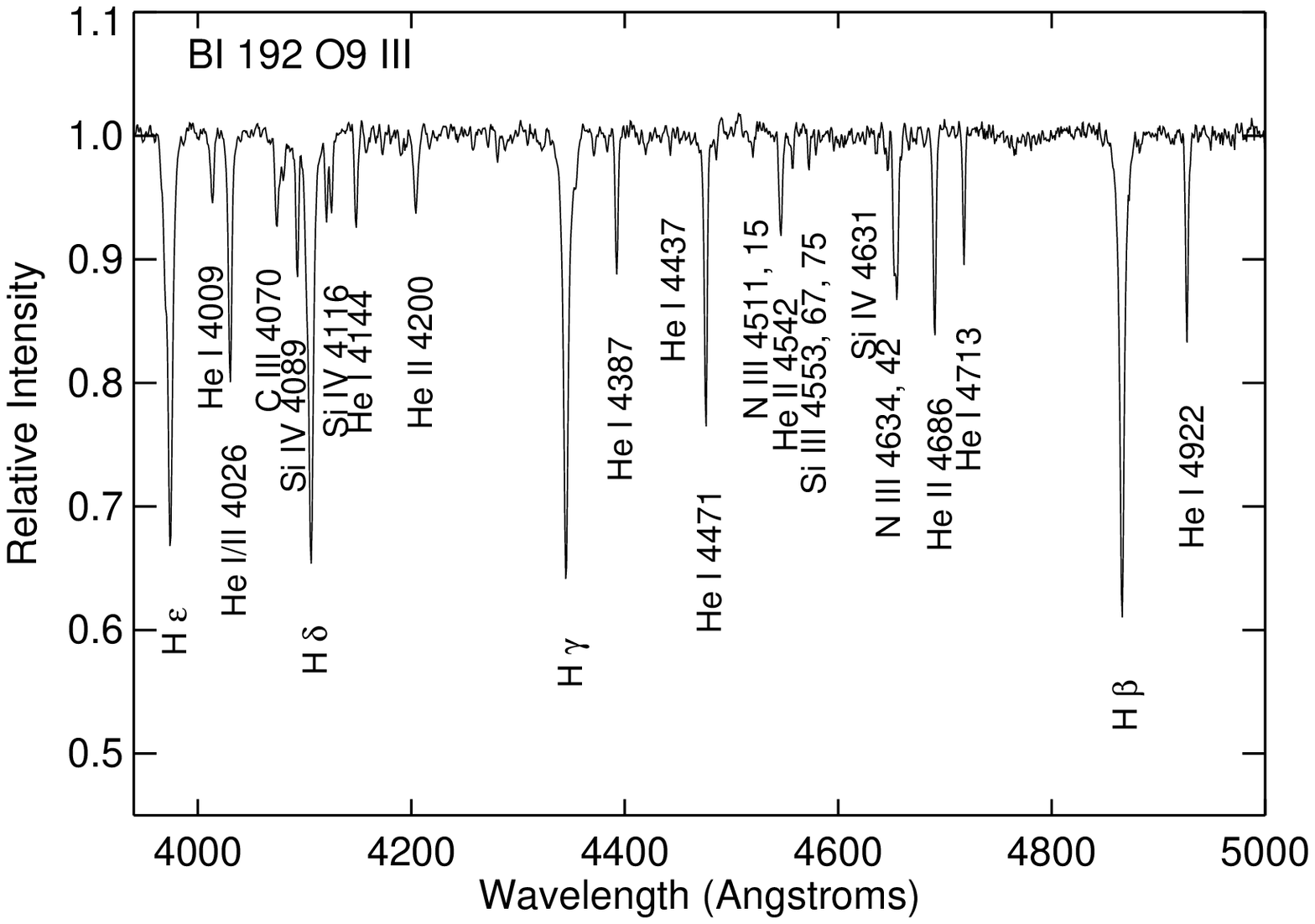}
\vskip 50pt
\plotone{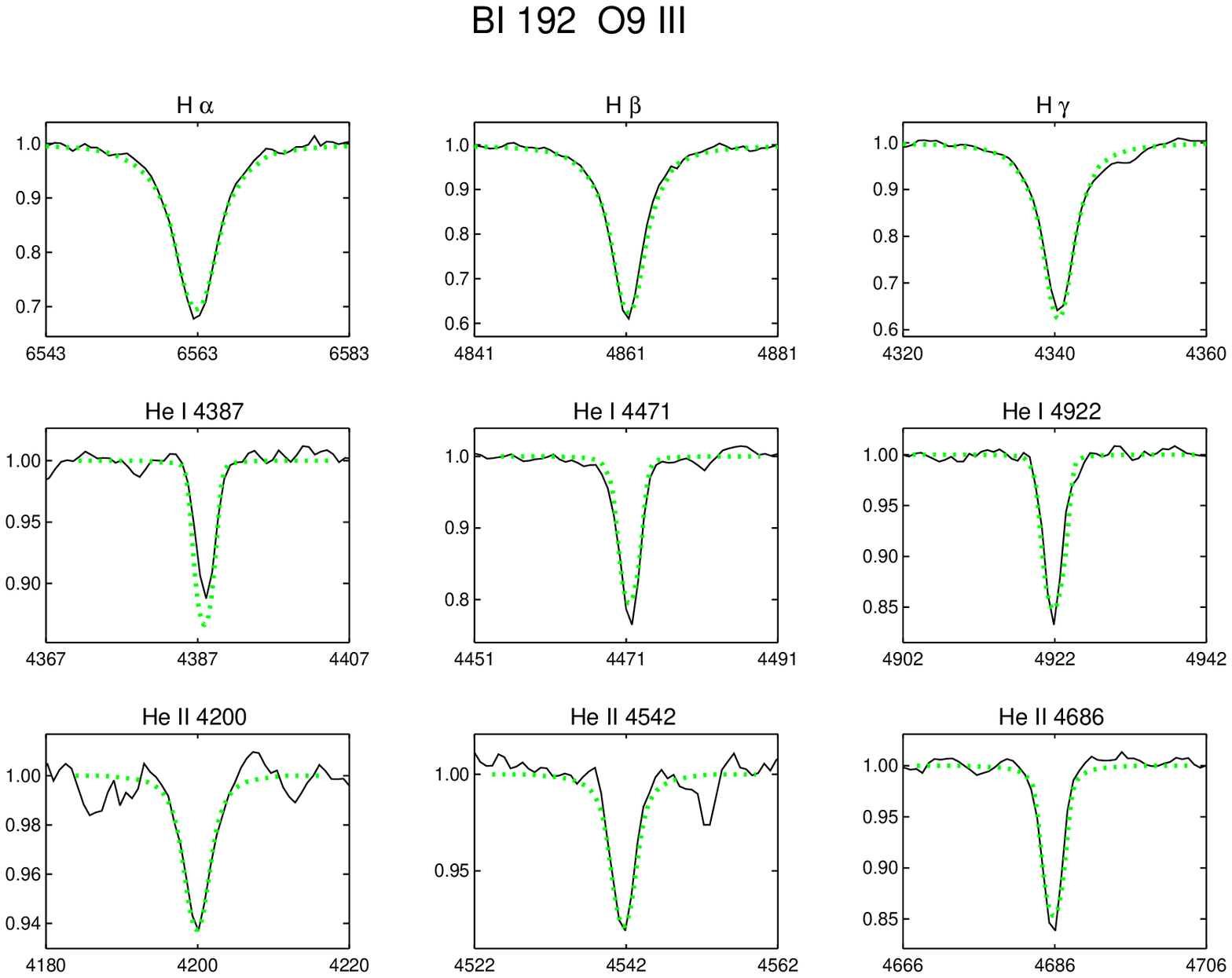}
\caption{\label{fig:BI192} BI 192. The upper figure shows a section of the blue spectrum of this star,
with the prominent lines identified.  The lower figure shows the fits (dotted) for the principle diagnostic lines.}
\end{figure}

 {\it BI 192}.  Visually, this is a mid-to-late O giant/supergiant (Fig~\ref{fig:BI192}, upper).  We measure EWs of 800 m\AA\ and 300 m\AA\ for He I $\lambda 4471$
 and He II $\lambda 4542$, respectively, leading to $\log W'=0.42$, an O9 subtype.  The absolute visual magnitude is -5.03, consistent
 with the star being a giant, as is indicated spectrally by the relative strengths of the Si IV and He I features.
 The fits are all quite good (Fig.~\ref{fig:BI192}, lower).

\begin{figure}
\epsscale{0.6}
\plotone{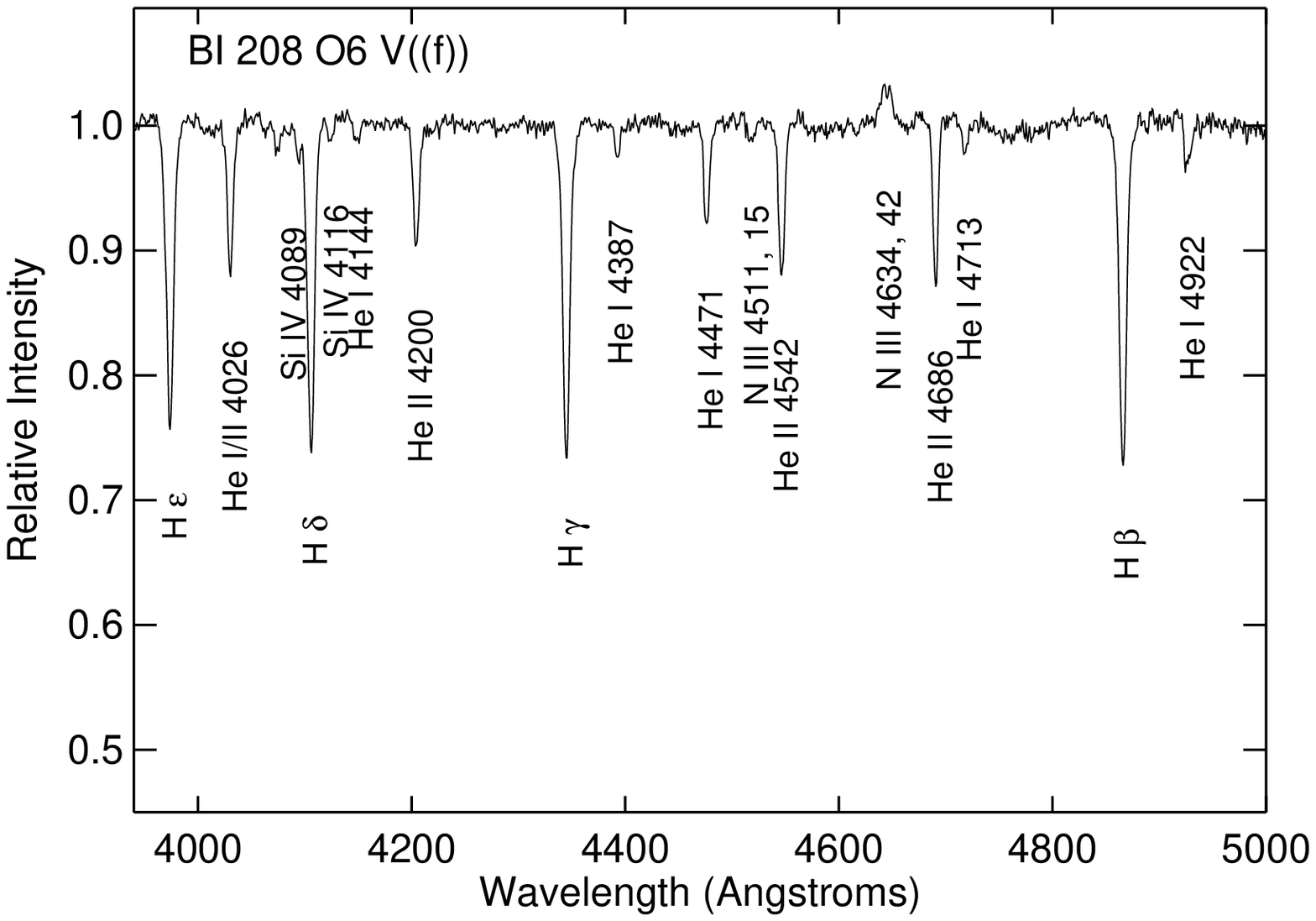}
\vskip 50pt
\plotone{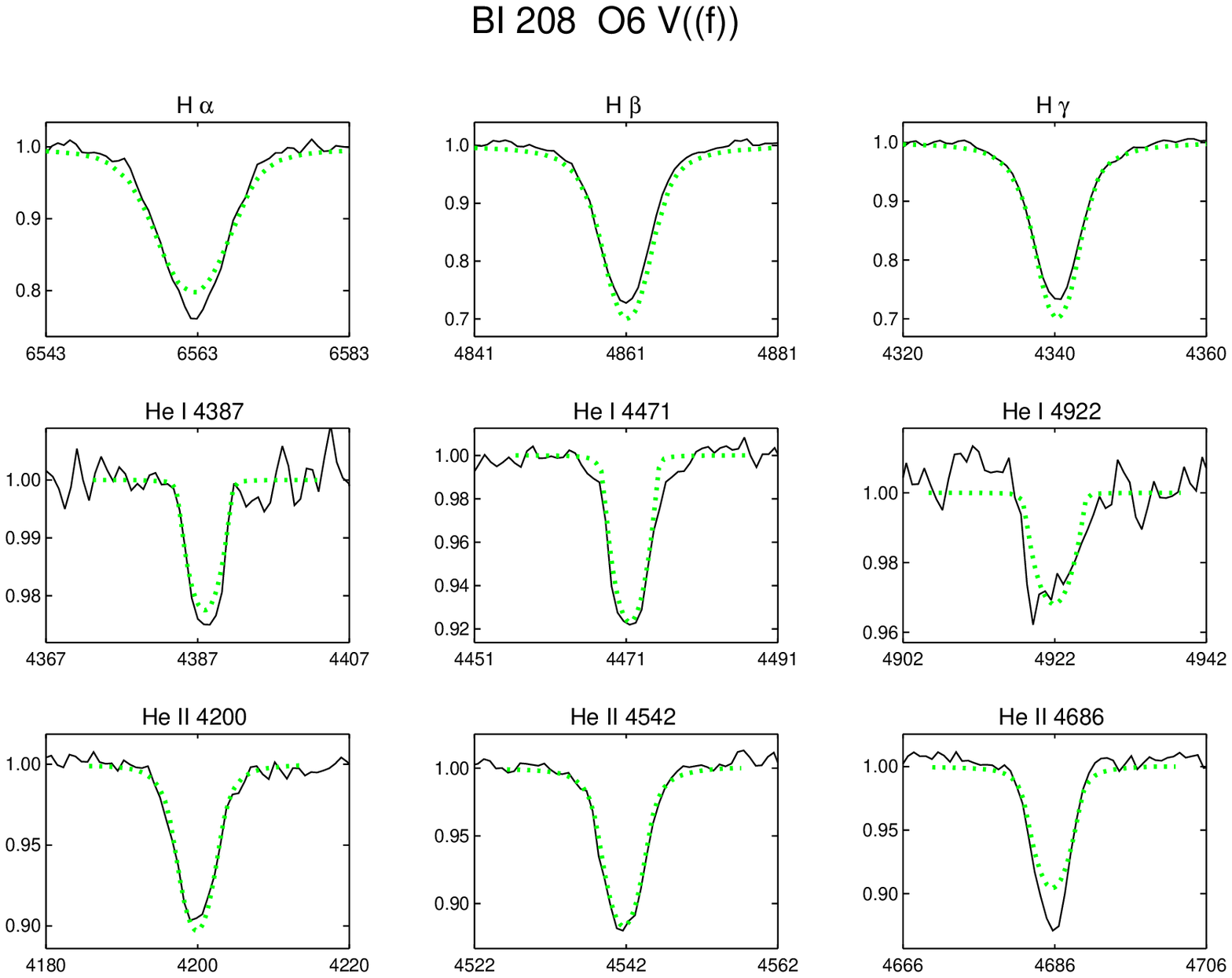}
\caption{\label{fig:BI208} BI 208. The upper figure shows a section of the blue spectrum of this star,
with the prominent lines identified.  The lower figure shows the fits (dotted) for the principle diagnostic lines.}
\end{figure}

 {\it BI 208}.  At first glance we would assign an O6-O6.5 subtype to this star, along with a luminosity class of V((f)), as He I $\lambda 4471$
 is marginally weaker than He II $\lambda 4542$, NIII $\lambda \lambda 4638,42$ is in emission but He II $\lambda 4686$ is strongly in
 absorption (Fig.~\ref{fig:BI208}, upper).   Our EW measurements of 510 m\AA\ and 830 m\AA\ for He I $\lambda 4471$
and  He II $\lambda 4542$ respectively, leading to $\log W'=-0.21$, making it just marginally an O6 subtype rather than an O6.5.  The
absolute visual magnitude -4.87 is consistent with the star being a dwarf, and so we classify the star O6 V((f)).

The fits for this star were quite good (Fig~\ref{fig:BI208}, lower). The terminal velocity measured by Prinja \& Crowther (1998) is listed as uncertain; it is quite small compared to
the expected 2.6$v_{\rm esc}$ (Table 2).  We ran additional models using a more likely value, 2000 km s$^{-1}$, to see what effect this would
have on our modeling of this star, and the differences were minor.  A slightly higher mass-loss rate was needed (0.5 rather than 0.3 in units of
$10^{-6} M_\odot$ yr$^{-1}$) but the fits were otherwise similar.  The only other peculiarity we note is a slightly larger than usual discrepancy (60 km s$^{-1}$) between the blue and red radial velocities.  In the absence of any other evidence of binarity we retain this star in our analysis.

 
 \section{Results and Discussion}
 \label{Sec-results}
 
 \subsection{Comparisons with Other Model Results}
 \label{Sec-compare}
 
We have attempted to model the spectra of 26 Magellanic Cloud O and early B stars, 14 in the SMC and 12 in the LMC.  We have succeeded
in 18 cases (69\%), with the rest likely being spectroscopic binaries or showing circumstellar features that rendered good fits impossible.

How do our results compare to others?  In Paper II we considered
 our effective temperature scale (effective temperature as a function of
spectral type) in comparison with those determined by others, finding that our FASTWIND-derived SMC  scale was warmer by
several thousand degrees than the effective temperatures determined for several NGC 346 stars by Bouret et al.\ (2003) using TLUSTY and
CMFGEN.   However, very few stars were in common between various studies at the time, making it difficult to compare the results directly.
The current study was largely motivated by this frustration.

In Table~\ref{tab:compA} we compare the model results for the stars analyzed in this study with those by others, and in Table~\ref{tab:compB} 
extend this to other stars in common in Papers I and II with others.
Several trends emerge from this comparison.  First, we find that the automatic fitting procedure with FASTWIND used by Mokiem et al.\ (2006, 2007) produced 
effective temperatures that were 1100 K warmer (on average)\footnote{This is the {\it median} difference; for some stars, the differences are even greater, such as for Sk$-70^\circ$ 69, where
Mokiem et al.\ (2007) find an effective temperature that is 2700 K higher as well as a higher surface gravity.}  {\it for the 11 stars in common
for which we found an acceptable fit}.  We attribute
this differences to the different weights given to fitting various lines: we relied heavily upon He I $\lambda 4387$ and 
(in the non-supergiants) He I $\lambda 4471$ in determining our best fits.

More significant perhaps is the fact that the automatic fitting procedure of Mokiem et al.\ (2006, 2007) always finds {\it some} fit which is
``best", resulting in their deriving physical properties of stars which we find to be clearly composite.  For example, in Paper I AzV 372 was
found to have He~I lines significantly broader than the He~II lines as well as a radial velocity shift between the He I and He II lines.
We expect from studies of Galactic O stars that at least 36\% are spectroscopic binaries (Garmany et al.\ 1980), with the true percentage of composites
probably higher.  Thus, it would be remarkable in an analysis of a large sample of O-type stars to always come up with an acceptable fit
(see discussion following de Koter 2008).   The LMC sample (Mokiem et al.\ 2007) were
``pre-filtered" to avoid stars that showed radial velocity variations in their
data, but no such selection was made for their SMC sample, and in neither case were
the goodness of the fits used other than to assign uncertainties to the physical properties. (We are indebted to Chris Evans for clarifying this situation for us.)  In any event, this problem could be avoided in future work by providing some objective ``goodness of fit" criteria
that distinguishes cases where the ``best" fit is not good, as per the general discussion in Press et al.\ (1992).

\begin{figure}
\epsscale{0.4}
\plotone{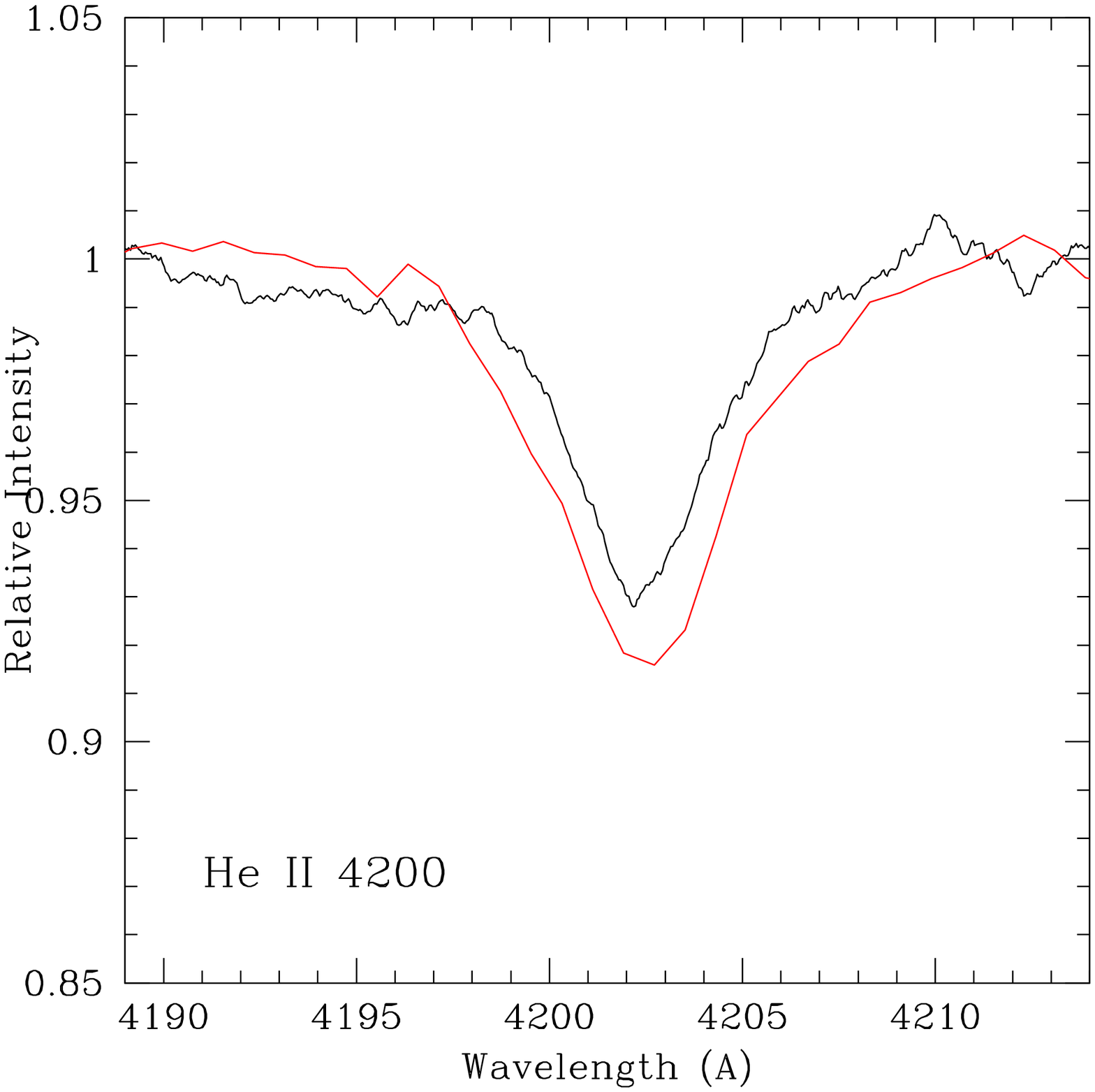}
\plotone{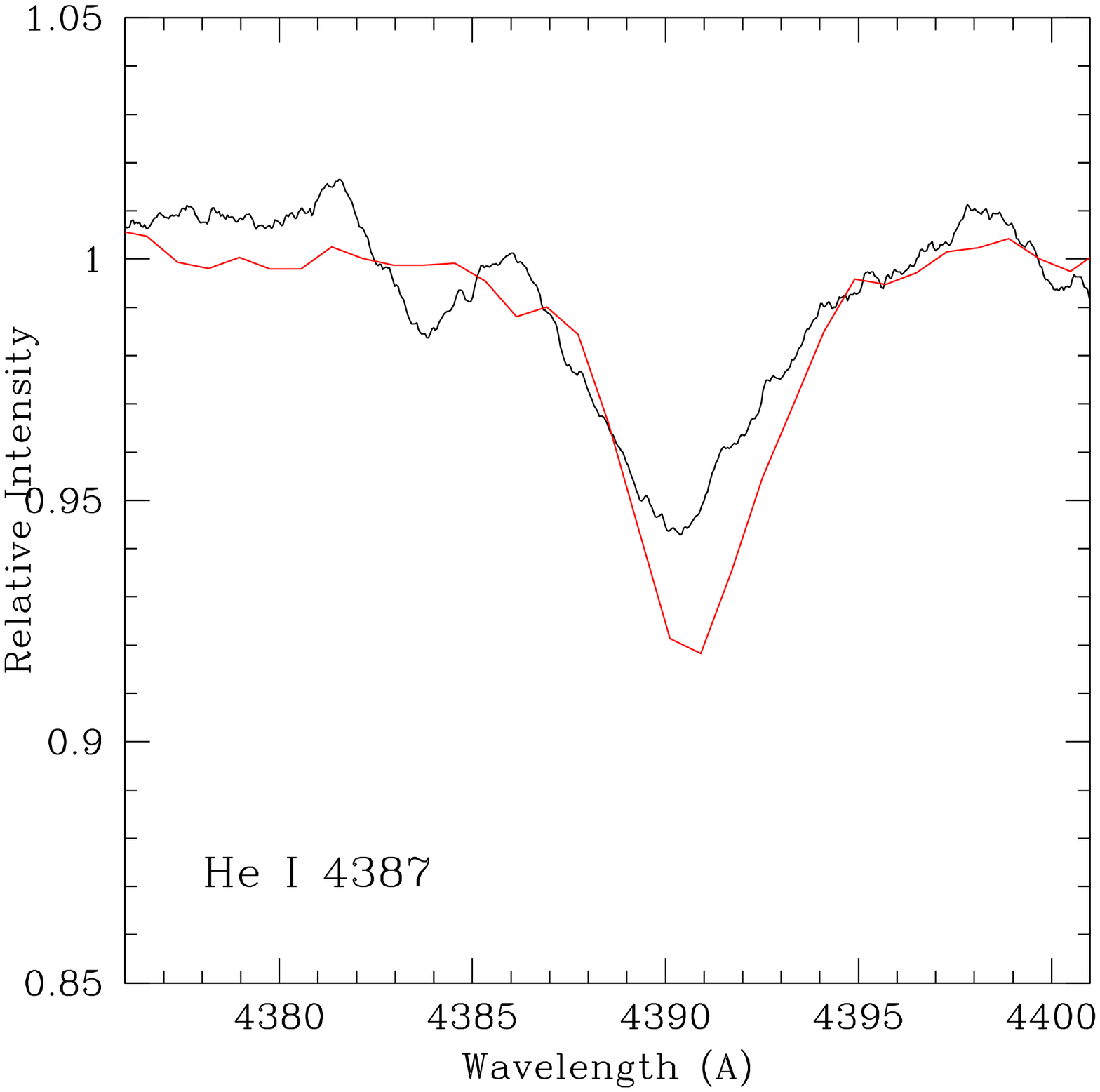}
\plotone{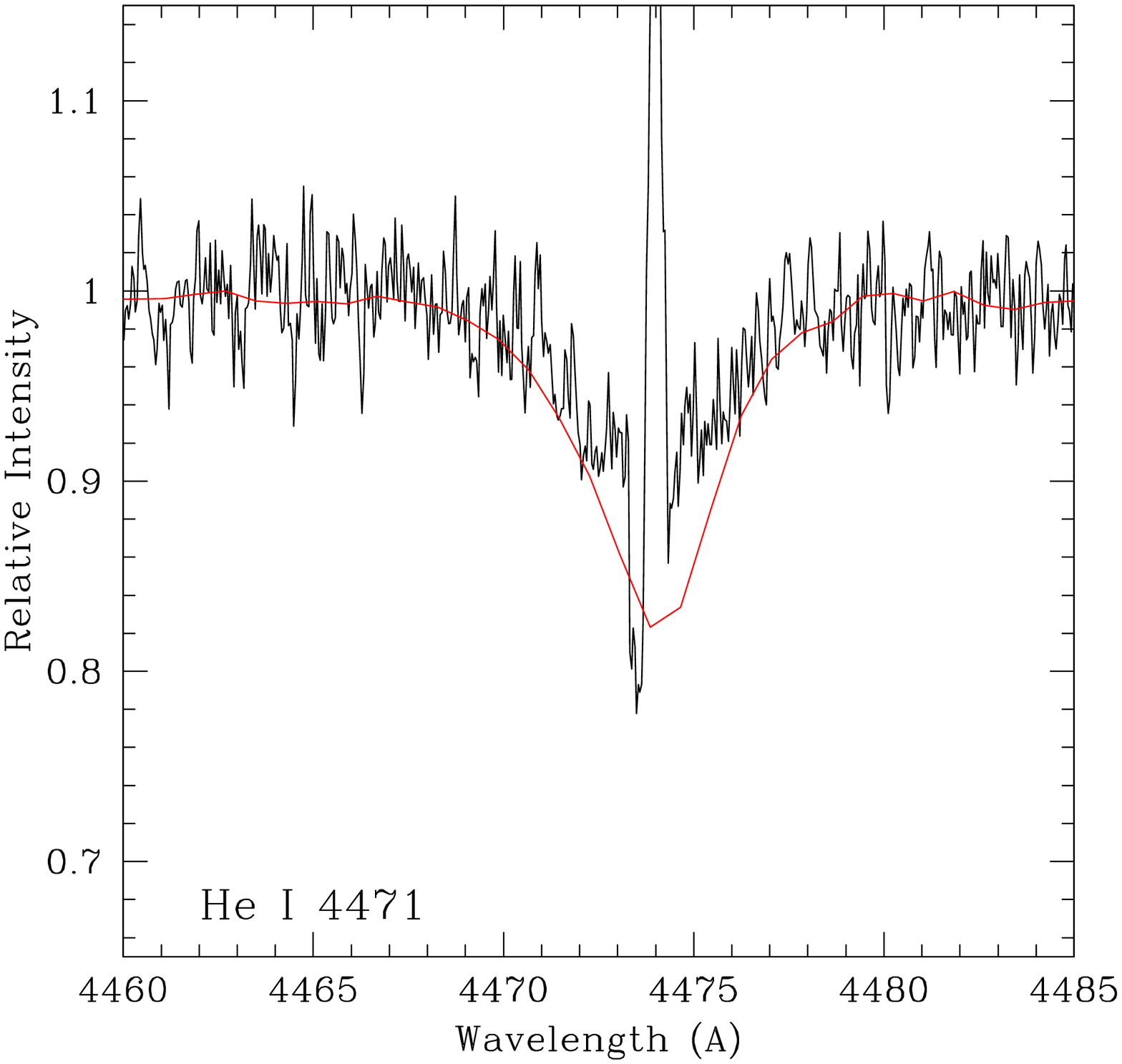}
\plotone{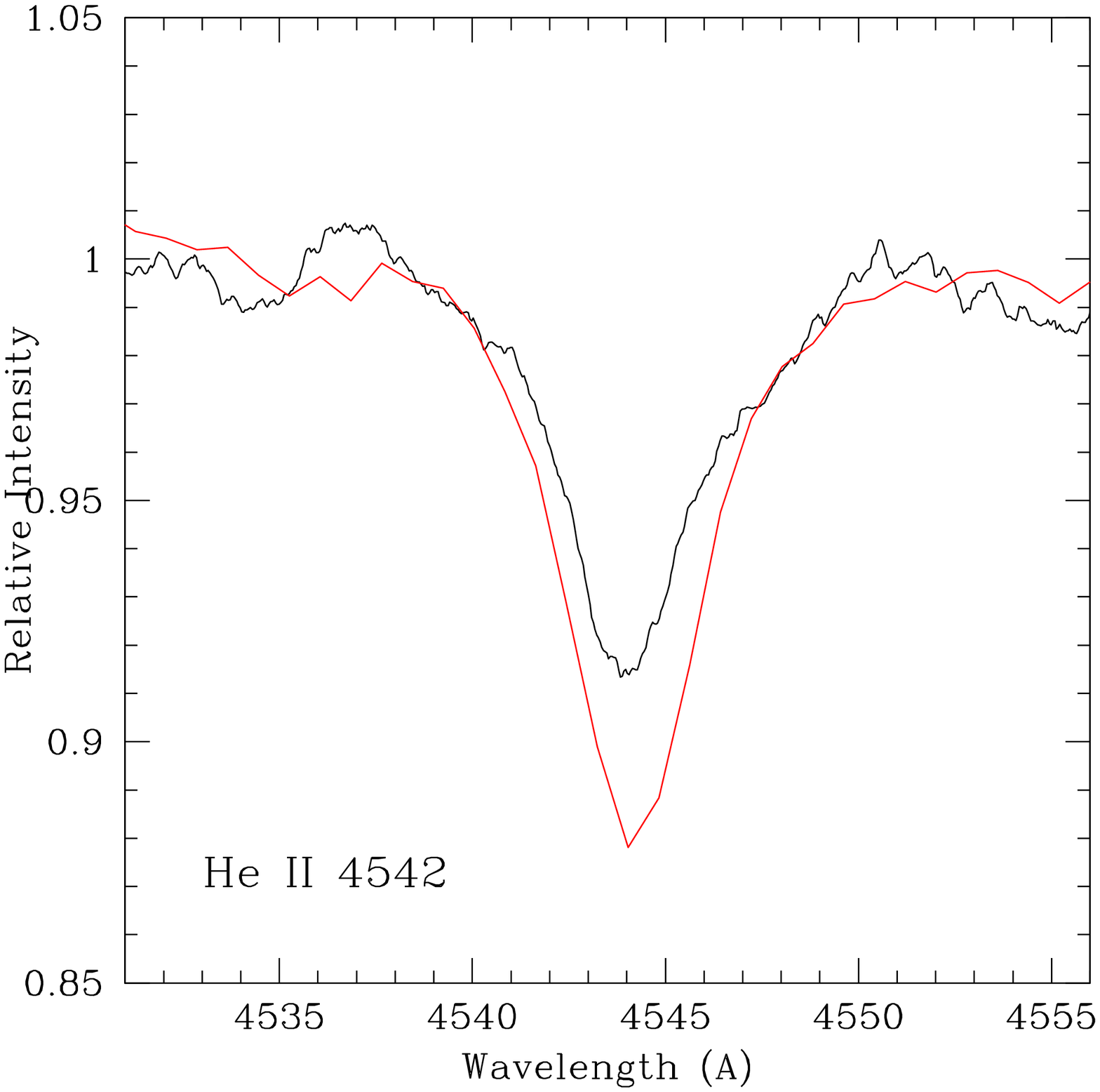}
\caption{\label{fig:heap}
A comparison between some of the He I and He II lines from the Magellan data used by us (red)
with the AAT data (black) used by Bouret et al.\ (2003) and Heap et al.\ (2006) for NGC 346-487. 
For all but the He I $\lambda 4471$ data, the AAT data have been smoothed to match the resolution
of the Magellan data.   The AAT data were kindly provided by C. J. Evans (2005, private communication).
}
\end{figure}

There are fewer stars (7 with good fits)  in common with our studies and the TLUSTY analysis of Bouret et al.\ (2003) and Heap et al.\ (2006), but here we find a much larger
range of differences (from $-3000$ K to +3500 K), with a median difference of 1000 K.  
Thus the differences we saw in Paper II between
our temperature calibration  and the results for their stars  were not due to a systematic effect
between the different modeling techniques, but rather due to issues with particular stars.
For example, as we noted above, Heap et al.\ (2006) retain the star NGC 346-368 in their determination of
their effective temperature scale, even though they (and we) conclude it is a binary.
In some cases the adopted spectral types
do not match: for NGC 346-368
they adopt the 
O4-5 V ((f)) type of Walborn et al.\  (2000), while we find a type of O6~V here. This type mismatch could be in part caused by the lack
of nebular subtraction in the Walborn et al.\ (2000) data, and in other cases could be the result of improper sky subtraction.
As noted in Paper II,
the AAT Echelle spectra used by Walborn et al.\ (2000) for spectral classification were incorrectly corrected for moonlight contamination: they describe
a procedure which removed the solar line absorption spectrum, but which would have left the moonlight continuum contribution contaminating
the spectrum.  Chris Evans (private communication, 2005) reports that he was subsequently able to correct the spectra for most stars used
in the analysis by Bouret et al.\ (2003) and Heap et al.\ (2006), but that there were insufficient sky pixels for sky subtraction for NGC 346-368
and NGC 346-487.  Yet these stars were included in their analysis.  How large a difference did this make in practice?  We give a comparison for NGC346-487 in
Fig.~\ref{fig:heap}.  The AAT lines are weaker, as would be expected from uncorrected night sky contamination.   The differences in EWs
are significant: in the AAT data we measure 240 m\AA\ for He II $\lambda 4200$,   270 
m\AA\  for He I $\lambda 4387$, and and 300 m\AA\  for
He II $\lambda 4542$, while in our own data we measure 530 m\AA, 350 m\AA, and 615 m\AA, respectively: factors of 2.2, 1.3, and 2.1 higher.
Thus, modeling involving these lines would invariably lead to discordant results
compared to ours. Clearly He I $\lambda 4471$ would not provide much
constraints in the AAT data.  Recall that Bouret et al.\ (2003) were unable to obtain a good fit to the optical spectrum of this star,
and had to rely primarily on their UV data. We derive very different results for this star, with our measurement 3000 K hotter
and a $\log g$ that is 0.2~dex higher, as shown in Table~\ref{tab:compA}.

What about the other stars?  The other star with no sky subtraction in their data, NGC 346-368, we find to be a binary.    For NGC 346-324, our
results are in very good agreement, with our temperature 500 K hotter.  For that star we measure EWs of 200 m\AA\ and 760 m\AA\ for
He I $\lambda 4471$ and He II $\lambda 4542$, respectively, compared to 250 m\AA\ and 760 m\AA\ in our own data, in reasonable agreement.
The difference is in the opposite sense than expected, though, in that our slightly stronger line should lead to a slightly cooler temperature,
everything else being equal.  For NGC 346-355 we are unable to measure He I in their spectra; we detect this line at the 50 m\AA\ level
thanks to our much higher SNRs.  For He II $\lambda 4542$ we measure an EW of 
685 m\AA\ in their data, while in ours we
measure 720 m\AA, again in substantial agreement.  The large difference in the derived effective temperature (our 49500 K value vs their
52500 value) is presumably due to the fact that we were able to detect He I, which constrained the effective temperature.  
Bouret et al.\ (2003) did also model UV data, which provided additional constraints; we plan such an extension to our work in
the future, as discussed further below.

\subsection{Effective Temperature Scale}

In Paper II we presented a revised effective temperature scale for O-type stars, based upon our analysis with FASTWIND of a sample of
O stars in the SMC and LMC,  as well as the study by Repolust et al.\ (2004) of Galactic stars.  We presented in tabular form our revised
scale for the SMC supergiants, SMC giants and dwarfs (which appeared to be indistinguishable), Galactic supergiants, and
Galactic giants and dwarfs.   Unsurprisingly we found
a substantial difference between the SMC and Galactic effective temperature calibrations, with the SMC scale significantly hotter
(3000-4000 K) for the early O's, with negligible differences by B0.  This is in accord with the expected effects of wind- and line-blanketing,
which will be less significant at lower metallicities.  The data were too scant to determine a scale for the LMC stars, although our data
showed that the temperatures for LMC stars were intermediate between those
for SMC and Milky Way stars, as would be expected from their metallicity.

\begin{figure}
\epsscale{0.5}
\plotone{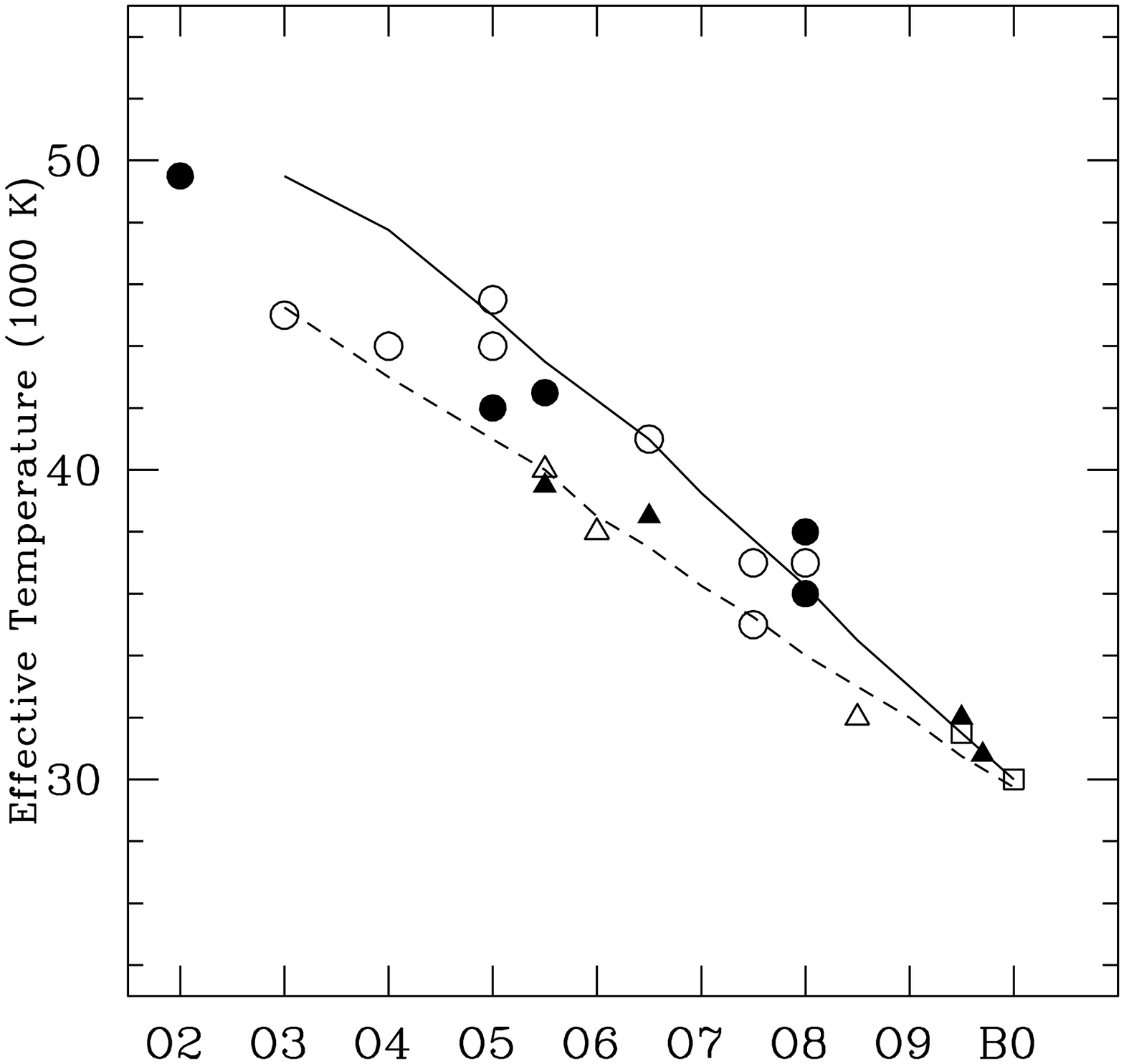}
\plotone{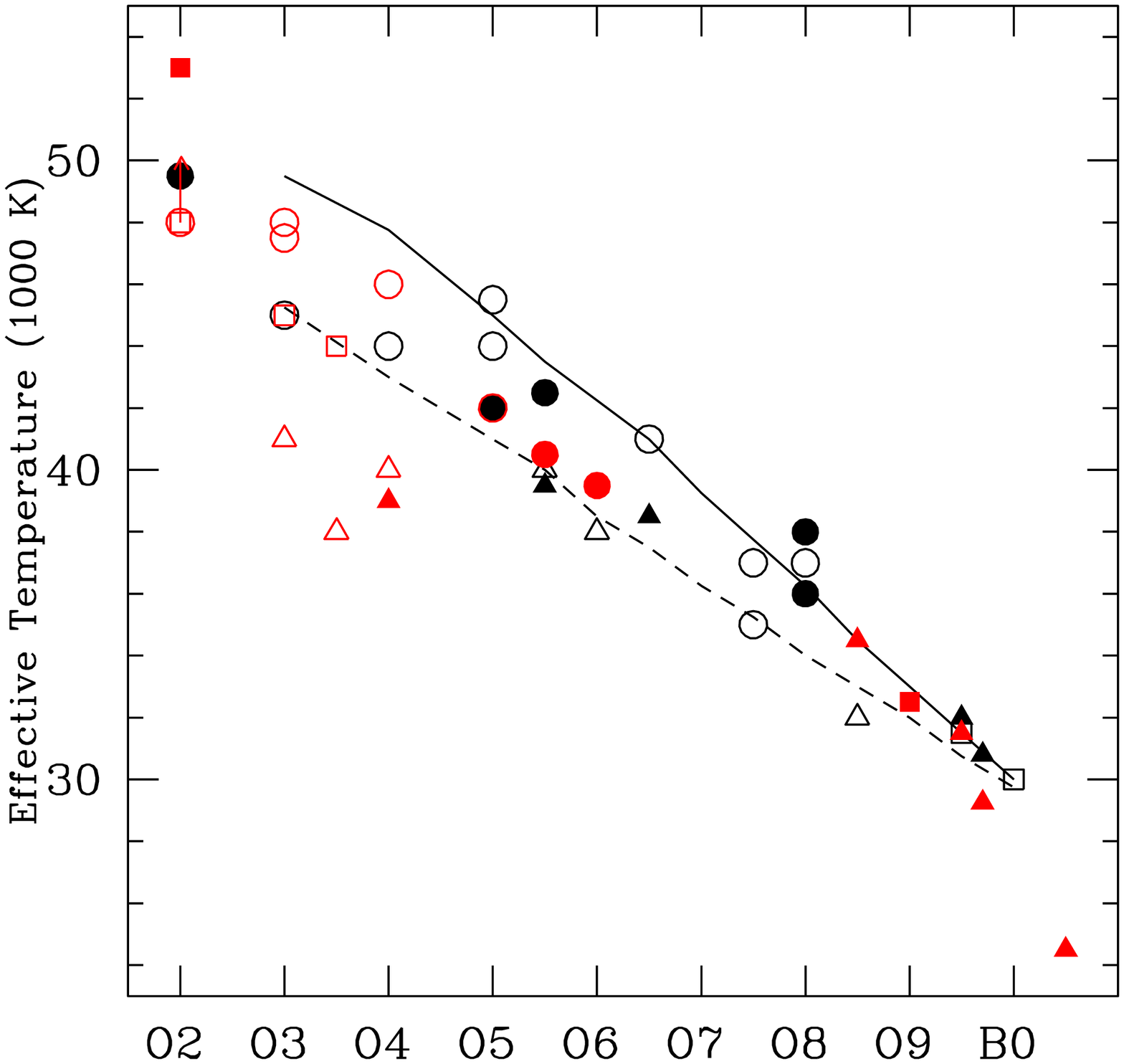}
\caption{\label{fig:teff}  Effective temperature as a function of spectral types.  {\it Upper.} The data for the SMC stars are shown here,
with filled symbols representing the new data from this paper, and open symbols representing the data from Papers I and II.  Circles are
dwarfs, boxes are giants, and triangles are supergiants.  The solid line shows the Paper II
calibration for SMC dwarfs and giants (corresponding to the circles and squares), and the
dashed line that for supergiants (i.e., corresponding to triangles).  {\it Lower.} Same as for the top, but now with the LMC data added
in red.}

\end{figure}

With the large sample of stars, we revisit this issue here.  In Fig~\ref{fig:teff} (upper) we show now all of the SMC data from Papers I, II, and the
present work, along with the SMC effective temperature calibration.  Filled symbols show data from the present paper in order to demonstrate
how the current study has improved our coverage of spectral types.   Dwarfs are indicated by circles, giants by squares, and supergiants by
triangles.  The proposed dwarf and giant effective temperature scale is shown by a solid line, and that of the supergiants by a dashed line.

We see that the new data are in reasonable accord with the adopted relationship.  If anything the dwarf and giant sequence may be a little
 hotter than it should be for the earlier types.   Note that even with the new data there are no constraints on the supergiant sequence
for stars earlier than O5 I.

What of the LMC stars?  In Fig.~\ref{fig:teff} (lower) we now add these stars to the mix using red to distinguish them from the SMC stars, but otherwise
retaining the same symbols.  
For simplicity we have excluded the O2-3.5 stars of uncertain types from Papers I and II.  We see that there is little to no difference
between LMC and SMC stars by the late O/early B types (even, surprisingly, for supergiants), but that for earlier types the LMC stars
{\it are} cooler than their LMC counterparts.    Again, this is what is expected.  We also see with the new data that giants may
indeed be intermediate in effective temperature between dwarfs and supergiants, at least for the earlier types.

What is needed to further refine these relationships?  For the LMC stars we need analysis of more mid- and late-type O dwarfs, and
mid-type supergiants.  For the SMC, analysis of early-type supergiants would be useful.
For now, we retain the effective temperature scale laid out in Table 9 of Paper II.

\subsection{The Mass Discrepancy Revisited}

Groenewegen et al.\ (1989) were the first to describe the ``mass discrepancy", which was subsequently studied extensively by
Herrero et al.\ (1992).   The ``spectroscopic mass" follows directly from the results of modeling the spectra, with
$M_{\rm spect}=(g_{\rm true}/g_\sun)R^2$, where the mass and the radius are expressed in solar units).  However, the effective temperature and luminosity from the modeling also imply a mass,
which we call the ``evolutionary mass" ($M_{\rm evol}$) from the evolutionary tracks.   The mass discrepancy refers to the systematic
difference between these two.  Early studies found that the spectroscopic mass was systematically less than the
evolutionary mass, by as much as a factor of two for supergiants.  
Improvements in the stellar atmosphere models have decreased or eliminated the size of the discrepancy for Galactic stars (Herrero 2003; Repolust et al.\ 2004; Mokiem et al.\ 2005).  
As we discuss in Paper II, the expected reason
is two-fold: the new atmosphere models include line blanketing and thus result in lower effective temperatures, decreasing the inferred
luminosities, and hence the deduced evolutionary mass will be less.  In addition, the new models result in larger
photospheric radiation pressure, so a higher surface gravity is needed to reproduce the Stark-broadened wings of the Balmer lines.
(To some extent, however, this is offset by the fact that the cooler temperatures require a lower surface gravity to fit the Balmer
lines; see Repolust et al.\ 2004.)
However, in Paper II we found that there was still a
mass discrepancy for the hottest O stars in the Magellanic Clouds.  Here we revisit the issue.

\begin{figure}
\epsscale{0.5}
\plotone{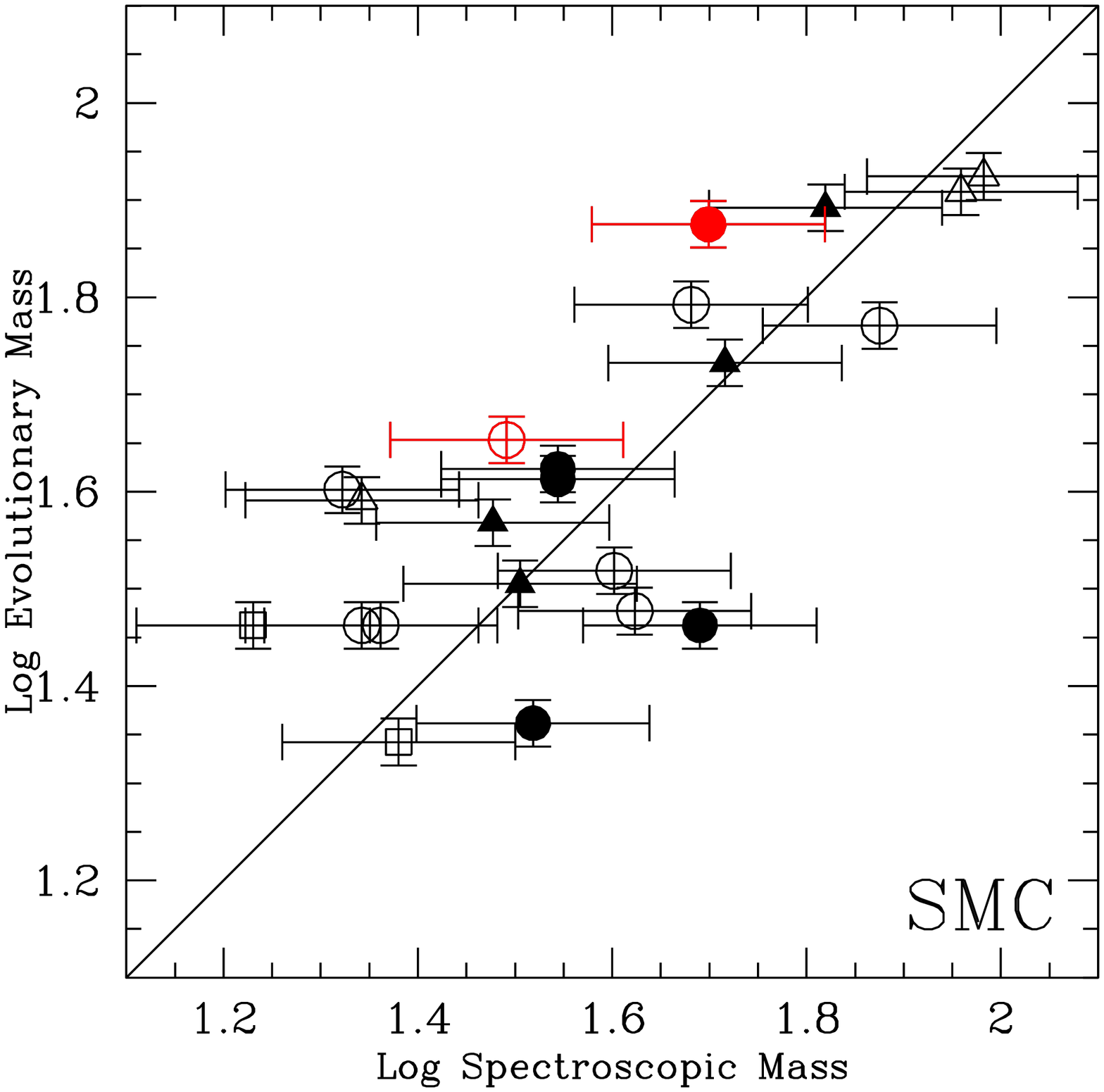}
\plotone{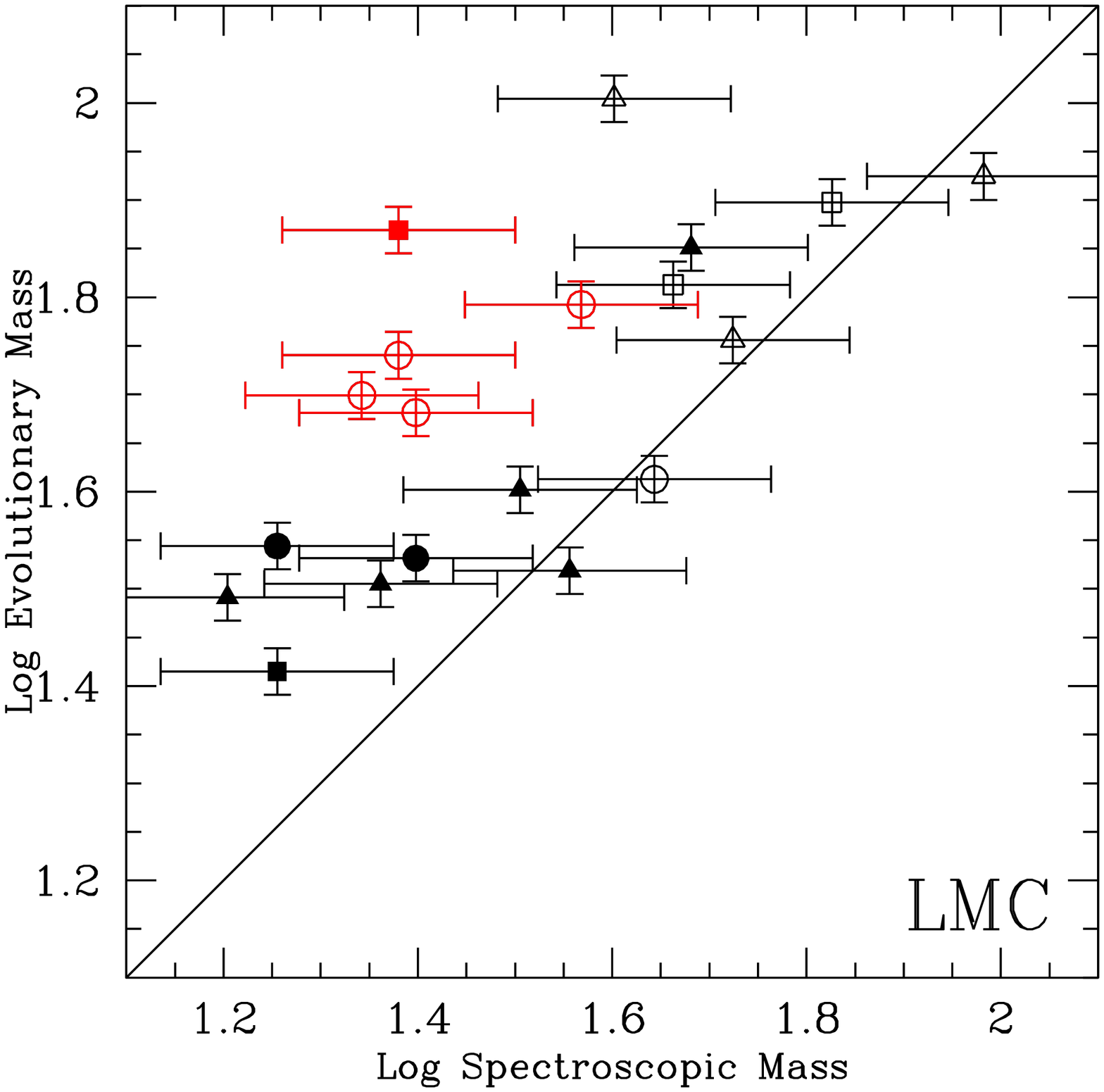}
\caption{\label{fig:mass}  The mass discrepancy for our sample.  Filled symbols denote the data new to this paper.  Circles represent
dwarfs, squares represent giants, and triangles represent supergiants.  We have indicated in red the stars for which $T_{\rm eff}$ is greater
than 45,000 K.}
\end{figure}

In Figure~\ref{fig:mass} we plot the evolutionary mass vs the spectroscopic mass.  Although
many points cluster along the 1:1 line, there are still a substantial number of points significantly
above the line.  The error bars are based on the same assumptions as in Paper II.  It is clear that the
mass discrepancy is still very much with us, at least in the Magellanic Clouds.  We have used the same
symbols as before: circles are dwarfs, squares are giants, and triangles are supergiants.  Filled
symbols denote the new results here, while open symbols come from Papers I and II.  (We have
included stars of uncertain spectral types but exclude any stars for which we derived only a lower
limit to their effective temperatures.)   In Paper II we suggested that the stars with a
significant mass discrepancy were those with the highest effective temperatures, $T_{\rm eff}>$ 45,000~K.
Here we denote such stars in red.  It is clear from the figure that this distinction is not really correct.
In part this is clarified by the addition of the new data here.  There are stars of relatively lower
mass with cooler temperatures which clearly show such an effect.  Furthermore, such stars are not
invariably supergiants, as was originally suggested for the Galactic example by Herrero et al.\ (1992).
We have searched for correlations between the size of the mass discrepancy with effective temperature and/or surface gravity, and find little connection.  As we suggest in Paper II, it may be that in a few
cases the answer is that a star has been affected by binary evolution; i.e., LH64-16, the giant
shown at the top of the LMC plot.  While the overall agreement in Fig.~\ref{fig:mass} is encouraging,
the figure shows that some improvement is still needed in either the atmosphere or evolutionary
models, or both.  We note that we have used the older, non-rotating evolutionary
models of Charbonnel et al.\ (1993) and Schaerer et al.\ (1993) in determining the
evolutionary masses, but the use of the newer, rotating 
models, such as Meynet \& Maeder (2005), (for which we do not have isochrones readily
available) would exacerbate the differences rather than reduce them, 
as argued in Paper II.

\section{Summary and Future Work}
We have analyzed the spectra of 26 O and early B stars in the Magellanic Clouds, obtaining satisfactory fits to 18 stars.  The
effective temperatures we derive are in accord with the effective temperature scale we presented in Paper II.  We emphasize that the
``Of" indicators (N III $\lambda 4634, 42$ and He II $\lambda 4686$ emission) do not track the luminosity of the star in 
precisely the same manner as at Galactic metallicities, as the causes of the emission are different.  With our very high
SNR data, we are able to detect the weak He I $\lambda 4471$ line in some of the earliest spectral types, providing strong
constraints on the effective temperatures and other physical properties of these stars.  We do find some significant differences
with the work of others: the automatic FASTWIND fitting procedure of Mokiem et al.\ (2006, 2007) results in effective temperatures
that are hotter than ours by 1100 K in the median, which is significant given that our estimated error is 500-1000~K.  More 
interesting
is the fact that Mokiem et al.\ (2006, 2007) find ``best fit" answers for stars for which we deemed no fit to be satisfactory.  On average,
the TLUSTY results of Bouret et al.\ (2003) and Heap et al.\ (2006) are 1000 K cooler than ours,  although we note
a  problem with some of their data due to moonlight continuum contamination.  There is still evidence of a discrepancy between
the masses derived by spectroscopic analysis, and those derived from the evolutionary tracks, in the sense that some stars have
a significantly lower spectroscopic mass. The problem does not seem to be
correlated with a single parameter (such as effective temperature), but is present for some stars and not for others.

In Paper II we argue that
the results from the literature suggest that there may be differences dependent upon what wavelength region is analyzed, and
even possibly what programs are used.  The trouble with such comparisons in the past is that they have been based on samples
with little or no overlap of individual stars.  We plan to now analyze the UV spectra of the same stars analyzed in Papers I, II, and the
present work to see how fits to the UV data compare with our
FASTWIND optical results.  In addition, we will analyze the same optical spectra by other means (e.g., CMFGEN, Hillier et al.\ 2003).

\acknowledgments
It is a pleasure to acknowledge the support of the Las Campanas Observatory in obtaining the
optical spectra.
Support for programs {\it HST} GO-9412, GO-9795, and AR-11270 was provided by NASA through grants from the Space Telescope Science Institute, which is operated by the Association of Universities for Research in Astronomy, Inc., under NASA contract NAS 5-26555.  A.M.Z.'s work was supported
through a National Science Foundation  REU grant, AST-0453611.  We are grateful to
Chris Evans for clarifying several issues for us in past studies of some of the stars
discussed here, as well as comments on an early draft.
 An anonymous referee made many useful suggestions which improved the paper.

\begin{deluxetable}{l l l c c c c c c c c l }
\tabletypesize{\tiny}
\rotate
\tablecaption{\label{tab:sample} Sample of Stars}
\tablewidth{0pt}
\tablehead{
\colhead{Star\tablenotemark{a}}
&\colhead{Type\tablenotemark{b}}
&\colhead{Galaxy}
&\colhead{$\alpha_{\rm 2000}$}
&\colhead{$\delta_{\rm 2000}$}
&\colhead{$V$}
&\colhead{$U-B$}
&\colhead{$B-V$}
&\colhead{$A_V(B-V)$\tablenotemark{c}}
&\colhead{$A_V(U-B)$\tablenotemark{c}}
&\colhead{$M_V$\tablenotemark{d}}
&\colhead{Comments} 
}
\startdata
AzV 15      &O6.5 I(f)        &SMC  &00 46 42.17 & -73 24 55.2 & 13.12 & -1.03 & -0.19 & 0.37 & 0.56 &-6.15 &\nodata \\
AzV 61      &O5.5 III((f))   &SMC  &00 50 01.77 & -72 11 26.0 & 13.54 & -1.05 & -0.18 & 0.43 & 0.60 &-5.79 & \nodata \\
AzV 75      &O5.5 I(f)      &SMC  &00 50 32.39 & -72 52 36.1 & 12.70 & -1.00 & -0.15 & 0.50 & 0.73 &-6.70 &\nodata \\
AzV 83      &O7 Iaf      &SMC  &00 50 52.05 & -72 42 14.8 & 13.31 & -0.96 & -0.12 & 0.59 & 0.78 &-6.18 & \nodata \\
NGC346-324  &O5 V        &SMC  &00 58 57.38\tablenotemark{e} & -72 10 33.7\tablenotemark{e} & 14.02\tablenotemark{f} & -1.05\tablenotemark{f} & -0.24\tablenotemark{f} & 0.25 & 0.60 &-5.13 &NGC 346 W6 \\
NGC346-342  &O5.5 III((f))      &SMC  &00 59 00.04\tablenotemark{e} & -72 10 37.9\tablenotemark{e} & 13.66\tablenotemark{f} & -1.05\tablenotemark{f} & -0.23\tablenotemark{f} & 0.28 & 0.52 &-5.52 &NGC 346 W4  \\
NGC346-355  &O2 V       &SMC  &00 59 00.75\tablenotemark{e} & -72 10 28.2\tablenotemark{e} & 13.50\tablenotemark{f} & -1.12\tablenotemark{f} & -0.23\tablenotemark{f} & 0.31 & 0.34 &-5.71 &NGC 346 W3 \\
NGC346-368  &O6 V        &SMC  &00 59 01.81\tablenotemark{e} & -72 10 31.3\tablenotemark{e} & 14.18\tablenotemark{f} & -1.08\tablenotemark{f} & -0.23\tablenotemark{f} & 0.28 & 0.43 &-5.00 &\nodata \\
NGC346-487  &O8 V        &SMC  &00 59 06.71\tablenotemark{e} & -72 10 41.3\tablenotemark{e} & 14.53\tablenotemark{f} & -1.01\tablenotemark{f} & -0.22\tablenotemark{f} & 0.28 & 0.56 &-4.65 &\nodata \\
AzV 223          &O9.5 II    &SMC  &00 59 13.41 & -72 39 02.2 & 13.66 & -0.98 & -0.14 & 0.40 & 0.43 &-5.64 &\nodata \\
NGC346-682  &O8 V        &SMC  &00 59 18.63\tablenotemark{e} & -72 11 10.0\tablenotemark{e} & 14.87\tablenotemark{f} & -1.02\tablenotemark{f} & -0.24\tablenotemark{f} & 0.22 & 0.52 &-4.25 &\nodata \\
AzV 232     &O7 Iaf+     &SMC  &00 59 31.97\tablenotemark{e} & -72 10 46.1\tablenotemark{e} & 12.31\tablenotemark{f} & -1.05\tablenotemark{f} & -0.19\tablenotemark{f} & 0.37 & 0.39 &-6.96 & Sk 80, NGC346-789 \\
AzV 327     &O9.7 I        &SMC  &01 03 10.49 & -72 02 14.2 & 13.03 & -1.04 & -0.16 & 0.37 & 0.43 &-6.24 &\nodata \\
AzV 388     &O5.5 V((f)) &SMC  &01 05 39.45 & -72 29 26.8 & 14.09 & -1.07 & -0.21 & 0.34 & 0.52 &-5.15 &\nodata \\
BI 9        &O7.5 III((f))   &LMC  &04 52 47.95 & -68 47 39.1 & 13.70 & -0.99 & -0.22 & 0.31 & 0.65 &-5.11 &\nodata \\
LMC054383   & O4-5 V((f))pec       &LMC  &05 02 09.94 & -70 32 58.1 & 14.10 & -1.08 & -0.19 & 0.43 & 0.52 &-4.83 &\nodata \\
Sk -70 60   &O4-5 V((f))pec       &LMC  &05 04 40.88 & -70 15 34.7 & 13.88 & -1.09 & -0.16 & 0.53 & 0.47 &-5.15 &\nodata \\
Sk -70 69   &O5.5 V((f))&LMC  &05 05 18.69 & -70 25 50.3 & 13.95 & -1.09 & -0.23 & 0.28 & 0.43 &-4.83 &\nodata \\
Sk -68 41   &B0.5 Ia     &LMC  &05 05 27.20\tablenotemark{e} & -68 10 02.8\tablenotemark{e} & 12.04\tablenotemark{g} & -0.93\tablenotemark{g} & -0.14\tablenotemark{g} & 0.25 & 0.47 &-6.71 &\nodata \\
Sk -69 124  &O9.7 I      &LMC  &05 25 18.35 & -69 03 11.3 & 12.77 & -1.03 & -0.16 & 0.34 & 0.21 &-6.07 &\nodata \\
BI 170      &O9.5 I      &LMC  &05 26 47.79 & -69 06 11.9 & 13.06 & -1.04 & -0.20 & 0.22 & 0.17 &-5.66 &\nodata \\
BI 173      &08.5 II(f)    &LMC  &05 27 10.05 & -69 07 56.3 & 12.96 & -1.02 & -0.16 & 0.39 & 0.52 &-5.93 &\nodata \\
LH64-16     &ON2 III(f*)   &LMC  &05 28 46.97 & -68 47 47.9 & 13.61 & -1.11 & -0.22 & 0.34 & 0.39 &-5.23 &\nodata \\
Sk -67 166  &O4 If       &LMC  &05 31 44.30 & -67 38 01.0 & 12.27\tablenotemark{g} & -1.01\tablenotemark{g} & -0.22\tablenotemark{g} & 0.31 & 0.73 &-6.54 & HD 269698\\
BI 192      &O9 III      &LMC  &05 32 00.00 & -67 32 55.4 & 13.75 & -1.03 & -0.22 & 0.28 & 0.43 &-5.03 &\nodata \\
BI 208      &O6 V((f))   &LMC  &05 33 57.38 & -67 24 20.3 & 13.94 & -1.09 & -0.22 & 0.31 & 0.39 &-4.87 &\nodata \\
\enddata
\tablecomments{Units of right ascension are hours, minutes, and seconds, and
units of declination are degrees, arcminutes, and arcseconds.  Coordinates and
photometry from Massey 2002 unless otherwise noted.}
\tablenotetext{a}{Star names are from the following catalogs: ``AzV" from Azzopardi \& Vigneuau 1975; 
``NGC 346" from Massey et al.\ 1989, and known to CDS as Cl* ``NGC 346 MPG"; ``Sk", from Sanduleak 1970;
``BI" from Brunet et al.\ 1975; ``LMC" from Massey 2002, and known to CDS as [M2002]; ``LH64-16", from
Lucke 1973; see also Massey et al.\ 2000.}
\tablenotetext{b}{From this paper.}
\tablenotetext{c}{$A_V(B-V)=3.1[(B-V)-(B-V)_0]$, and $A_V(U-B)=4.3[(U-B)-(U-B)_0]$, where 
$(B-V)_0$ and $(U-B)_0$ are the intrinsic colors for stars of a given spectral type, taken from
Massey 1998b.}
\tablenotetext{d}{Based upon $V-3.1E(B-V)-$DM, where the distance modulus (DM) is assumed
to be 18.9 for the SMC and 18.5 for the LMC.}
\tablenotetext{e}{Coordinates newly determined in this paper.}
\tablenotetext{f}{Stars in NGC 346 have photometry from Massey et al.\ 1989.}
\tablenotetext{g}{Photometry taken from Buscombe \& Foster 1995.}
\end{deluxetable}

\begin{deluxetable}{l l l c c c}
\tabletypesize{\footnotesize}
\tablecaption{\label{tab:vinf} Adopted Terminal Velocities $v_\infty$}
\tablewidth{0pt}
\tablecolumns{6}
\tablehead{
\colhead{Star}
&\colhead{Type}
&\colhead{Galaxy}
&\colhead{$v_\infty$}
&\colhead{Ref.}
&\colhead{$2.6 v_{\rm esc}$} \\
&&&\colhead{[km s$^{-1}$]}&&\colhead{[km s$^{-1}$]}
}
\startdata
AzV 15      &O6.5 I(f)        &SMC  & 2125 & 1 & 2100\\
AzV 61      &O5.5 III((f))   &SMC & 2025 & 1 & \nodata\\
AzV 75      &O5.5 I(f)      &SMC  & 2120 & 2  & 1860 \\
AzV 83      &O7 Iaf      &SMC  & 930 & 1 & \nodata \\
NGC346-324  &O5 V        &SMC &2300 & 3 & 2450\\
NGC346-342  &O5.5 III((f))      &SMC  &1945 & 4 & \nodata\\
NGC346-355  &O2 V       &SMC  & 2800 & 3 & 2200\\
NGC346-368  &O6 V        &SMC & 2100 & 3 & \nodata \\
NGC346-487  &O8 V        &SMC &$\geq$1100 &3 & 3500\tablenotemark{a}\\
AzV 223          &O9.5 II    &SMC  & 1680 & 5& 1950\\
NGC346-682  &O8 V        &SMC  & \nodata &5& 2900 \\
AzV 232     &O7 Iaf+        &SMC  & 1400 & 6& \nodata  \\
AzV 327     &O9.7 I             &SMC  &1500 & 1 & 1460 \\
AzV 388     &O5.5 V((f)) &SMC & 1935 & 4 & 2440 \\
BI 9        &O7.5 III((f))   &LMC   & 1900 & 4 & \nodata\\
LMC 054383   &O4-5V((f))pec      &LMC  & 2380 & 5 & \nodata \\
Sk -70 60   &O4-5V((f))pec        &LMC  & 2150 & 4 & \nodata \\
Sk -70 69   &O5.5 V((f)) &LMC  & 2300: & 4 & 1720\\
Sk -68 41   &B0.5 Ia     &LMC  & 865 & 4 & 1300 \\
Sk -69 124  &O9.7 I      &LMC  & 1430 & 4 & 1240\\
BI 170      &O9.5 I      &LMC  & 1370 & 4 & 1120\\
BI 173      &08.5 II(f)    &LMC    & 1950: &  4 & 1630\\ 
LH64-16     &ON2 III(f*)  & LMC & 3250 & 5 & 1475 \\
Sk -67 166  &O4 If       &LMC  & 735 & 4\\
BI 192      &O9 III      &LMC  & 1100: & 4 & 1670  \\
BI 208      &O6 V((f))   &LMC & 980: & 4 & 2110\\
\enddata 
\tablerefs{
(1) Evans et al.\ 2004;
(2) Paper I;
(3) Bouret et al.\ 2003;
(4) Prinja \& Crowther 1998;
(5) This paper;
(6) Puls et al.\ 1996.}
\tablenotetext{a}{Adopted.}
\end{deluxetable}

\clearpage
\begin{deluxetable}{l l l c c c r r r r r c c c c c r c l}
\rotate
\pagestyle{empty}
\setlength{\tabcolsep}{0.06in}
\tabletypesize{\tiny}
\tablecaption{\label{tab:results} Results of Model Fits}
\tablewidth{0pt}
\tablecolumns{19}
\tablehead{
\colhead{Star}
&\colhead{Type}
&\colhead{Galaxy}
&\colhead{$v_r$}
&\colhead{$v \sin i$}
&\colhead{$T_{\rm eff}$}
&\colhead{$\log g_{\rm eff}$}
&\colhead{$\log g_{\rm true}$}
&\colhead{$R/R_\odot$}
&\colhead{$M_V$}
&\colhead{BC}
&\colhead{$\log L/L_\odot$}
&\colhead{$m_{\rm spect}$}
&\colhead{$m_{\rm evol}$}\tablenotemark{a}
&\colhead{$\dot{M}$\tablenotemark{b}}
&\colhead{$\beta$}
&\colhead{$v_\infty$}
&\colhead{He/H\tablenotemark{c}} 
&\colhead{Comments} \\
&\colhead{}
&\colhead{}
&\colhead{km s$^{-1}$} 
&\colhead{km s$^{-1}$}
&\colhead{(K)} 
&\colhead{[cgs]} 
&\colhead{[cgs]} 
&\colhead{}
& \colhead{(mag)} 
&\colhead{(mag)} 
&\colhead{}
&\colhead{$M_\odot$} 
&\colhead{$M_\odot$}
&
& \colhead{} 
& \colhead{km s$^{-1}$}
&
& \colhead{}
}
\startdata
AzV 15      &O6.5 I(f)        &SMC  & 130 & 180 & 38500 &3.6 & 3.63 & 18.3 & -6.15 & -3.66 & 5.82 & 52 & 54 & 1.8 & 0.8 & 2125& 0.10 & Previously modeled\\
AzV 61      &O5.5 III((f))   &SMC  & \nodata & \nodata & \nodata & \nodata & \nodata & \nodata & -5.79 &  \nodata & \nodata & \nodata & \nodata & \nodata & \nodata & 2025 & \nodata & Disk signature\\
AzV 75      &O5.5 I(f)      &SMC  & 140 & 180 & 39500 & 3.5 & 3.52 & 23.4 & -6.70 & -3.75 & 6.08 & 66 &78 & 3.0 & 0.8 & 2120 & 0.10 & Previously modeled \\
AzV 83      &O7 Iaf      &SMC    & \nodata & \nodata & \nodata  & \nodata & \nodata & \nodata & -6.18 & \nodata & \nodata & \nodata & \nodata & \nodata & \nodata & 926 & \nodata & Prev.\ modeled but binary?\\
NGC346-324  &O5 V        &SMC & 170 & 150 & 42000 & 3.9 & 3.91 & 10.8 & -5.13 & -3.91 & 5.51 & 35 & 41 & 0.6 & 0.8 & 2300 & 0.10 & Previously modeled \\
NGC346-342  &O5.5 III((f))      &SMC & \nodata & \nodata & \nodata & \nodata & \nodata & \nodata & -5.52 & \nodata & \nodata & \nodata & \nodata & \nodata & \nodata & 1945 & \nodata & Ecl. binary\\
NGC346-355  &O2 V       &SMC  & 190 & 140 & 49500 & 3.9 & 3.92 & 12.8 & -5.71 & -4.42 & 5.95 & 50 & 75 & 2.0 & 0.8 & 2800 & 0.10 & Previously modeled\\
NGC346-368  &O6 V        &SMC &  \nodata & \nodata & \nodata & \nodata & \nodata & \nodata & -5.00 & \nodata &\nodata & \nodata & \nodata & \nodata & \nodata &2100 & \nodata & Prev.\ modeled but binary?\\
NGC346-487  &O8 V        &SMC &  170 & 160 & 38000 & 4.2 &  4.21 & 9.1 & -4.65 & -3.58 & 5.19 & 49 & 29 &0.1 & 0.8 & 3500 & 0.10 & Prev.\ modeled \\
AzV 223          &O9.5 II    &SMC  & 190 & 140 & 32000 & 3.5 & 3.52 & 16.4 & -5.64 & -3.12 & 5.40 & 32 & 32 & 0.5 & 0.8 &1680 & 0.10\\
NGC346-682  &O8 V        &SMC  & 180 & 150 & 36000 & 4.1 & 4.11 & 7.8 & -4.25 & -3.41 & 4.97 & 33 & 23 & 0.05 & 0.8 & 2900 & 0.10& Prev.\  modeled\\
AzV 232     &O7 Iaf+        &SMC  & \nodata & \nodata & \nodata & \nodata & \nodata & \nodata & -6.96 &  \nodata &\nodata & \nodata & \nodata & \nodata & \nodata & 1400 & \nodata & Prev.\ modeled but binary?\\
AzV 327     &O9.7 I             &SMC  & 190 & 150 & 30800 & 3.2 & 3.23 & 22.1 & -6.24 & -3.01 & 5.60 & 30 & 37 &0.7 & 0.8 & 1500  & 0.15 &Previously modeled \\
AzV 388     &O5.5 V((f)) &SMC & 190 & 190 & 42500 & 3.9 & 3.92 & 10.8 & -5.15 & -3.94 & 5.53 & 35 &42 & 0.3 & 0.8 & 1935 & 0.10 & Prev.\ modeled \\
BI 9        &O7.5 III((f))   &LMC  & \nodata & \nodata & \nodata & \nodata & \nodata & \nodata & -5.46 &  \nodata &\nodata & \nodata & \nodata & \nodata & \nodata &  1900 & \nodata & No good fit---binary?\\
LMC054383   &O4-5V((f))pec        &LMC & \nodata & \nodata & \nodata & \nodata & \nodata & \nodata  &-4.83& \nodata&\nodata & \nodata & \nodata & \nodata & \nodata & 2380& \nodata & Disk signature \\
Sk -70 60   &O4-5 V((f))pec &LMC & \nodata & \nodata & \nodata & \nodata & \nodata & \nodata & -5.15&\nodata &\nodata & \nodata & \nodata & \nodata & \nodata & 2150& \nodata & Disk signature \\     
Sk -70 69   &O5.5 V((f)) &LMC  & 285 & 165 & 40500 & 3.7 & 3.73 & 9.6 & -4.83 & -3.79 & 5.35 & 18 & 35& 0.8 & 0.8 &   2300: & 0.10& Prev.\ modeled \\
Sk -68 41   &B0.5 Ia     &LMC  & 270 & 150 & 24500 &  2.9 & 2.95 & 33.4 & -6.71 & -2.44 & 5.56 & 36 & 33 & 0.9 & 2.0 &  865 & 0.10  \\
Sk -69 124  &O9.7 I      &LMC  & 190 & 160 & 29250 & 3.1 &  3.15 & 21.3 & -6.07 & -2.88 & 5.48 & 23 & 32 & 0.7 & 1.2 &1430 & 0.10 \\
BI 170      &O9.5 I      &LMC  & 290 & 150 & 31500 & 3.15 & 3.20 & 16.6 &  -5.66 & -3.06 & 5.39 & 16 & 31 & 0.8 & 1.2 & 1370 & 0.15 \\
BI 173      &08.5 II(f)    &LMC    & 285 & 200 & 34500 & 3.4 & 3.45 & 17.6 & -5.93 &  -3.31 & 5.60 & 32 & 40 &1.6 & 0.8 & 1950: &  0.10\\ 
LH64-16     &ON2 III(f*)  & LMC & 290 & 150 & 53000 & 3.8 & 3.82 & 9.9 & -5.23 & -4.62 & 5.84 & 24 & 74 &4.0 & 0.8 & 3250 & 0.50 & Prev.\ modeled \\
Sk -67 166  &O4 If       &LMC  & 290 & 180 & 39000 & 3.4 &  3.43  & 22.1 & -6.54 & -3.73 & 6.01 & 48 &71 & 4.5 & 0.8 & 735 & 0.20& Prev.\ modeled\\
BI 192      &O9 III      &LMC  & 300 & 140 & 32500 & 3.5 & 3.53 &12.1 & -5.03 & -3.13 & 5.17 & 18 & 26 & 0.1 & 0.8 & 1100: & 0.10 \\
BI 208      &O6 V((f))   &LMC & 320 &240 & 39500 & 3.8 & 3.85 & 9.88 & -4.87 & -3.71 & 5.33 & 25 & 34 & 0.3 & 0.8 & 980: & 0.10 \\
\enddata
\tablenotetext{a}{From the nonrotating models of Charbonnel et al.\ (1993; SMC) and Schaerer et al.\ (1993; LMC).}
\tablenotetext{b}{Units of $10^{-6} M_\odot$ yr$^{-1}$.  The true mass-loss rates will be these
values corrected by $\sqrt{1/f}$, where $f$ is the unknown ``clumping" factor, probably of order
6-10 or less; see the recent workshop proceedings by Hamann et al.\ (2008).}
\tablenotetext{c}{By number.}
\end{deluxetable}

\clearpage

\begin{deluxetable}{l l  l l l l l l l }
\tabletypesize{\tiny}
\tablecaption{\label{tab:compA} Comparisons  of Current Sample with Previous Modeling}
\tablewidth{0pt}
\tablecolumns{9}
\tablehead{
\colhead{Type}
&\colhead{$T_{\rm eff}$}
&\colhead{$\log g_{\rm eff}$}
&\colhead{$log L/L_\odot$\tablenotemark{a}}
&\colhead{$\dot{M}$\tablenotemark{b}}
&\colhead{$\beta$}
&\colhead{He/H\tablenotemark{c}} 
&\colhead{Program} 
&\colhead{Ref.} 
}
\startdata
\cutinhead{AzV 15}
O6.5 I(f) & 38500 & 3.6 & 5.82 & 1.8 & 0.8 & 0.10 & FASTWIND & This study\\
O7 II       &  39400 & 3.7 & 5.82 &  1.1 & 0.8 & 0.10 & FASTWIND & 6\\
O6.5 II(f) & 37000 & 3.5 & 5.6 & \nodata & \nodata & 0.10 & TLUSTY & 1\\
\cutinhead{AzV 75}
O5.5 I(f) & 39500 & 3.5 & 6.08 & 3.0 & 0.8 & 0.10 & FASTWIND & This study\\
O5.5 I(f)  & 40000 & 3.6 & 6.18 & 3.5 & 0.8 & 0.10 & FASTWIND & Paper I \\
O5 III(f+)  & 38500 & 3.5 & 5.9 & \nodata & \nodata & 0.10 & TLUST & 1 \\
\cutinhead{AzV83}
O7 Iaf & \nodata & \nodata & \nodata & \nodata & \nodata & \nodata &FASTWIND & This study---binary?\\
O7 Iaf+ & 35000 & 3.25 & 5.5 & \nodata & \nodata & 0.20 & TLUSTY & 1\\
O7 Iaf+  & 32800 & 3.25 & 5.46 & 2.3\tablenotemark{d} & 2.0 & 0.20 & CMFGEN & 2 \\
\cutinhead{NGC 346-324}
O5 V & 42000 & 3.9 & 5.51 & 0.6 & 0.8 & 0.10 & FASTWIND   & This study\\
O4 V ((f+))  & 42800 & 3.9 & 6.02 & 0.2 & 0.8 & 0.08 & FASTWIND & 6---Binary \\
O4 V ((f)) & 41500 & 4.0 & 5.44 & 0.3 & 1.0 & 0.10 & TLUSTY /CMFGEN& 3 \\
O4 V ((f)) & 40000 & 3.7 & 5.46 & $\le$0.3& 1.0 & 0.10 & Non-blanketed models & 4 \\
\cutinhead{NGC346-355}
O2 V & 49500 & 3.9 & 5.95 & 2.0 & 0.8 & 0.10 & FASTWIND & This study \\
ON2 III(f*) & 52500 & 4.0 & 5.96 & 2.5 & 0.8 & 0.10 & TLUSTY/CMFGEN & 3 \\
ON2 III(f*) & 52500 & 4.0 & 5.99 & 1.6 & 1.0 & 0.10 & CMFGEN & 5 \\
O3 III f* &   55000 & 3.9 & 6.02 & 2.3 & 0.8 & 0.10 & Non-blanketed models & 4 \\
\cutinhead{NGC346-368}
O6 V & \nodata & \nodata & \nodata & \nodata & \nodata & \nodata &FASTWIND & This study---binary?\\
O4-5 V((f)) & 40000 & 3.8 & 5.34 & 0.2 & 1.0 & 0.10 & TLUSTY/CMFGEN & 3---Binary?\\
\cutinhead{NGC346-487}
O8 V & 38000 & 4.2 & 5.19 & 0.1      & 0.8 & 0.10 & FASTWIND & This study \\
O8 V & 35000 & 4.0 & 5.08 & $<$0.1 & 1.0 & 0.10 & TLUSTY/CMFGEN & 3  \\ 
\cutinhead{NGC346-682}
O8 V & 36000 & 4.1   & 4.97 & 0.05 & 0.8 & 0.10 & FASTWIND & This study\\
O9 V & 36800 & 4.2  &  4.95 & 1.1   & 0.8 & 0.10 & FASTWIND & 6 \\
O9-O9.5 V &  32500 & 3.75 & 5.0 & \nodata & \nodata & 0.1 & TLUSTY & 1 \\
\cutinhead{AzV 232}
O7 Iaf+ & \nodata & \nodata & \nodata & \nodata & \nodata & \nodata &FASTWIND & This study---binary?\\
O7 Iaf+ & 34100 & 3.4 & 6.02 & 6.0 & 1.2 & 0.24 & FASTWIND & 6\\
O7 Iaf+ & 32000 & 3.1 & 5.85 & 4.5 & 1.65 & 0.2 & CMFGEN   & 7 \\
O7 Iaf+ &  37500 & 3.2 & 6.11 & 5.5 & 1.4 & 0.20 & Non-blanketed models & 4 \\
\cutinhead{AzV 327}
O9.7 I & 30800 & 3.2 & 5.6 & 0.7 & 0.8 & 0.15 & FASTWIND & This study \\
O9.5 II-Ibw & 30000 & 3.25 &  5.3 & \nodata & \nodata & 0.1 & TLUSTY & 1 \\
\cutinhead{AzV 388}
O5.5 V ((f)) & 42500 & 3.9   & 5.53     & 0.3   & 0.8 & 0.10  & FASTWIND & This study \\
O4 V &           43000 & 3.9  & 5.55 & 0.33 & 0.8 & 0.09 & FASTWIND & 6 \\
O4 V  &          48000 & 3.70 & 5.66 & 0.17 & 1.0 & 0.10 & Non-blanketed models & 4 \\
\cutinhead{Sk$-70^\circ$69}
O5.5 V((f)) & 40500 & 3.7 & 5.35 & 0.8 & 0.8 & 0.10 & FASTWIND & This study\\
O5 V  &         43200 & 3.9 & 5.41 & 1.1 & 0.78 & 0.17 & FASTWIND & 8 \\
\cutinhead{LH64-16}
ON2 III(f*) & 53000 & 3.8 & 5.84 & 4.0 & 0.8 & 0.50  & FASTWIND & This study \\
ON2 III(f*)  & 54500 & 3.9 & 5.86 & 4.0 & 0.8 & 0.60 & FASTWIND & Paper I \\
ON2 III(f*)  &  55000 & 4.0  & 5.90 & 1.6 & 1.0 & 0.25 & CMFGEN & 5 \\
\cutinhead{Sk $-67^\circ 166$}
O4 If      &  39000 & 3.8  & 6.01 &  4.5  & 0.8  & 0.20 & FASTWIND & This paper \\
O4 Iaf+ & 40300 & 3.7 & 6.03   & 9.3 & 0.9   & 0.28 & FASTWIND & 8 \\
O4 If*    & 47500 & 3.6 & 6.24  & 13.0 & 0.7 & 0.10 & Non-blanketed models& 4\\
\enddata
\tablerefs{
(1) Heap et al.\ 2006;
(2) Hillier et al.\ 2003;
(3) Bouret et al.\ 2003;
(4) Puls et al.\ 1996;
(5) Walborn et al.\ 2004;
(6) Mokiem et al.\ 2006;
(7) Crowther et al.\ 2002;
(8) Mokiem et al.\ 2007.
}
\tablenotetext{a}{Corrected to a true distance modulus of 18.9 (60.3~kpc) for the SMC.  Previous studies of LMC stars have assumed
the same true distance modulus as we do, 18.5 (50.1~kpc).}
\tablenotetext{b}{In units of $10^{-6} M_\odot$ yr$^{-1}$.  These are all for ``unclumped" (homogeneous) stellar winds.}
\tablenotetext{c}{By number.}
\tablenotetext{d}{Corrected to an ``unclumped" wind mass-loss rate by the factor of 3.3.}
\end{deluxetable}

\begin{deluxetable}{l l  l l l l l l l }
\pagestyle{empty}
\tabletypesize{\tiny}
\tablecaption{\label{tab:compB} Comparisons of Other Stars with Previous Modeling}
\tablewidth{0pt}
\tablecolumns{9}
\tablehead{
\colhead{Type}
&\colhead{$T_{\rm eff}$}
&\colhead{$\log g_{\rm eff}$}
&\colhead{$log L/L_\odot$}
&\colhead{$\dot{M}$\tablenotemark{a}}
&\colhead{$\beta$}
&\colhead{He/H\tablenotemark{b}} 
&\colhead{Program} 
&\colhead{Ref.} 
}
\startdata

\cutinhead{AzV 14}
O5 V & 44000 & 4.0 & 5.85 & 0.1 & 0.8 & 0.10& FASTWIND & Paper I\\
O5 V & 45300 & 4.1 & 5.86 & 0.3 & 0.8 & 0.10 & FASTWIND & 1\\

\cutinhead{AzV 26}
O6 I(f) & 38000 & 3.5 & 6.14 & 2.5 & 0.8 &0.10 &  FASTWIND & Paper I\\
O7 III &  40100 & 3.8 & 6.17 & 1.7 & 0.8 & 0.09 &  FASTWIND & 1 \\

\cutinhead{AzV372}
Binary & \nodata & \nodata & \nodata & \nodata & \nodata & \nodata & FASTWIND & Paper 1---Binary \\
O9Iabw & 31000 & 3.2 & 5.83 & 2.0 & 1.3 & 0.11 & FASTWIND & 1 \\

\cutinhead{AzV 469}
O8.5  I(f)& 32000 & 3.1 & 5.64 & 1.8 & 0.8 & 0.20 & FASTWIND & Paper I\\
O8 II((f)) & 34000 & 3.4 & 5.70 &  1.1 & 1.2 & 0.17 & FASTWIND & 1\\

\cutinhead{BI 237}
O2 V((f*)) & 48000 & 3.9 & 5.77 & 2.0 & 0.8 & 0.10 & FASTWIND & Paper II \\
O2 V((f*)) & 53200 & 4.1 & 5.83 & 0.8 & 1.3 & 0.10 & FASTWIND & 2 \\

\cutinhead{BI 253}
O2 V ((f*)) & $>$48000 & 3.9 & $>$5.77 & 3.5 & 0.8 & 0.10 & FASTWIND & Paper II\\
O2 V ((f*)) &       53800 & 4.2 &       5.93 & 1.9  & 1.2 &  0.09 & FASTWIND & 2 \\
\enddata
\tablerefs{
(1) Mokiem et al.\ 2006;
(2) Mokiem et al.\ 2007.
}
\tablenotetext{a}{In units of $10^{-6} M_\odot$ yr$^{-1}$.  These are all for ``unclumped" (homogeneous) stellar winds.}
\tablenotetext{b}{By number.} 
\end{deluxetable}

\clearpage

\clearpage

\clearpage

\clearpage

\clearpage

\clearpage

\clearpage

\clearpage


\clearpage

\clearpage

\clearpage

\clearpage

\clearpage

\clearpage
\clearpage

\clearpage


\end{document}